\documentclass[twocolumn, nolinenumbers]{aastex631}
\usepackage[whole]{bxcjkjatype}
\usepackage{soul}
\usepackage{amsmath}
\usepackage{afterpage}
\usepackage{tabularx}
\usepackage{ltablex}
\usepackage{ulem} 
\usepackage{gensymb}
\usepackage{cleveref}
\usepackage{hyperref}

\newcommand{\Msun}{{\rm M}_{\odot}}

\newcommand{\cm}{{\mathrm{\,cm}}}

\newcommand{\second}{{\mathrm{\,s}}}

\newcommand{\km}{{\mathrm{\,km}}}

\newcommand{\kms}{{\km\second^{-1}}}

\newcommand{\gm}{{\mathrm{\,g}}}

\newcommand{\gram}{{\gm}}

\newcommand{\massden}{{\gram\cm^{-3}}}

\newcommand{\K}{{\mathrm{\,K}}}

\let\oldhat\hat
\renewcommand{\vec}[1]{\boldsymbol{#1}} 
\renewcommand{\hat}[1]{\oldhat{\boldsymbol{#1}}}

\newcommand{\rhostiff}{\rho_{\rm stiff}}
\newcommand{\nstiff}{n_{\rm stiff}}

\received{October 18, 2024}
\revised{January 6, 2025}
\accepted{January 21, 2025}

\shorttitle{Protostar Formation Through Magnetized Collapse}
\shortauthors{Das et al. (2025)}

\begin{document}

\title{How Does a Protostar Form by Magnetized Gravitational Collapse?
}

\correspondingauthor{Indrani Das, Hsien Shang}
\email{idas@asiaa.sinica.edu.tw, idas2@uwo.ca, shang@asiaa.sinica.edu.tw}

\author[0000-0002-7424-4193]{Indrani Das}
\affiliation{Institute of Astronomy and Astrophysics, Academia Sinica, Taipei 106216, Taiwan}

\author[0000-0001-8385-9838]{Hsien Shang (尚賢)}
\affiliation{Institute of Astronomy and Astrophysics, Academia Sinica,  Taipei 106216, Taiwan}

\author[0000-0001-5557-5387]{Ruben Krasnopolsky}
\affiliation{Institute of Astronomy and Astrophysics, Academia Sinica, Taipei 106216, Taiwan}


\begin{abstract}
Star formation through the dynamical magnetized collapse remains an active area of astrophysical research.
We carry out a comprehensive exploration on the magnetized gravitational collapse of a non-rotating self-gravitating initially spherically symmetric prestellar cloud core using two-dimensional nonideal magnetohydrodynamic simulations incorporating ambipolar diffusion and Ohmic dissipation. 
Our study encompasses a broader range of equations of state (EOSs) in the form of $P(\rho) \propto \rho^{\Gamma}$, with the aim of constraining the choice of EOSs for allowing star formation.
Our results reveal that the collapse with a $\Gamma$ no stiffer than $4/3$, complemented by magnetized virial theorem, allows the dynamical contraction of the prestellar core to happen continuously where a central   
point mass forms and steadily builds up its mass from the infalling envelope, with a mass accretion rate of a scale of the order of $c_{\rm s}^3/G$.  
The choice of an isothermal EOS most naturally facilitates the collapse as a magnetic analog of the inside-out collapse. In addition to that, our study exhibits that the nonisothermal magnetized collapse models with a $\Gamma$ no stiffer than 4/3 qualitatively demonstrate similar infall features to those of an isothermal EOS.
Furthermore, the collapse models with a $\Gamma$ stiffer than $4/3$ fail to ensure the sufficient cooling to allow the direct mass growth of the central point mass, thus delaying the infall.
Our work can offer deeper insights in understanding the significance of EOSs on the magnetized gravitational collapse, enabling star formation.
\end{abstract}





\section{Introduction} 
\label{sec:intro}

Magnetically supported interstellar molecular clouds are known to be the birthplaces of stars. 
The molecular cloud cores without protostars are often called as prestellar cores (gravitationally bound starless core) and considered younger than those associated with protostars, which are named as protostellar cores \citep[][and references therein]{Andre+2014}. 
When the prestellar cores become massive enough (typically number density of molecular hydrogen ${\rm n_{{H}_2}} \gtrsim 10^4 \, {\rm cm}^{-3}$), they start to contract under their self-gravity due to insufficient magnetic and/or thermal and/or turbulent support present within. 
In the most studied scenario of star formation, the removal of magnetic fields occurs through the action of ambipolar diffusion, wherein the neutrals drift relative to the ions that are anchored to the magnetic fields \citep{MestelSpitzer56,Mouschovias1976a,Mouschovias1976b,Nakano1979,Shu1983,LizanoShu1989,BasuMouschovias94,CiolekMouschovias1994,AdamsShu2007}
and Ohmic dissipation, wherein the
inductive effect of the plasma is restricted by the collisions encountered by the charged carriers primarily with neutral particles \citep[][]{Shu+2006,Dapp+2012} that leads to the continued contraction of the prestellar cloud core with ever growing central concentration, thus collapsing into a growing (point mass like)
protostar and essentially approaching ballistic conditions.

Theoretical studies of gravitational collapse, intended to
model the collapse of a dense prestellar core to form a
star, have been made by several authors over the decades, both analytically \citep[e.g.,][]{Shu1977,GalliShu1993a,GalliShu1993b,Allen+2003a,Allen+2003b,Galli+2006,AdamsShu2007} as well as numerically using the hydrodynamic (HD) \citep[e.g.,][etc.]{Hunter1969,Penston1966R, Pentson1969,Larson1969}, magnetohydrodynamic (MHD) \citep[e.g.,][etc.]{Tomisaka1995, Tomisaka1996,Nakamura+1995,Tomisaka1998,Tomisaka2002,Allen+2003a,Allen+2003b,
Machida+2004,Machida+2005a,Machida+2005b,Hennebelle2003,Machida+2008a,Dapp+2012,Wurster2016,Zhao+2020,Wurster+2021,Wurster+2021a,XuKunz2021A,XuKunz2021B}, radiation-hydrodynamic (RHD) \citep[e.g.,][etc.]{Larson1969,Masunaga+1998,MasunagaInutsuka2000,Tomida+2010a}, radiation magnetohydrodynamic (RMHDs) \citep[e.g.,][etc.]{Tomida+2010b}, and smoothed particle radiation nonideal MHD \citep[etc.]{Wurster+2021} simulations.

The consequences of infall demonstrates that the inflow in the dense central regions proceeds nearly at a scale of free-fall until encountering a strong radiating shock upon the impact with the surface of the protostar at the center, i.e, the point mass like object present at the center of the dynamical collapse flow. 
Achieving a comprehensive overview on star-formation requires solving the fundamental collapse problem governing magnetohydrodynamics, equations of state (EOSs) for the gas including different polytropic indices, astrochemistry, and radiative transfer across multiple scales, ranging from the molecular cloud scale down to the close vicinity of the protostar, simultaneously within a single mathematical framework, yet remains challenging owing to the limitations stemming from the complexity of the analytical approaches and computational modelling.

\cite{Shu1977} 
(hereafter, SIS collapse) provides the basis for the development of the current archetype of the gravitational collapse for the formation of low-mass stars.
The analytical treatment of SIS model addresses
the gravitational inside-out collapse of a nonmagnetic singular isothermal sphere by applying the self-similarity method to the simplest form of the gas-dynamic flow. 
The term ``inside-out'' refers that the dense innermost regions of the envelope freely-falling well before the more extended successive layers of envelope do so. 
The self-similarity solutions of SIS model describe all self-gravitating isothermal clouds that slightly exceed the maximum limit allowable for hydrostatic equilibrium , tend to develop $r^{-2}$ density profiles in the nearly static outer envelope as long as the initial conditions allow the early phases of the flow to evolve nearly at sonic speed \cite[see][]{BodenheimerSweigart1968}, and as the inside-out collapse progresses, upon the formation of a point mass
at the center, an expansion wave having the density profile of the form $\propto r^{-3/2}$ propagates outward at the sound speed and eventually extends up to the outer boundary of the collapsing prestellar core.  
In this case the mass accretion rate 
into the central point mass is
constant in time, implying $\dot{M}_\ast \propto c_s^3/G$, where $c_s$ and $G$ are the local sound speed and the gravitational constant, respectively.

Magnetic fields  play a pivotal role in the gravitational collapse of such magnetized molecular cloud cores. 
In the current paradigm of magnetized collapse in the formation of low-mass stars \citep{Shu+1987R,Shu+1993}, the central region of a dense core begins to contract quasistatically through the process of ambipolar diffusion. 
The study of a magnetized isothermal cloud collapse carried out by \cite{GalliShu1993a,GalliShu1993b} showed that strong magnetic pinching forces deviate the infalling gas to the equatorial plane to form a flattened quasi-equilibrium structure around the star, often called as pseudodisk. 
The observed prestellar/protostellar candidate cores are recognized to be more flattened than the SIS model.  
As observed from the hourglass morphology and a substantially flattened density structures with scales of $\sim$1000-2000~au for the list of candidate objects, for example, L1448 IRS 2 \citep{Kwon+2019}, L1157 \citep{Looney+2007,Stephens+2013}, B335 \citep{Maury+2018}, HH211 \citep{Lee+2019} indicates the pseudodisk-like configuration of protostellar envelopes at large scales where the magnetic field dynamically
dominates the gas dynamics \citep[see][and references therein]{Vaisala+2023}.

Additionally, in magnetized collapse, 
the (marginally) submagnetosonically evolving mass shells in the outer region of the cloud core are characterized by the densities with the power-law of $\propto r^{-2}$ before the runaway collapse begins in the inner regions
(with the same power law index as the outer regions of the SIS), indicating the collapse has not produced sufficient
flattening on the largest scales.
Eventually the magnetic support weakens as the fields diffuse outwards, the central region reaches an unstable quasi-equilibrium state which is often considered as the initial conditions for the dynamical collapse.
Once the core predominantly becomes gravitationally unstable, it undergoes a runaway magnetized inside-out collapse of centrally condensed regions of dense cloud cores, often termed as 
magnetized ``point mass formation''
where in principle, the central density can grow infinitely large due to the loss of support from the magnetic fields.
The epoch of this 
point mass like protostar formation
bridges the transition from the prestellar (precollapse) states to the protostellar (postcollapse) states. 
Subsequently, \cite{LiShu1996,ShuLi1997,LiShu1997,Li1998a,Li1998b,Allen+2003a,Allen+2003b,Galli+2006} studied the isothermal collapse of magnetized clouds exploring different configurations to address the effects of magnetic fields on self-gravitational collapse.

Later on, \cite{AdamsShu2007} reexamined the effects of ambipolar diffusion in the runaway evolution of the cloud core leading to the
point mass formation and found that the theoretically predicted mass infall (or accretion) rate can be several times higher than the standard value of $c_s^3/G$ (corresponding to the pure hydrostatic states at $t=0$ according to the SIS model), roughly consistent with the measured values in Class 0 sources \citep[see][and further references therein]{Fischer+2023,Manara+2023}. 
From their theoretical study, it is shown that such a quantitative difference in the mass infall rate for a magnetized collapse primarily results from the magnetically modulated gravitational factor due to the flattened geometry of the core.

Our current work focuses on exploring the concept of direct mass infall into the central point mass, a key physical process in the collapse problem that drives the protostar mass growth, with an emphasis on constraining the choice of EOSs (refer to Sec.\ \ref{sec:paper}). 
In our present work, we demonstrate the formation of a protostar, which is numerically represented by the mass of a sink particle without any attempt to resolve the protostar. 
We use the terms `point mass' and `sink particle' interchangeably to convey the same meaning. 
In this study, we will explore the magnetized collapse cases for a large magnetically supercritical prestellar core, yielding mass infall rates those are close to the hydro regime  
\citep[e.g., SIS collapse model of][]{Shu1977} than those of the magnetically modulated rates as estimated by \cite{AdamsShu2007}. 
The magnetic modulations are suppressed only on the mass infall rates, while keeping the other fundamental magnetic features intact, in terms of the pinching magnetic fields lines, pseudodisk formation, etc.

\subsection{Organization of the paper}
 
In Sec.\ \ref{sec:paper}, we demonstrate the motivation of this work and its significance within the framework of the magnetized virial theorem (Sec.\ \ref{sec:virial}). 
In Sec.\ \ref{sec:setup}, we present the detailed numerical setup of this study including the initial and boundary conditions.  
We 
cover an impressively large parameter space of interest (see Table \ref{tab:MODELS}) keeping the computational cost of such simulations reasonable while maintaining appreciable numerical accuracy. 
In Sec.\ \ref{sec:IsothermalCollapse}, we discuss our results on the isothermal magnetized collapse. 
Thereafter in Sec.\ \ref{sec:nonisothermal_collapse}, we present our detailed study of magnetized collapse models with harder EOSs.  
In Sec.\ \ref{sec:discussions}, we discuss our results in relation to previous studies elucidating the significance of choice of an EOS in allowing star formation.  
Finally, in Sec.\ \ref{sec:conclusions}, we summarize the key results of this work along with the physically motivated interpretation.

\section{Theoretical Considerations of this paper} \label{sec:paper}
Our current work focuses on the magnetized gravitational collapse of a spherically symmetric non-rotating prestellar core under the presence of ambipolar diffusion and Ohmic dissipation to explore how a protostar forms.  
The choice of a non-rotating prestellar core is considered 
as we focus on the  intrinsic radial properties caused by the interplay among the gravity, magnetic field, and the nonideal MHD effects.

\subsection{Motivation}
We carry out our numerical exploration of gravitational collapse with a notion of isothermal inside-out collapse inspired by the SIS model of \citep{Shu1977}, but with the presence of magnetic fields incorporating two nonideal MHD effects, which is in partly consistent with the initial analytical implementation of ambipolar diffusion aided isothermal collapse calculations by \cite{AdamsShu2007}.
Once the magnetized collapse commences, magnetically supported flattened quasi-equilibrium pseudodisk forms with infall signatures.

In addition to that, we extend our study of magnetized collapse for a wide range of power-law barotropic EOSs, 
of the form $P(\rho) \propto \rho^{\Gamma}$. 
A value of $\Gamma = 1$ corresponds to an isothermal EOS. 
An EOS either with a stiffer $\Gamma$ than $\Gamma=4/3$
and with a choice of low $\rhostiff$ leads to the formation of a condensed sphere within the central region of infalling collapse flow due to hardening of EOS, which 
prevents the matter from 
falling directly into 
the point mass like protostar 
and 
causes a reduction in the mass accretion rate into that point mass, as found in the numerical studies by \citep[][and series of their papers]{Larson1969, Tomisaka2002, Machida+2004}. 
However, the collapse models with 
a $\Gamma$ as soft as of $\Gamma=1$ \citep[e.g.,][]{Shu1977,AdamsShu2007} or with a $\Gamma$ no harder than 4/3 (see Sec.\ \ref{sec:virial}) or even a sufficiently high $\rhostiff$ \citep[e.g.,][]{Li+2013} allows the collapse to happen continuously without the formation of such a condensed sphere in the central region of infalling collapse flow, which aids the point mass like protostar to grow and 
steadily build up its mass from the infalling envelope, in which the mass accretion rate into the 
central point mass
has a scale of the order of $c_{\rm s}^3/G$.  
Interestingly, from the magnetized virial theorem (see Sec.\ \ref{sec:virial}) it can be shown that only a $\Gamma$ no harder than $4/3$ is able to provide sufficient cooling in allowing the direct 
mass growth of the central point mass. 
The choice of an appropriate EOS is crucial 
in aiding 
the collapse and in interpreting the amount of energy radiated away during star formation.

\subsection{Magnetized virial theorem for a self-gravitating collapsing cloud} \label{sec:virial}
The virial theorem is a powerful theoretical tool to obtain the stability conditions for a self-gravitating cloud core under the conditions of magnetized collapse.

If a star-forming cloud of mass $M$ and of size $R$, confined volume of $V$ by a rarefied and hot external medium of pressure $\mathcal{P}_{\rm ext}$, is statically supported against its self-gravitation, by a combination of internal pressure and magnetic fields, the scalar virial theorem for such a self-gravitating magnetized cloud can be described in the following form
\begin{equation}
    \mathcal{W} + 2 \mathcal{K} + 2 \mathcal{U} + \mathcal{M} = \frac{1}{2} \frac{d^2I}{dt^2} + \mathcal{P}_{\rm ext} \oint \vec{x} . \hat{n} dA - \oint \vec{x} . \stackrel{\leftrightarrow}{T} \hat{n} dA
    \label{eq:frankvirial}
\end{equation}
where, 
\begin{equation}
    \mathcal{W} = - \int_{V} \rho x_i \frac{\delta \mathcal{V}}{\delta x_i} dV \,  ,
\label{eq:eqW}
\end{equation}
\begin{equation}
    \mathcal{K} =  \int_{V} \frac{1}{2} \rho |{\bf u}^2| dV \, ,
\label{eq:eqK}
\end{equation}
\begin{equation}
    \mathcal{U} = \frac{3}{2} \int_{V} P dV \, ,
\label{eq:eqU}
\end{equation}
\begin{equation}
    \mathcal{M} = \int_{V} \frac{|{\bf B}|^2}{8\pi} dV \, ,
\label{eq:eqM}
\end{equation}
where, 
$\mathcal{W}$, $\mathcal{K}$, $\mathcal{U}$, $\mathcal{M}$ represent the self-gravitational energy ($\mathcal{V}$ being the gravitational field) of the gas, the kinetic energy due to the bulk-motion (${\bf u}$) of the gas, thermal energy of the gas due to random motion, and the magnetic energy in volume $V$, respectively \citep[See further on p. 328-332 in \S\ 24 of][]{ShuBookVol2}. 
The first term on the RHS of Equation \ref{eq:frankvirial} presents one-half the second order time-derivative of moment of inertia of the system. 
The middle term on the RHS of Equation \ref{eq:frankvirial} presents the net flow from the contribution of the surface pressure term where $P_{\rm ext}$ refers to the pressure on surface area A. 
The last term on the RHS of Equation \ref{eq:frankvirial} presents the contribution from the net flow of Maxwell stress tensor $ \stackrel{\leftrightarrow}{T}$ out of the closed surface \citep[see Eq. 24.7 for the expression of $\stackrel{\leftrightarrow}{T}$ in][]{ShuBookVol2}.

For a polytropic spherically symmetric cloud composed of ideal gas, having an EOS in the form $P(\rho) \propto \rho^{\Gamma}$, 
\begin{equation}
    \mathcal{U} \propto \int_{V} P dV \propto \int_{V} \rho^{\Gamma} dV \, .
\label{eq:eqC}
\end{equation}
Performing dimensional analysis on Equation \ref{eq:eqC} we obtain, 
\begin{equation}
    \mathcal{U} \equiv \delta_{\Gamma} \left(\frac{M}{R^3} \right)^{\Gamma} V \, ,
\label{eq:eqUGamma}
\end{equation}
where $\delta_{\Gamma}$ is the proportionality constant.

Now, collecting the respective terms as addressed in Equations. \ref{eq:eqW}-\ref{eq:eqM} and using \ref{eq:eqUGamma}, the solution of the Equation \ref{eq:frankvirial} can be written as follows
\begin{equation}
    P_{\rm ext} = \frac{1}{4\pi} \left[-\alpha \frac{GM^2}{R^4} + \beta \frac{\Phi_{\rm mag}}{R^4}+ \delta_{\Gamma} \left(\frac{M}{R^3}\right)^{\Gamma} \right] \, ,
\label{eq:Gammavirial}    
\end{equation}
where the first, second, and third term on RHS of Equation \ref{eq:Gammavirial} represent the contribution due to the gravitational, magnetic, and thermal term, respectively.
Note that, for an isothermal ($\Gamma=1$) magnetized self-gravitating cloud, Equation \ref{eq:Gammavirial} reduces to the following 
\begin{equation}
  P_{\rm ext} = \frac{1}{4\pi} \left[-\alpha \frac{GM^2}{R^4} + \beta \frac{\Phi_{\rm mag}}{R^4}+ 3c_s^2 \left(\frac{M}{R^3}\right) \right]
    \label{eq:Gamma1Virial} \, ,
\end{equation}
as shown by Equation 24.17 in \cite{ShuBookVol2}.

Furthermore we delve into the magnetic virial theorem to better understand the contribution from magnetic pressure \citep[See further on p. 336 in \S\ 24 of][]{ShuBookVol2}. 
In a scenario where a magnetized cloud of size $R$ is compressed more-or-less spherically while conserving magnetic flux, meaning $\Phi_{\rm mag} \propto B R^2$, hence the magnetic fields strength increases with decreasing radius, as $B \propto R^{-2}$. 
Now, under such magnetized compression, the density of the cloud scales as $\rho \propto M/R^3$ (following mass conservation), resulting in a power-law behaviour between magnetic stresses and matter density, having the form of $B^2 \propto \rho^{4/3}$. 
In simpler terms, we can realize this picture as a magnetic fields behaving in three-dimensional compression akin to a gas with $\Gamma = 4/3$.

For such a polytropic self-gravitating magnetized cloud, an analog to Eq.\ \ref{eq:Gammavirial} can be derived as follows
\begin{equation}\label{eq:magvirial}
    P_{\rm ext} = \frac{1}{4\pi} \left[\frac{\alpha G}{R^4}(M^2_{\Phi} - M^2) + \delta_{\Gamma} \frac{M^{\Gamma}}{R^{3\Gamma}} \right] \, ,
\end{equation}
where, $\alpha$ is a nondimensional constant of the order of magnitude unity, magnetic critical mass $M_{\phi} \approx 0.13 G^{-1/2} \Phi_{\rm mag}$, magnetic flux $\Phi_{\rm mag}= \pi B R^2$. 
For a cloud with a mass $M< M_{\Phi}$ has a magnetically subcritical mass for gravitational collapse if the magnetic flux remains frozen to the matter. 
On the other hand, a cloud with a mass $M> M_{\Phi}$ is referred to as magnetically supercritical clouds, which can be prone to collapse; however, depending on the choice of an EOS.

The virial formulation shows that the critical mass for collapse depends on the EOS and the degree of magnetization. 
Hence, it is important to realize the significance of the exponents of the gravitational, magnetic, and thermal term to constrain the choice of an EOS in aiding the collapse.
From the magnetized virial theorem (Eq.\ \ref{eq:Gammavirial}), 
it is seen that for the magnetized collapse geometry, the virial exponent in the scaling of the gravitational term and magnetic term are the same as each of these two terms has a dependence of $\propto R^{-4}$. 
However, for the thermal virial exponent to be the same as the  scaling of $\propto R^{-4}$ 
requires the selection of $\Gamma=4/3$ only, as can be shown from Eq. \ref{eq:Gammavirial}. This leads to the criticality at $\Gamma=4/3$
that corresponds to a transition from a regime of instability (referring to $\Gamma < 4/3$) to stability (referring to $\Gamma > 4/3$).
This criticality of $\Gamma=4/3$, 
where all the terms in Eq.\ \ref{eq:magvirial} are of the same order ($\propto R^{-4}$), allows an appropriate definition of a critical mass for the collapse. 
Thus, the limiting case of $\Gamma=4/3$ is of significant astrophysical interest, as the stability criterion in this virial balance is determined by a critical mass that depends on an EOS. 
This critical mass is 
directly connected to a series of fundamental mass limits for objects having this kind of EOS, most notably the well-known Chandrasekhar mass limit for a white dwarf \citep{Chandrasekhar1931}.

For a non-magnetized cloud  ($\Phi_{\rm mag}=0$) with a $\Gamma$ softer than 4/3, the thermal term exhibits a dependence shallower than $R^{-4}$, leading to the collapse inevitably (see Eq.\ \ref{eq:Gammavirial}). 
For example, the SIS collapse illustrates one such framework of the nonmagnetic collapse with a $\Gamma=1 < 4/3$, for which the thermal term evolves as $\propto R^{-3}$, at a shallower rate than the self-gravity term (see Eq.\ \ref{eq:Gamma1Virial}). 
Now, for a magnetized cloud with a $\Gamma$ softer than 4/3, as the thermal term evolves subdominantly than the other two terms, these clouds are prone to collapse once the magnetic field fails to provide sufficient support against the self-gravity, 
implying $M> M_{\Phi}$ (see Eq.\ \ref{eq:magvirial}).  
Refer to Sec.\ \ref{sec:IsothermalCollapse} and \ref{sec:polytropes} for further discussions on such magnetized collapse cases.  
On the other hand, for a star-forming cloud or star or even a condensed sphere with a $\Gamma$ harder than 4/3, the thermal term becomes the most dominant as it exhibits a  dependence steeper than $R^{-4}$. 
For example, for a specific choice of an EOS with a $\Gamma = 5/3$, thermal term contains a $R^{-5}$ dependence, which is steeper than the self-gravity and the magnetic term. 
Therefore, such a choice of an EOS gives rise to formation of condensed spheres at the center of infalling collapse flow as a result of stable solutions. 
For such magnetized collapse cases with a $\Gamma>4/3$ (refer to Sec.\ \ref{sec:polytropes} and \ref{sec:gamma5by3}), the support from the magnetic field is not relevant as long as the thermal pressure alone can decide the fate of the collapse.

Interested readers may further refer to an additional discussion on the virial theorem \citep{Chandrasekhar1939} under an equilibrium configuration for a non-magnetized cloud in Appendix  \ref{sec:VirialChandra}, in which the SIS collapse \citep{Shu1977} can be understood as a theoretical configuration for an unstable non-magnetic solution with $\Gamma=1$.


\section{Numerical Setup} \label{sec:setup}

\begin{table*}[ht!]
\caption{Parameterized nomenclature for the models: model name, softening/stiffening index of the EOS ($\Gamma$), the stiffening density threshold ($\rho_{\rm stiff}$), initial sound speed (${\rm c_s}$), initial magnetic parameter ($\mu_{\rm param}$), measure of ambipolar diffusion (AD) in terms of neutral-ion collisional coupling parameter ($\tilde{\gamma}_{\rm AD}$), Ohmic resistivity ($\eta_{\rm OD}$), and the occurrence of the transient condensed spheres as an outcome of the collapse. 
}
\vspace{-0.3cm}
\centering
\begin{longtable}
{cccccccc}
\hline 
\multicolumn{1}{c}{Model} & \multicolumn{1}{c}{$\Gamma$} & \multicolumn{1}{c}{$\rho_{\rm stiff}$} & \multicolumn{1}{c}{$c_s$} & \multicolumn{1}{c}{$\mu_{\rm param}$} & \multicolumn{1}{c}{$\tilde{\gamma}_{\rm AD}$} & \multicolumn{1}{c}{$\eta_{\rm OD}$}& \multicolumn{1}{c}{Occurrence of the transient}\\
  & ($P \propto \rho^{\Gamma}$)  & (${\rm g}\, {\rm cm}^{-3}$)  & $({\rm km}\, {\rm s}^{-1})$ & & ($3.15 \, \times$) & (${\rm cm}^2 \, {\rm s}^{-1}$) & {condensed spheres} \\
 \hline

Model-G1A &  1 & - & 0.2 & 8 & $10^{-3}$ & $10^{17}$  & No sphere\\ 
Model-G1B &  1 & - & 0.2 & 8 &  $10^{-3}$ & $10^{18}$ & No sphere\\
Model-G1C &  1 & - & 0.2 & 8 &  $10^{-4}$ & $10^{17}$ & No sphere\\
Model-G1D &  1 & - & 0.2 & 8 &  $10^{-4}$ & $10^{18}$ & No sphere\\
Model-G1E &  1 & - & 0.3 & 8 &  $10^{-3}$ & $10^{17}$ & No sphere\\ 
Model-G1F &  1 & - & 0.2 & 4 &  $10^{-3}$ & $10^{17}$ & No sphere\\ 
\hline  
Model-G1.2S15A &  1.2 & $10^{-15}$ & 0.2 & 8 & $10^{-3}$ & $10^{17}$ & No sphere\\
Model-G1.2gA & 1.2 & Global & 0.2 & 8 & $10^{-3}$ & $10^{17}$ & No sphere\\
Model-G1.33S15A &  1.33 & $10^{-15}$ & 0.2 & 8 & $10^{-3}$ & $10^{17}$ & No sphere\\
Model-G1.33gA &  1.33 & Global & 0.2 & 8 & $10^{-3}$ & $10^{17}$ & No sphere\\
Model-G1.4S15A &  1.4 & $10^{-15}$ & 0.2 & 8 & $10^{-3}$ & $10^{17}$ & Condensed sphere\\
Model-G1.4gA &  1.4 & Global & 0.2 & 8 & $10^{-3}$ & $10^{17}$ & Condensed sphere\\
Model-G1.5S15A &  1.5 & $10^{-15}$ & 0.2 & 8 & $10^{-3}$ & $10^{17}$ & Condensed sphere\\
Model-G1.5gA &  1.5 & Global & 0.2 & 8 & $10^{-3}$ & $10^{17}$ & Condensed sphere\\
\hline 
Model-G1.67S15A &  1.67 & $10^{-15}$ & 0.2 & 8 &  $10^{-3}$ & $10^{17}$ & Condensed sphere\\
Model-G1.67S15B &  1.67 & $10^{-15}$ & 0.2 & 8 &  $10^{-3}$ & $10^{18}$ & Condensed sphere\\
Model-G1.67S15C &  1.67 & $10^{-15}$ & 0.2 & 8 &  $10^{-4}$ & $10^{17}$ & Condensed sphere\\
Model-G1.67S15D &  1.67 & $10^{-15}$ & 0.2 & 8 &  $10^{-4}$ & $10^{18}$ & Condensed sphere\\
Model-G1.67S15F &  1.67 & $10^{-15}$ & 0.2 & 4 & $10^{-3}$ & $10^{17}$ & Condensed sphere\\
\vspace{0.25cm}

Model-G1.67S14A &  1.67 & $10^{-14}$ & 0.2 & 8 & $10^{-3}$ & $10^{17}$ & Condensed sphere\\
Model-G1.67S14B &  1.67 & $10^{-14}$ & 0.2 & 8 & $10^{-3}$ & $10^{18}$ & {Condensed sphere}\\
Model-G1.67S14C &  1.67 & $10^{-14}$ & 0.2 & 8 & $10^{-4}$ & $10^{17}$ & {Condensed sphere}\\
Model-G1.67S14D &  1.67 & $10^{-14}$ & 0.2 & 8 & $10^{-4}$ & $10^{18}$ & {Condensed sphere}\\
Model-G1.67S14E &  1.67 & $10^{-14}$ & 0.3 & 8 & $10^{-3}$ & $10^{17}$ & {Condensed sphere}\\
Model-G1.67S14F &  1.67 & $10^{-14}$ & 0.2 & 4 & $10^{-3}$ & $10^{17}$ & {Condensed sphere}\\

Model-G1.67S13A &  1.67 & $10^{-13}$ & 0.2 & 8 & $10^{-3}$ & $10^{17}$ & {Condensed sphere}\\
Model-G1.67S13B &  1.67 & $10^{-13}$ & 0.2 & 8 & $10^{-3}$ & $10^{18}$ & {Condensed sphere}\\
Model-G1.67S13C &  1.67 & $10^{-13}$ & 0.2 & 8 & $10^{-4}$ & $10^{17}$ & {Condensed sphere}\\
Model-G1.67S13D &  1.67 & $10^{-13}$ & 0.2 & 8 & $10^{-4}$ & $10^{18}$ & {Condensed sphere}\\
Model-G1.67S13F &  1.67 & $10^{-13}$ & 0.2 & 4 & $10^{-3}$ & $10^{17}$ & {Condensed sphere}\\

\hline 
Model-G2S15A &  2 & $10^{-15}$ & 0.2 & 8 &  $10^{-3}$ & $10^{17}$ & {Condensed sphere}\\
Model-G2S15B &  2 & $10^{-15}$ & 0.2 & 8 &  $10^{-3}$ & $10^{18}$ & {Condensed sphere}\\
Model-G2S15C &  2 & $10^{-15}$ & 0.2 & 8 &  $10^{-4}$ & $10^{17}$ & {Condensed sphere}\\
Model-G2S15D &  2 & $10^{-15}$ & 0.2 & 8 &  $10^{-4}$ & $10^{18}$ & {Condensed sphere}\\
Model-G2S15F &  2 & $10^{-15}$ & 0.2 & 4 & $10^{-3}$ & $10^{17}$ & {Condensed sphere}\\

Model-G2S14A &  2 & $10^{-14}$ & 0.2 & 8 & $10^{-3}$ & $10^{17}$ & {Condensed sphere}\\
Model-G2S14B &  2 & $10^{-14}$ & 0.2 & 8 & $10^{-3}$ & $10^{18}$ & {Condensed sphere}\\
Model-G2S14C &  2 & $10^{-14}$ & 0.2 & 8 & $10^{-4}$ & $10^{17}$ & {Condensed sphere}\\
Model-G2S14D & 2 & $10^{-14}$ & 0.2 & 8 & $10^{-4}$ & $10^{18}$ & {Condensed sphere}\\
Model-G2S14E &  2 & $10^{-14}$ & 0.3 & 8 & $10^{-3}$ & $10^{17}$ & {Condensed sphere}\\
Model-G2S14F & 2 & $10^{-14}$ & 0.2 & 4 & $10^{-3}$ & $10^{17}$ & {Condensed sphere}\\

Model-G2S13A &  2 & $10^{-13}$ & 0.2 & 8 & $10^{-3}$ & $10^{17}$ & {Condensed sphere}\\
Model-G2S13B &  2 & $10^{-13}$ & 0.2 & 8 & $10^{-3}$ & $10^{18}$ & {Condensed sphere}\\
Model-G2S13C &  2 & $10^{-13}$ & 0.2 & 8 & $10^{-4}$ & $10^{17}$ & {Condensed sphere}\\
Model-G2S13D &  2 & $10^{-13}$ & 0.2 & 8 & $10^{-4}$ & $10^{18}$ & {Condensed sphere}\\
Model-G2S13F &  2 & $10^{-13}$ & 0.2 & 4 & $10^{-3}$ & $10^{17}$ & {Condensed sphere}\\

\hline
\end{longtable}
\label{tab:MODELS}
\end{table*}

We solve the magnetohydrodynamic equations in a simulation setup using ZeusTW code \citet{Krasnopolsky+2010} on a non-uniform grid in spherical polar coordinates ($r, \theta, \phi=0$) incorporating two nonideal magnetic diffusion terms: ambipolar diffusion and Ohmic dissipation. 
We perform a series of two-dimensional (axisymmetric)
nonideal MHD simulations to study the collapse of a non-rotating self-gravitating spherically symmetric prestellar core. 

\subsection{Computational grid and boundary conditions}\label{sec:comGrid}
As described in \cite{Krasnopolsky+2012}, we choose a 2D non-uniform grid of $192 \times 128$.
In the radial direction, the inner and outer
boundaries are located at $r_{\rm min} = 7.5 \times 10^{12}$ and $r_{\rm max}= 2\times 10^{17}$ cm (0.5~au and 13369.2~au), respectively.
The radial cell size is smallest near the inner boundary. 
In the polar direction, we choose a relatively large cell size ($ 3.75^{\circ}$) near the polar axes 
and it decreases smoothly to a minimum of $\sim 0.5 ^{\circ}$ near the equator, where a magnetically supported pseudodisk may form.

The boundary conditions in the radial direction, we impose the standard outflow boundary conditions. 
Material leaving the inner radial boundary is collected as a sink particle
point mass at the center.
It acts on the matter in the computational domain through gravity. 
In our present work, we demonstrate the formation of a protostar, which is numerically represented by the mass of a sink particle at the center, without any attempt to resolve the protostar.
On the polar axes, the boundary condition is chosen to be reflective, consistent with the axisymmetry.

\subsection{Initial Conditions} \label{sec:ICs}
Low-mass prestellar cores in nearby star-forming regions are observed to have relatively simple dynamical structures \citep{BerginTafalla2007R}.
We idealize such prestellar cores as initially uniform
spheres of $r_{\rm max}= 2\times10^{17} \, {\rm cm}$ in radius and $2.95 \, \Msun$ in mass at rest \citep[see][for a similar numerical setup]{HennebelleFromang2008}, adopting a mean mass per neutral species $m_n = 2.33 \, m_{\rm H}$ (accounting for molecular hydrogen with about $20$ percent ${\rm He}$ by number). 
We obtain number density $n$ as $\rho/2.33{m_{\rm H}}=\rho/(3.881\times 10^{-24}) \, {\rm cm}^{-3}$, where $\rho$ is the density. 
We choose the initial density profile to have $\rho \propto r^{-2}$ with a density level of $\rho_0 = 4.77 \times 10^{-19} \ {\rm g} \, {\rm cm}^{-3}$ and a flattening radius of $r_c=10^{15} \, {\rm cm}$ has a form as follows
\begin{equation}
    \rho(r) = \rho_0 \left[\rho_f \left(1 + \left(\frac{r}{r_c}\right)^2 \right) \right]^{-e_{\rho}/2} \, ,
    \label{eq:rhoInitial}
\end{equation}
where $\rho_{f} = 1+ (r_{\rm max}/r_c)^2$, and $e_{\rho} = 2$.  
The prestellar core is initially spherical, can be considered as a similar representative of a typical ${\rm NH}_3$ quasi-spherical core \citep{Jijina+1999}. 
The initial temperature of the gas is set to $T \sim \, 10 \, {\rm K}$, giving an initial isothermal sound speed $c_s \equiv (k_{\rm B} T/m_{\rm n})^{1/2} = 0.2 \, {\rm km} \, {\rm s}^{-1}$.

Molecular clouds are supported against self-gravity in part by magnetic fields.  
\cite{MestelSpitzer56} 
introduced a mass scale associated with the amount of magnetic flux $\Phi$ frozen into it and threaded through a self-gravitating, electrically conducting cloud, which is known as the magnetic critical mass,
$M_{\Phi} = \Phi/(2\pi \sqrt{G})$ \citep{Shu+1999R}. 
Here, the selection of the coefficient $1/2\pi$ defines $M_{\Phi}$ as the maximum mass that can be supported when the magnetic field alone opposes the cloud's self-gravity, leading to a highly flattened configuration. 
The cloud core of mass $M$ greater or lesser than $M_{\Phi}$ is called as magnetically supercritical (prone to collapse against self-gravity) or subcritical (stable against self-gravity).  
Hence, it is important to measure the mass-to-flux (or mass-to-magnetic flux) ratio, $M/\Phi$ and compare it to the critical ratio, $(2 \pi G^{1/2})^{-1}$, i.e., to measure the dimensionless ratio $M/M_{\Phi}$. 
Alternatively, the dimensionless mass-to-flux ratio can also be defined as a spatially constant value that has the following form,
\begin{equation}
    \lambda = 2\pi \sqrt{G} \frac{\Sigma}{B_z} \, ,
    \label{eq:lambdaEq}
\end{equation}
where $B_z$ is total $B$ in Gaussian units distributed vertically at the midplane and $\Sigma$ is the integrated column density along the vertical direction of the cloud.
Under the flux-freezing, the ratio $\Sigma/B_z$ is also temporally constant. 
In our numerical setup, we employ the initial B-fields following the definition of spatially constant mass-to-flux ratio as described in Equation \ref{eq:lambdaEq}. 
In ZeusTW code, the calculations of B-fields are implemented in Lorentz-Heaviside units.
The initial prestellar core is threaded by an initially vertical magnetic fields corresponding to the magnetic parameter, $\mu_{\rm param}$, used in the code, which is scaled to 
the Equation (\ref{eq:lambdaEq}), i.e., the
definition of $\lambda$ 
(Appendix \ref{sec:lambdaAPP}).

Molecular clouds are weakly ionized yet retain a relatively good (though imperfect) coupling between plasma and neutrals due to the enhanced Langevin cross section for ion-neutral collisions \citep[see][\S\ 27]{ShuBookVol2}.
In our numerical setup, the strength of the initial ambipolar diffusion is considered in terms of the neutral-ion collisional coupling constant $\gamma_{\rm in}$. 
Ohmic resistivity ($\eta_{\rm OD}$) is regarded as a measure of Ohmic dissipation.
The magnetic induction equation in Lorentz-Heaviside units as used in our nonideal MHD simulations has the following form as follows

\begin{equation}
\begin{aligned}
    & \frac{\partial \vec{B}}{\partial t} +  {\vec{\nabla}} \times \left(\vec{B} \times \vec{v} \right) = \\
                & \vec{\nabla} \times \left\{\frac{\vec{B}}{\gamma_{in} \mathcal{C} \rho^{3/2}} \times \left[\vec{B} \times \left(\vec{\nabla} \times \vec{B} \right)  \right] \right\}  
            - \> \vec{\nabla} \times \left(\eta_{\rm{OD}} \> \vec{\nabla} \times \vec{B} \right)  \, ,
\end{aligned}
\label{eq:inductionequn}   
\end{equation}
that includes two nonideal MHD effects: ambipolar diffusion and Ohmic dissipation.
The current density $\vec{J}$ can be written as 
\begin{equation}
    \vec{J} = {\vec \nabla} \times \vec{B} \, , 
\end{equation}
and the ion abundance is expressed by the local relation $\rho_{\rm i} = \mathcal{C} \rho^{1/2}$ \citep{ShuBookVol2}. 
Here, $\mathcal{C}$ presents a constant if cosmic rays provide the dominant source of ionization, but it quickly rises to much larger values near the surfaces of molecular clouds because of the ultraviolet ionization of elements like carbon and others \citep{Mckee1989}. 
Please refer to Appendix \ref{sec:ambipolarApp} for further details on the numerics of ambipolar diffusion used in ZeusTW code. 
In our numerical exploration we choose to use the constant Ohmic resistivity of $10^{17}~{\rm cm}^2~{\rm s}^{-1}$ and $10^{18}~{\rm cm}^2~{\rm s}^{-1}$. 
However, Ohmic resistivity/diffusivity may vary by a few orders of magnitude depending on the density \citep[e.g., NICIL, see further in][]{Wurster2016,Wurster+2021a}, however exploring these variations is beyond our current scope.

\subsection{Nomenclature of the numerical models} \label{sec:models}
We briefly describe the nomenclature of the numerical models of this study.  
Please refer to Table \ref{tab:MODELS} for further details.
Models labeled with ``A'', ``B'', ``C'', ``D'' mean different combination of nonideal MHD effects. 
Models labeled with ``E'' refers to a different initial sound speed set at $0.3 \, {\rm km}\, {\rm s}^{-1}$. 
The model names labeled with ``F'' refers to the models with an initial magnetic parameter of $\mu_{\rm param} = 4$ that corresponds to a stronger magnetic field than the other models. 
All other models (i.e., those without the ``F'' label) are explored with $\mu_{\rm param} = 8$ (see Appendix \ref{sec:lambdaAPP} for its corresponding value of the dimensionless mass-to-magnetic flux ratio $\lambda$ as described in Equation \ref{eq:lambdaEq}). 
Next, the models labeled with ``G1'', ``G1.2'', ``G1.33'', ``G1.4'', ``G1.5'', ``G1.67'', ``G2'' (going from softer to harder EOS) represent the EOS with different   
barotropic index $\Gamma$ ($P \propto \rho^{\Gamma}$) applied to the respective models. 
Models labeled with ``S15'', ``S14'', ``S13'' correspond to the different values of $\rho_{\rm stiff}$. 
Models labeled with ``g'' such as ``Model-G1.2gA'', ``Model-G1.33gA'', ``Model-G1.4g'', ``Model-G1.5g'' represent the magnetized collapse models for EOSs with global $\Gamma$.

\section{Results from the Isothermal magnetized Collapse}
\label{sec:IsothermalCollapse}
In this section, we present the numerical results from the isothermal nonideal MHD collapse of a non-rotating prestellar cloud core.


\begin{figure*}[!ht]
\centering
\gridline{\fig{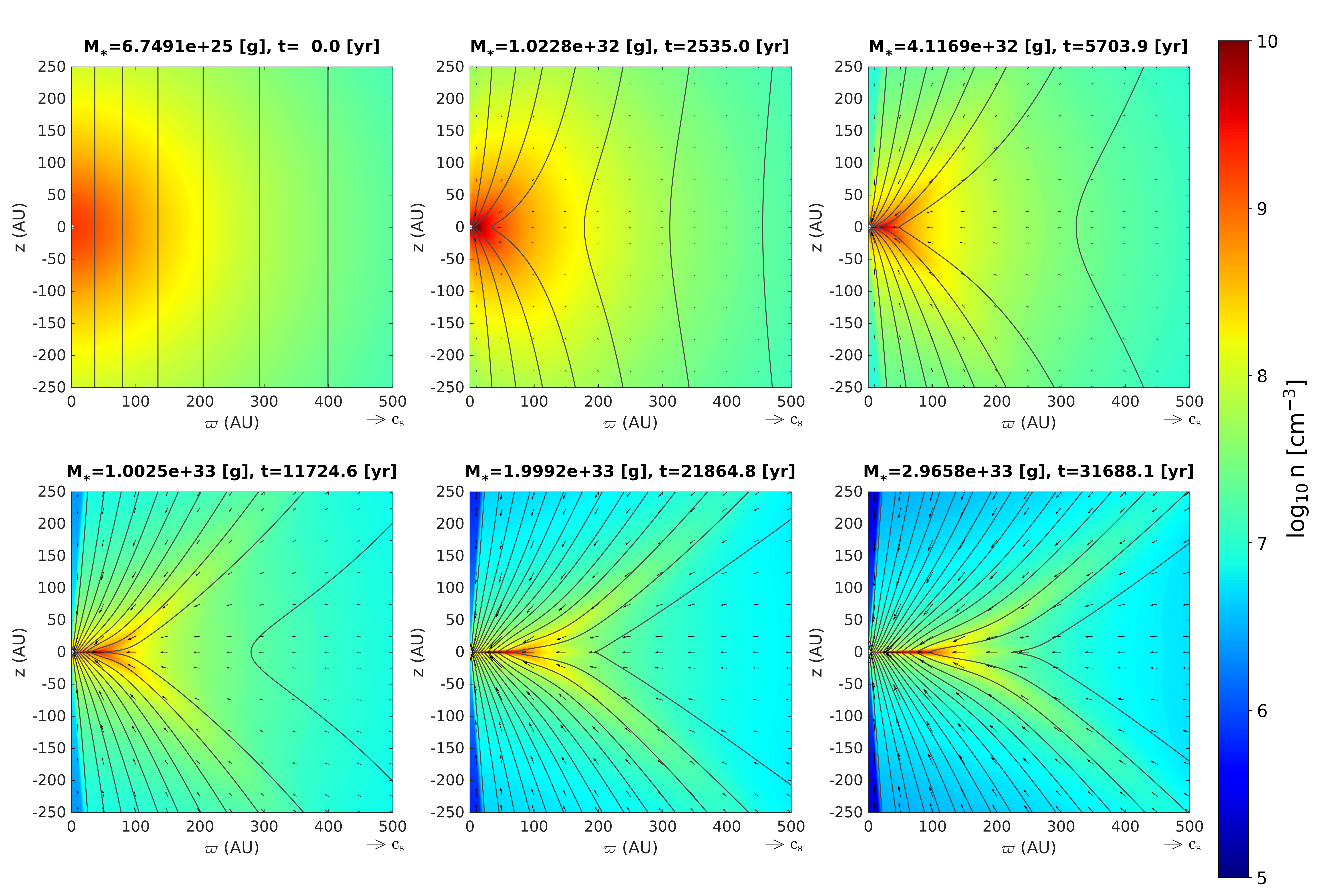}{\linewidth}{}
}
\vspace*{-0.8cm}
\caption{Time sequence of the magnetized isothermal collapse for Model-G1A.  
A two-dimensional intensity map of number density with overlaid magnetic flux lines and poloidal velocity vectors on scales of $500 \,  {\rm au} \times 500 \, {\rm au}$.  
}
\label{fig:E0-A1-s500}
\end{figure*}

\begin{figure*}[!ht]
\centering
\gridline{\fig{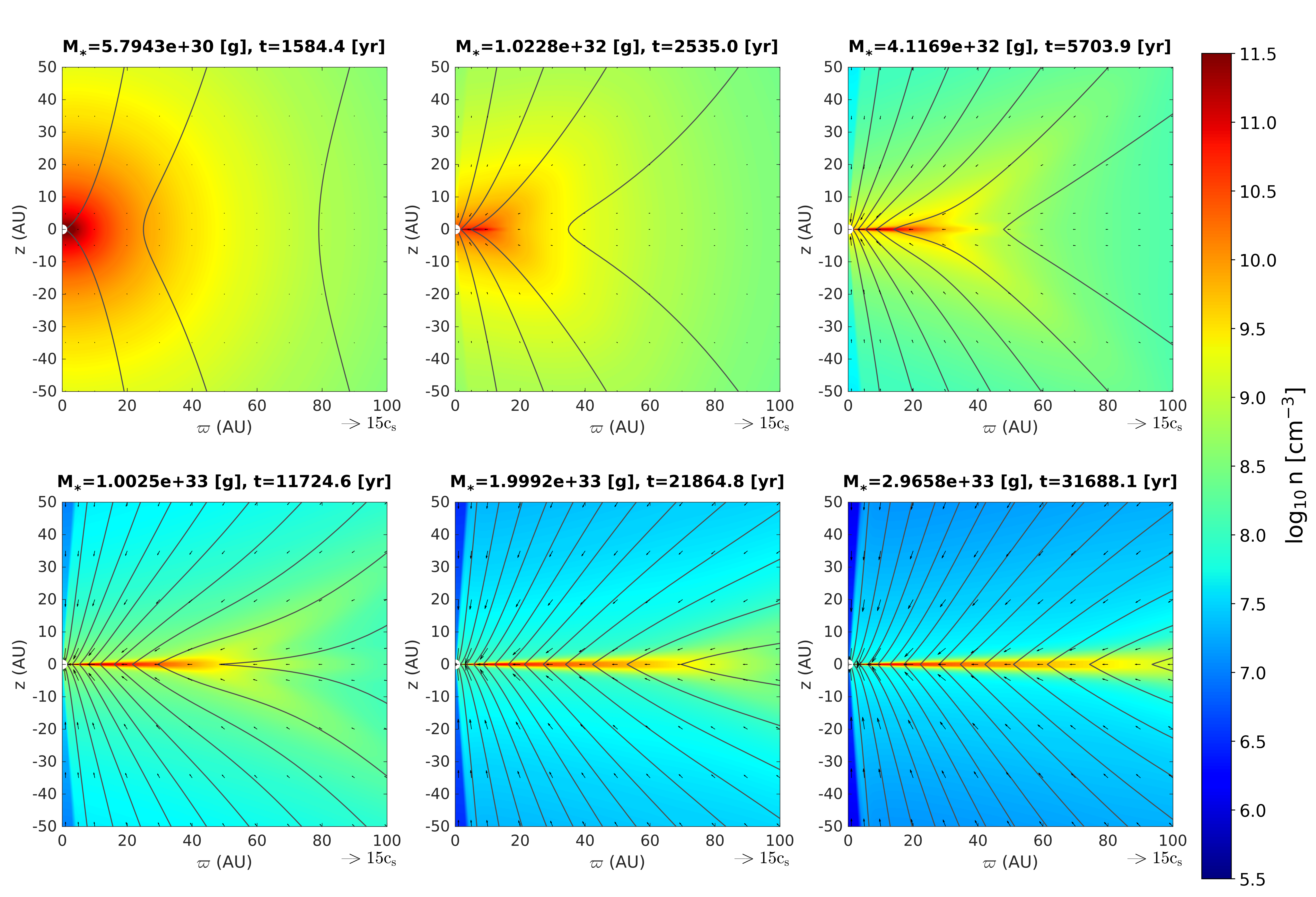}{\linewidth}{}
}
\vspace*{-0.8cm}
\caption{Time sequence of the magnetized isothermal collapse for Model-G1A.  
A two-dimensional intensity map of number density with overlaid magnetic flux lines and poloidal velocity vectors on scales of $100 \,  {\rm au} \times 100 \, {\rm au}$. 
}
\label{fig:E0-A1-s100}
\end{figure*}

\begin{figure*}
\plotone{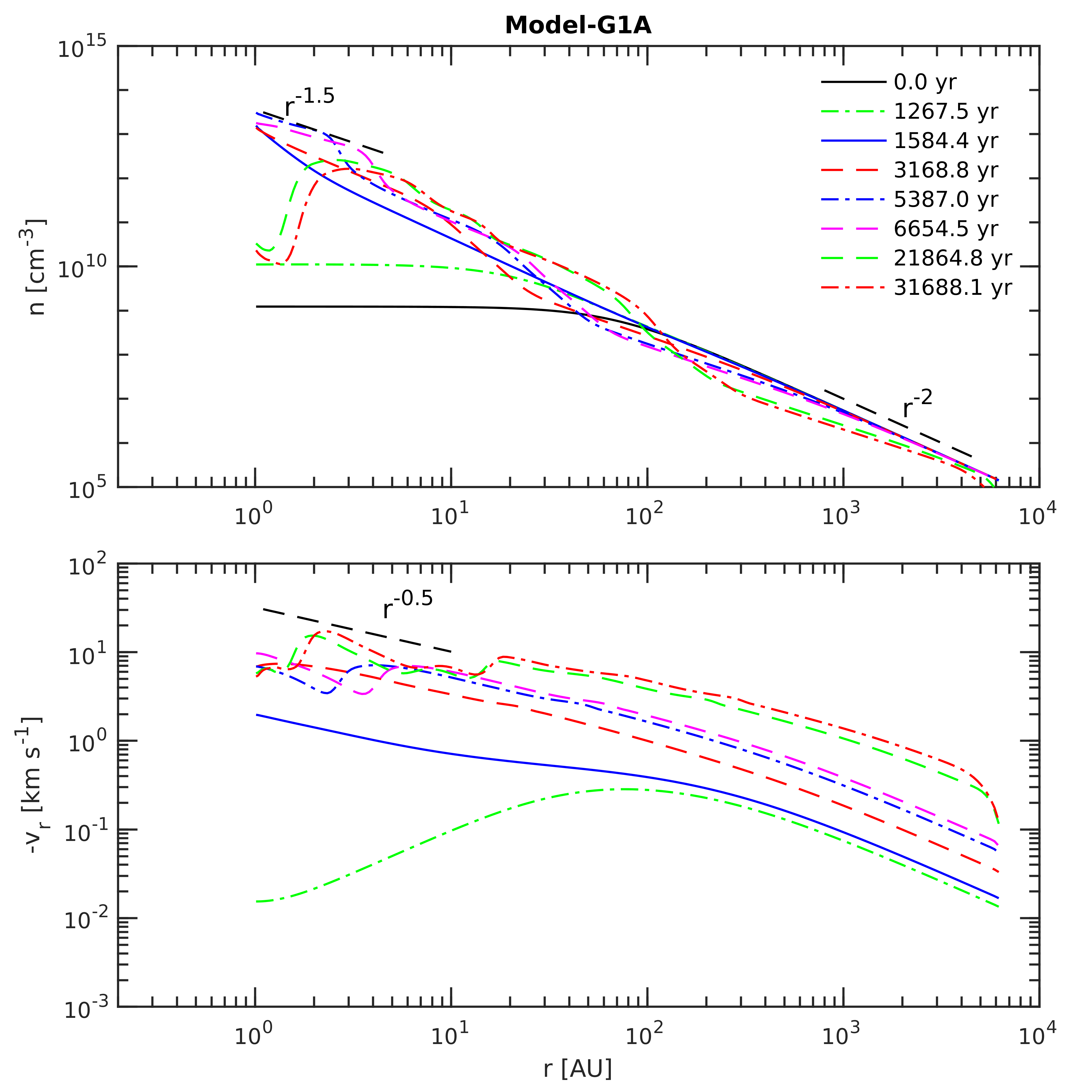}
\caption{Radial profiles of the number density (top panel) and radial component of infall velocity (bottom panel) at the equatorial plane corresponding to different evolutionary time instances for Model-G1A. }
\label{fig:1D-densVr}
\end{figure*}

\begin{figure*}[!htb]
\gridline{\fig{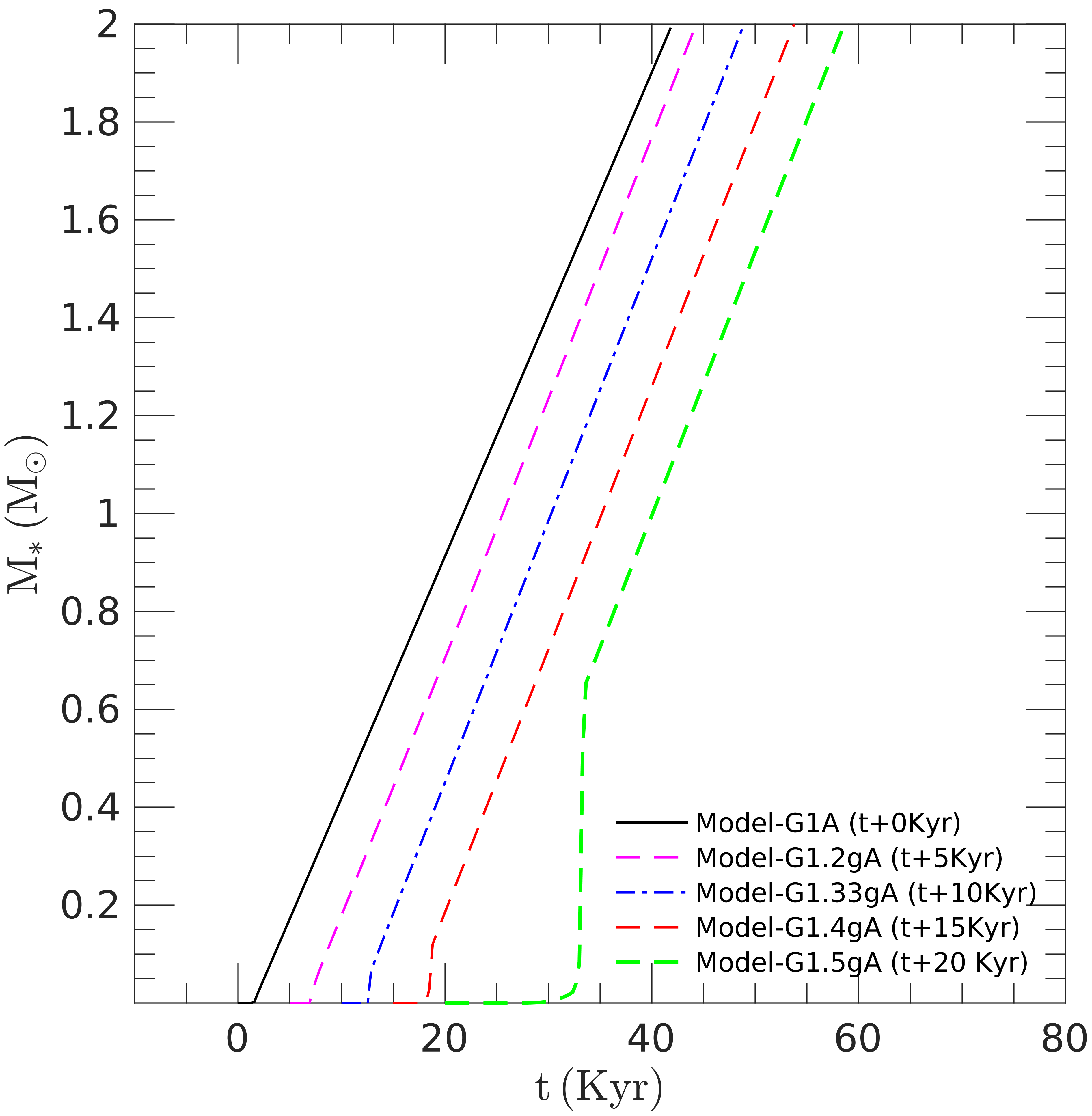}{0.45\textwidth}{(a)}
\fig{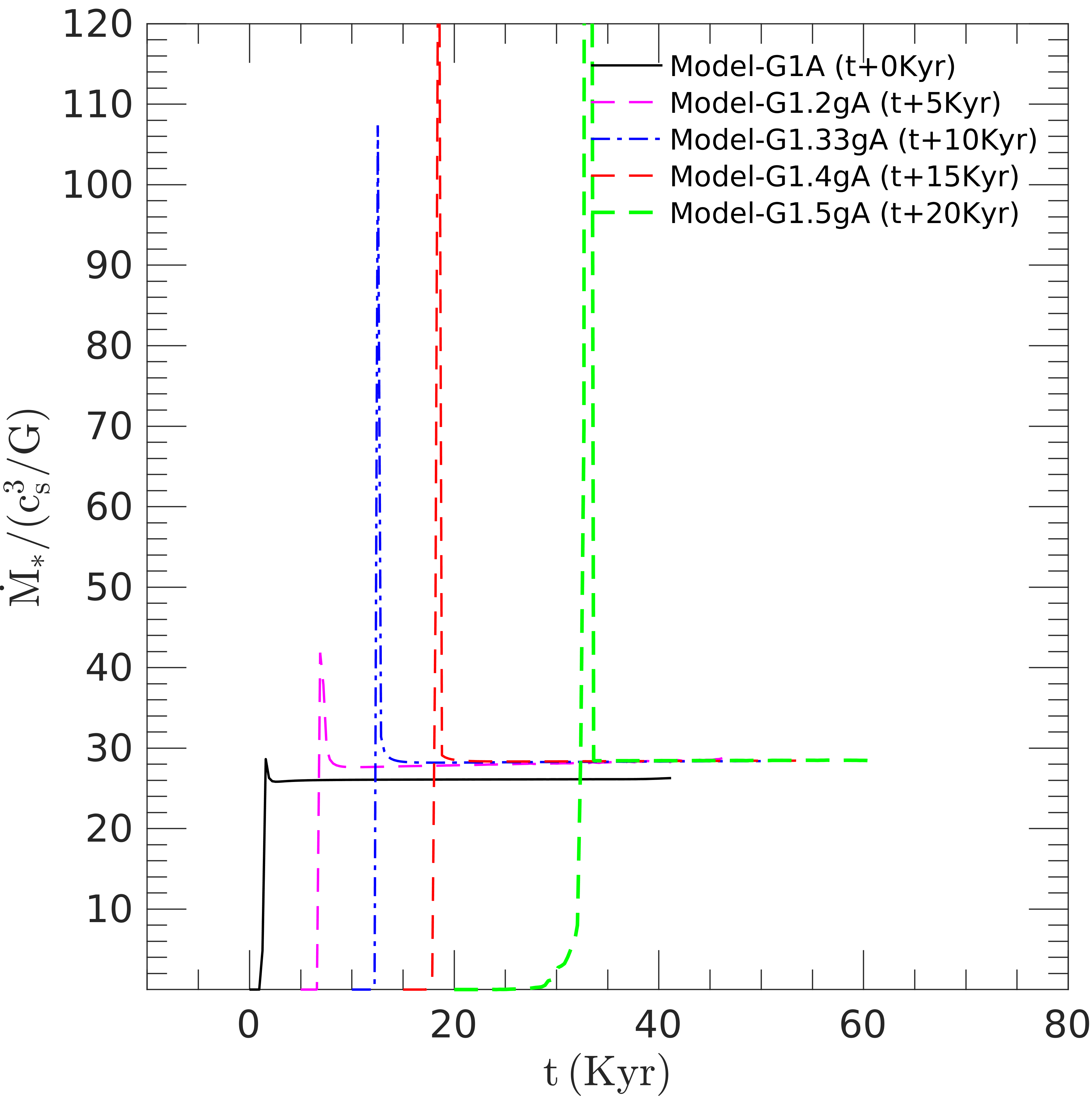}{0.45\textwidth}{(b)}
}
\caption{Left: Mass of the central sink particle point mass 
($M_{\ast}$) with time;  
Right: Mass accretion rate in the units of ${{\rm c}^3_s}/G$ with time for {Model-G1A, Model-G1.2gA, Model-G1.33gA, Model-G1.4gA, and Model-G1.5gA}. Here, $c_s^3/G = 1.6 ({\rm T}/ 10{\rm K})^{3/2} \, {\rm M}_{\odot} \, {\rm Kyr}^{-1}$. Artificial time offsets that are mentioned in the label, are added to distinguish the respective temporal profiles visually.} 
\label{fig:masscompare-G1}
\end{figure*}

\begin{figure*}[ht!]
\gridline{\fig{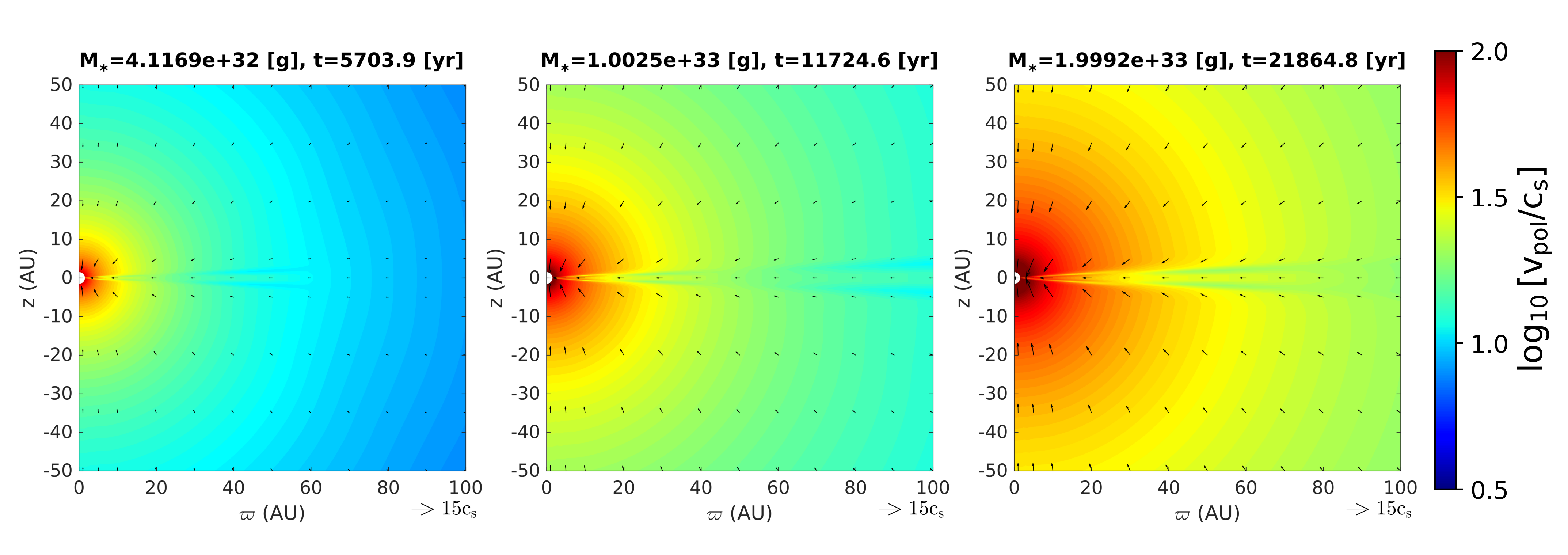}{\linewidth}{}
}
\vspace{-0.5cm}
\caption{Time sequence of the poloidal velocity ${\rm v_{pol}}$ normalized to the initial sound speed ($v_{\rm pol}/c_s$, given $c_s=0.2 \kms$) with overlaid poloidal velocity vectors in black color
for the magnetized isothermal collapse with Model-G1A on scales of $100 \,{\rm au} \times 100 \, {\rm au}$ at significant time instances.}
\label{fig:G1A_vpol_s100}
\end{figure*}

\begin{figure*}[ht!]
\gridline{\fig{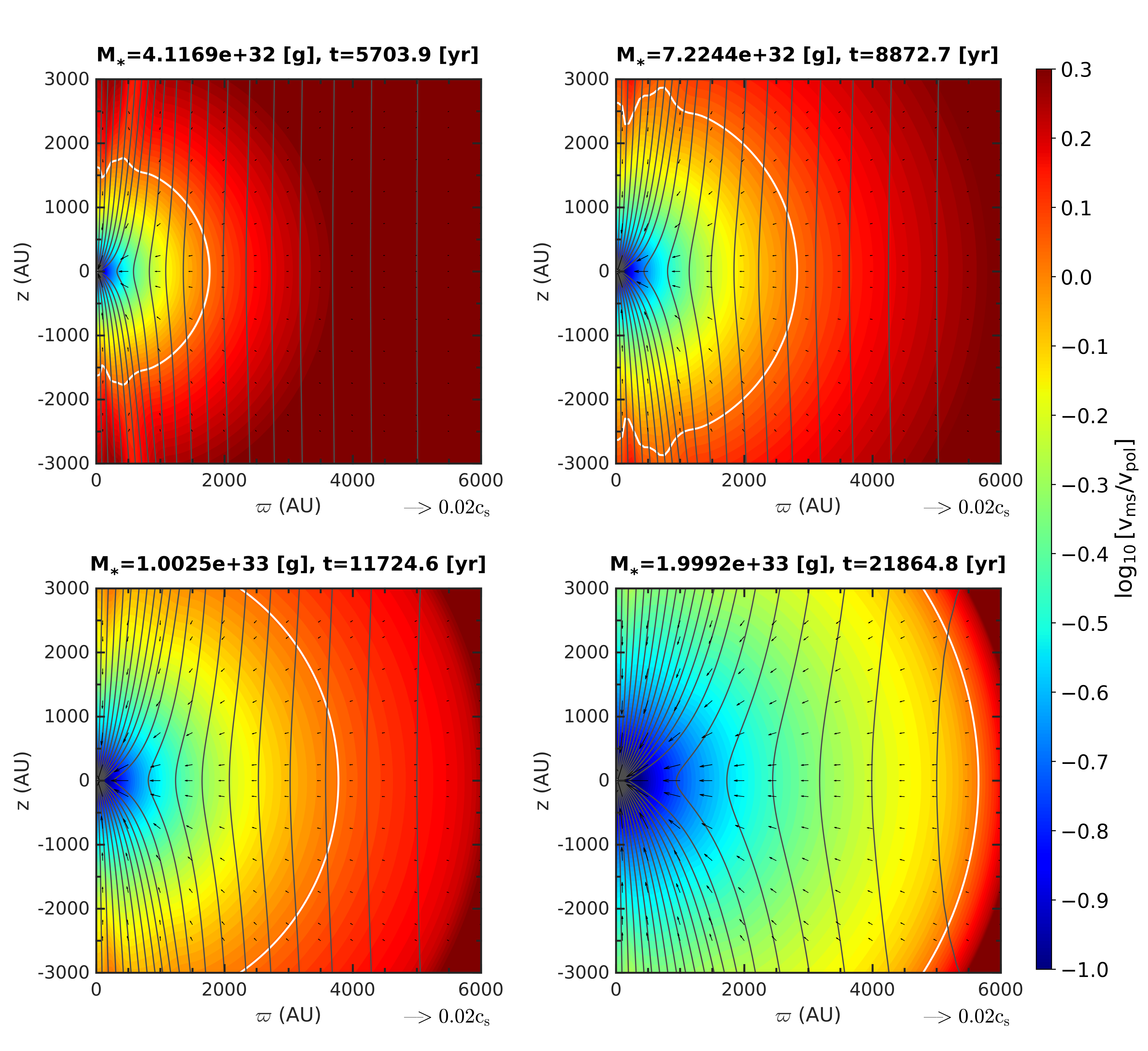}{\linewidth}{}
}
\vspace{-0.5cm}
\caption{Dynamical evolution of the ratio of magnetosonic velocity to poloidal infall velocity ($v_{\rm ms}/v_{\rm pol}$, given $v_{\rm ms}=(v_{\rm A}^2 +c_s^2)^{1/2}$ where $v_{\rm A}$ is the Alfv\'en velocity and $c_s=0.2 \kms$) over the time with overlaid poloidal velocity vectors in black color
for the magnetized isothermal collapse with Model-G1A on scales of $6000 \,{\rm au} \times 6000 \, {\rm au}$ at significant time instances. The white line corresponds to ${\rm log_{10}\, [v_{\rm ms}/v_{\rm pol}]}=0$, implying $v_{\rm ms}=v_{\rm pol}$.}
\label{fig:G1A_vmsbypol_s6000}
\end{figure*}

\begin{figure*}[ht!]
\gridline{\fig{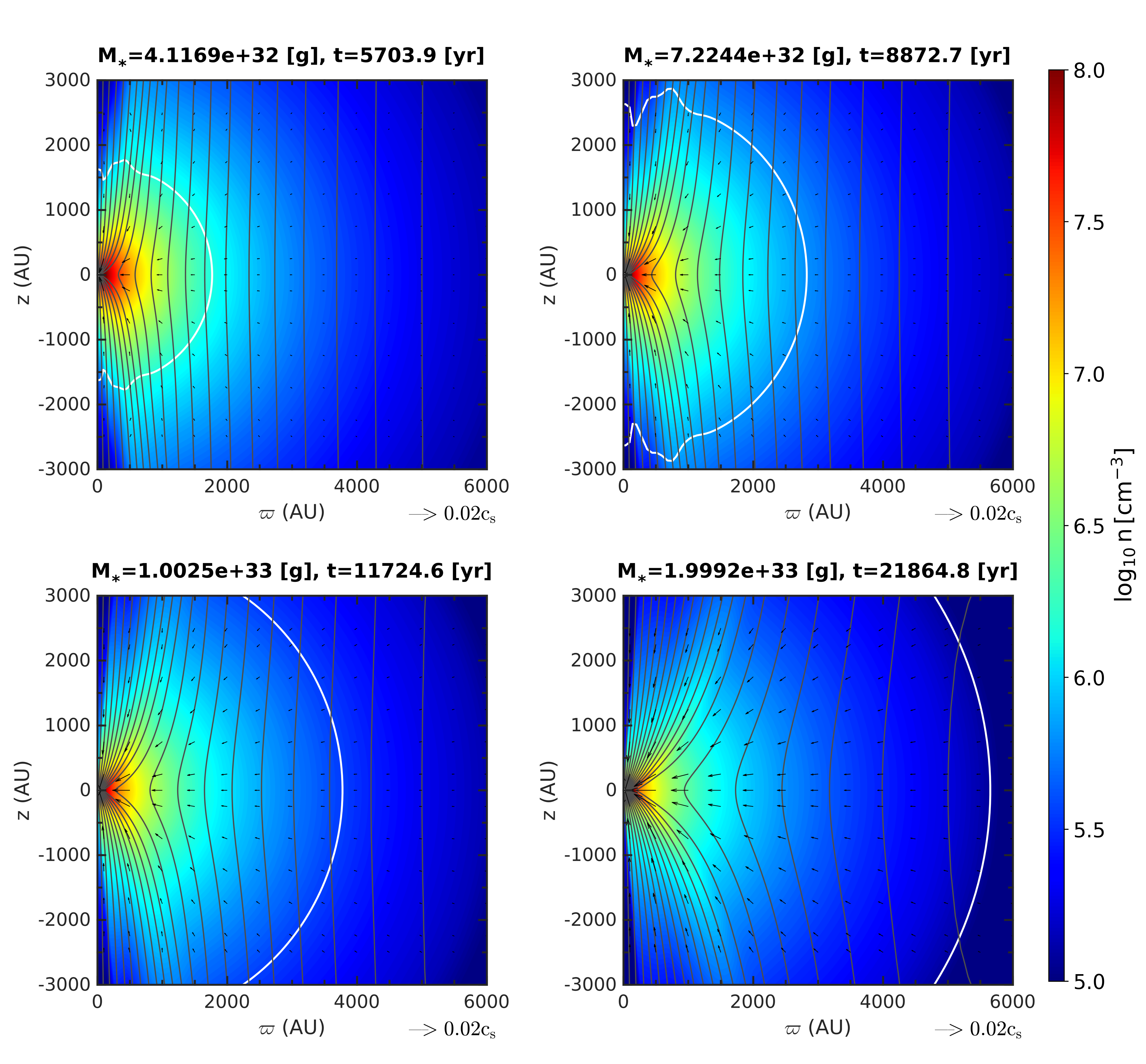}{\linewidth}{}
}
\vspace{-0.5cm}
\caption{Dynamical evolution of an intensity map of number density with overlaid magnetic flux lines and poloidal velocity vectors on scales of $6000 \,  {\rm au} \times 6000 \, {\rm au}$
for the magnetized isothermal collapse with Model-G1A on scale $6000 \,{\rm au} \times 6000 \, {\rm au}$ at significant time instances. The white line corresponds to ${\rm log_{10}\, [v_{\rm ms}/v_{\rm pol}]}=0$, implying $v_{\rm ms}=v_{\rm pol}$.}
\label{fig:G1A_dens_s6000}
\end{figure*}

\begin{figure*}[ht!]
\gridline{\fig{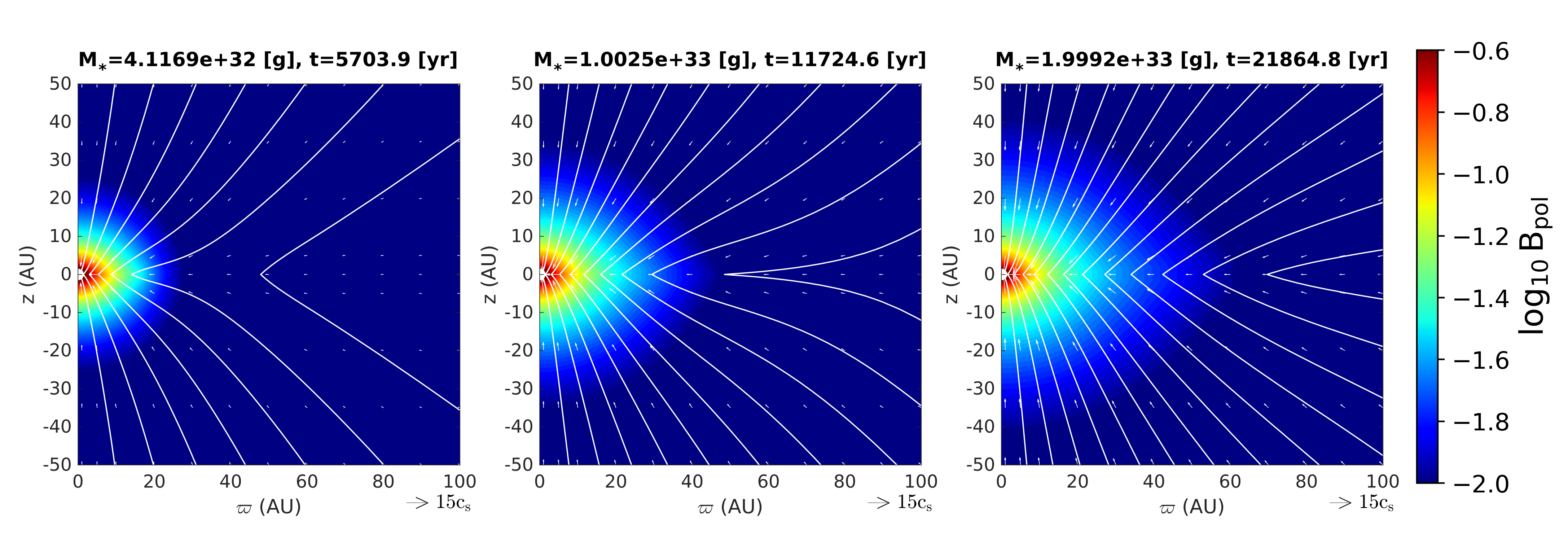}{\linewidth}{}
}
\vspace{-0.5cm}
\caption{Dynamical evolution of poloidal magnetic fields (expressed in Lorentz-Heaviside units) with overlaid poloidal velocity vectors as well as the magnetic flux lines in white color for the magnetized isothermal collapse with Model-G1A on scales of $100 \, {\rm au} \times 100 \, {\rm au}$ at significant time instances. 
}
\label{fig:G1A_Bpol_s100}
\end{figure*}

\begin{figure*}[ht!]
\gridline{\fig{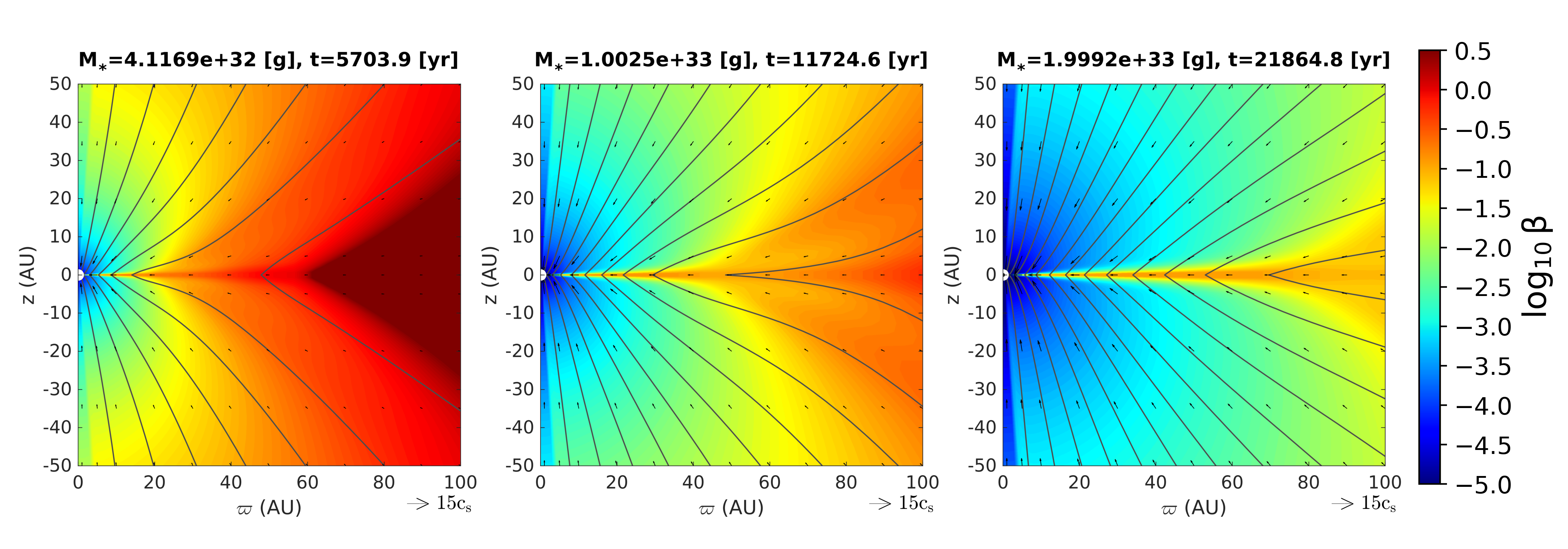}{\linewidth}{}
}
\vspace{-0.5cm}
\caption{Dynamical evolution of the  
plasma-$\beta$ with overlaid poloidal velocity vectors as well as the magnetic flux lines in black color  
for the magnetized isothermal collapse with Model-G1A on scales of $100 \,  {\rm au} \times 100 \, {\rm au}$ at significant time instances.
}
\label{fig:G1A_plasmabeta_s100}
\end{figure*}

\begin{figure*}[!ht]
\centering
\includegraphics[height=\textheight]
{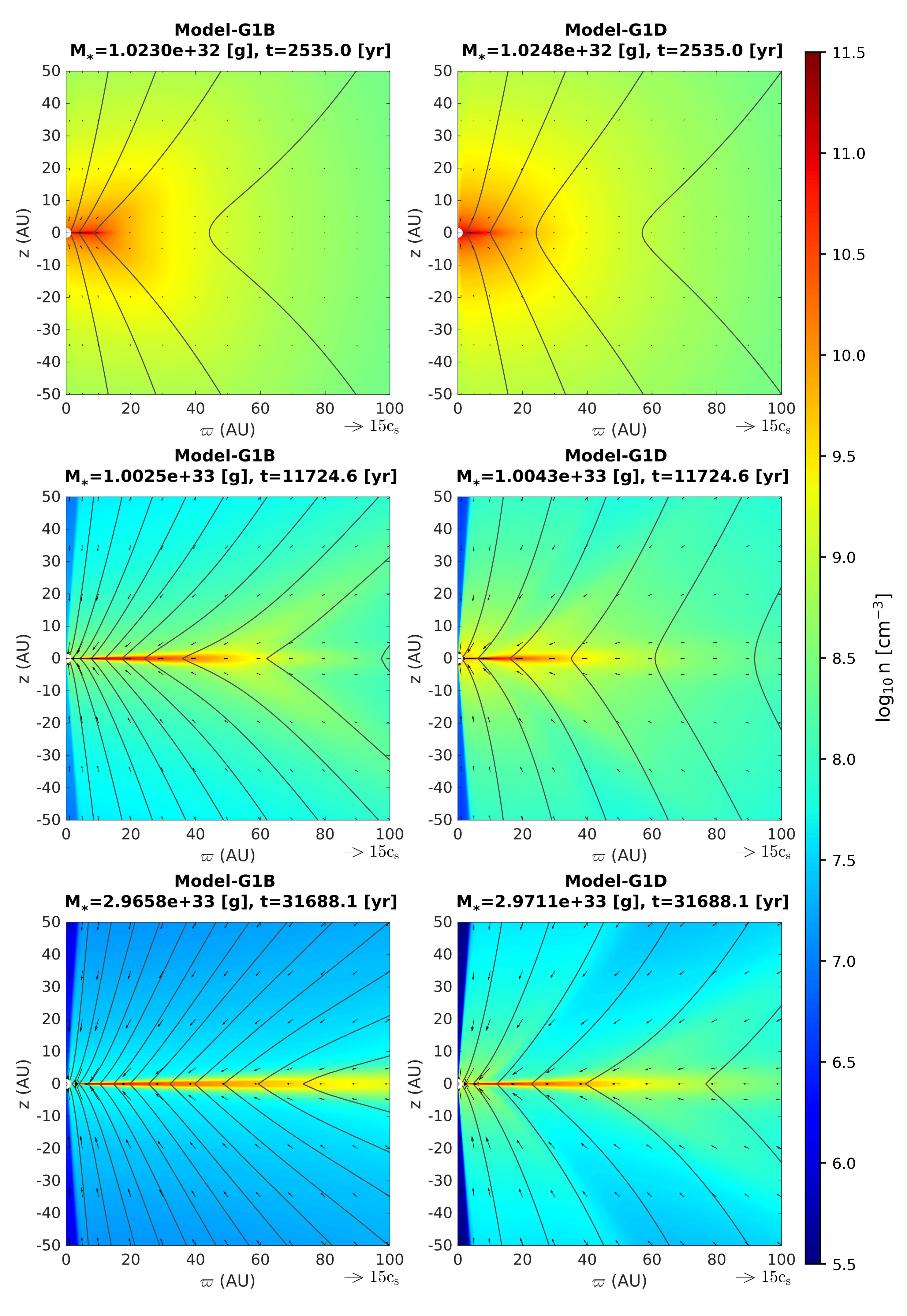}
\vspace*{-0.8cm}
\caption{Time sequence of the magnetized isothermal collapse for Model-G1B (left column) and Model-G1D (right column) on scales of $100 \,  {\rm au} \times 100 \, {\rm au}$ at selected epochs. A two-dimensional intensity map of number density with overlaid magnetic flux lines and poloidal velocity vectors. 
The ambipolar diffusivity is higher in Model-G1D than in Model-G1B.}
\label{fig:E0-BD-s100}
\end{figure*}

\begin{figure*}[!ht]
\gridline{\fig{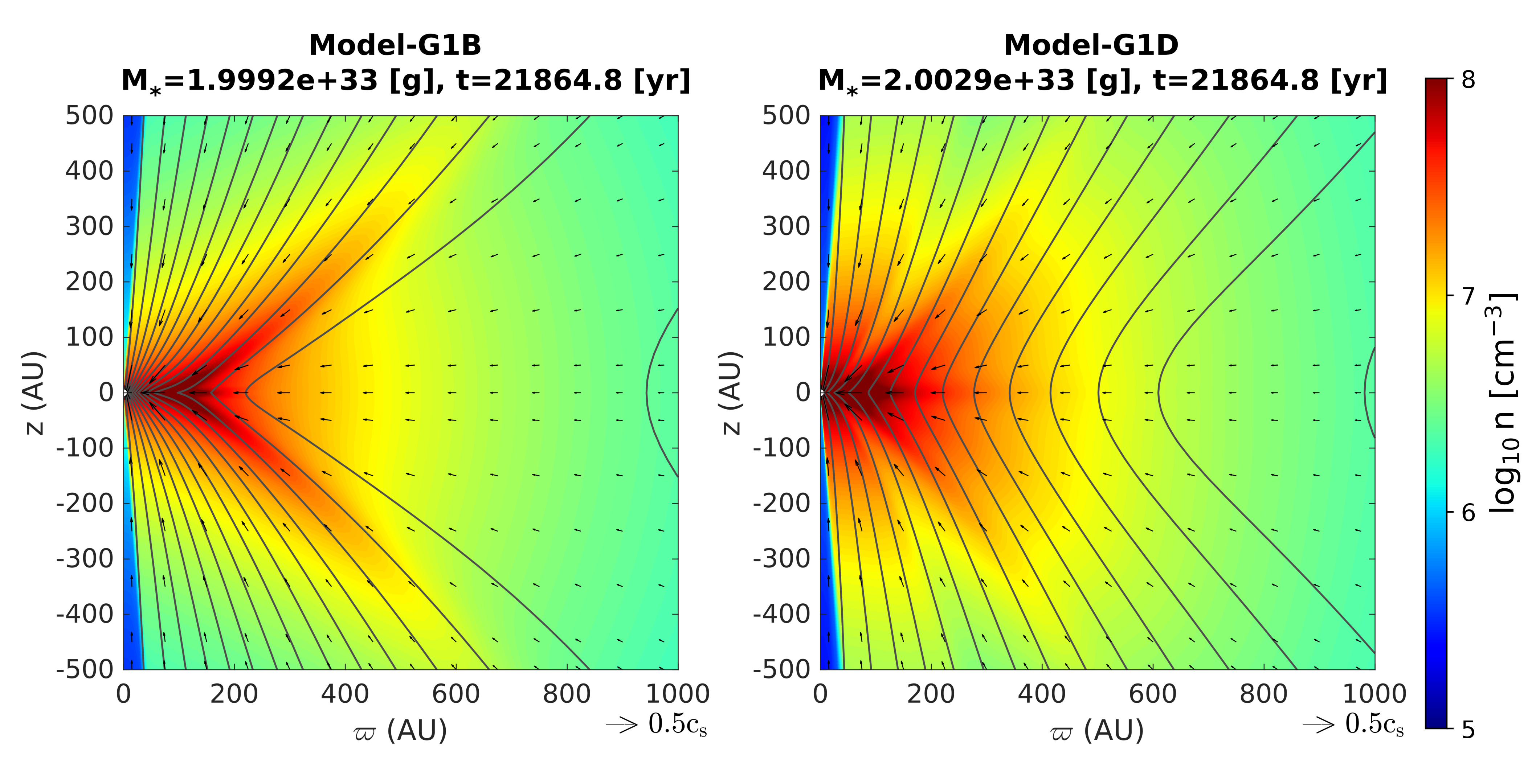}{\linewidth}{}
}
\vspace{-0.7cm}
\caption{A two-dimensional intensity map of number density with overlaid magnetic flux lines and poloidal velocity vectors on scales of $1000 \,  {\rm au} \times 1000 \, {\rm au}$ for Model-G1B  (left panel) and Model-G1D (right panel, with the higher ambipolar diffusivity than in Model-G1B) at a same time instance when the 
central point mass ($M_*$) reaches $\sim 1/3$ of the initial prestellar core mass.}
\label{fig:ModelG1BD-dens}
\end{figure*}

\begin{figure*}[!ht]
\gridline{\fig{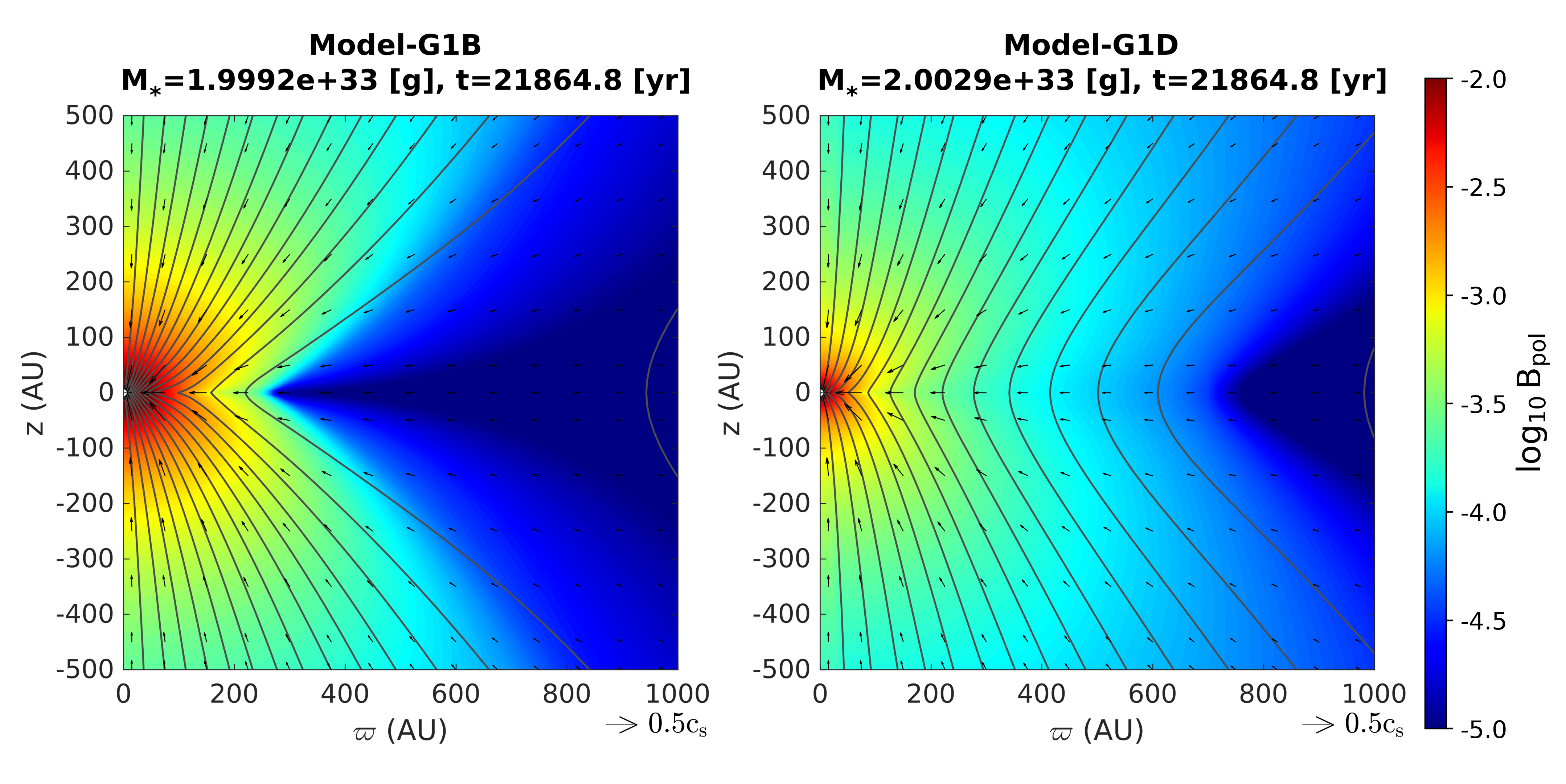}{\linewidth}{}
}
\vspace{-0.7cm}
\caption{A two-dimensional intensity map of poloidal magnetic field strength with overlaid magnetic flux lines and poloidal velocity vectors on scales of $1000 \,  {\rm au} \times 1000 \, {\rm au}$ for Model-G1B (left panel) and Model-G1D (right panel, with the higher ambipolar diffusivity than in Model-G1B) at a same time instance when the  
central point mass ($M_*$) reaches $\sim 1/3$ of the initial prestellar core mass.}
\label{fig:ModelG1BD-Bpol}
\end{figure*}

\begin{figure*}[!ht]
\gridline{\fig{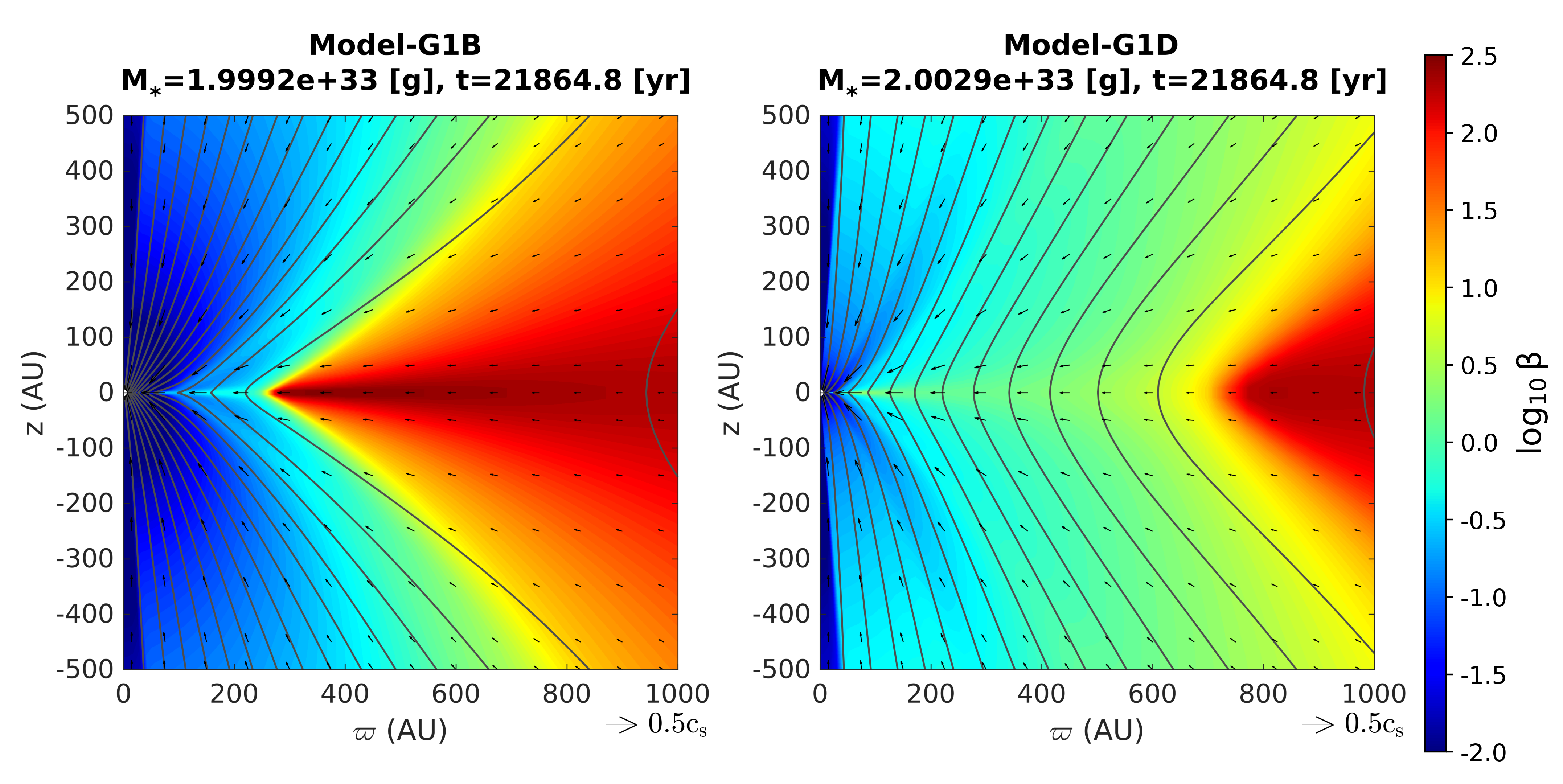}{\linewidth}{}
}
\vspace{-0.7cm}
\caption{A two-dimensional intensity map of plasma-$\beta$ with overlaid magnetic flux lines and poloidal velocity vectors on scales of $1000 \,  {\rm au} \times 1000 \, {\rm au}$ for Model-G1B  (left panel) and Model-G1D (right panel, with the higher ambipolar diffusivity than in Model-G1B) at a same time instance when the 
central point mass ($M_*$) reaches $\sim 1/3$ of the initial prestellar core mass.}
\label{fig:ModelG1BD-plasmabeta}
\end{figure*}

\begin{figure*}[!ht]
\centering
\gridline{\fig{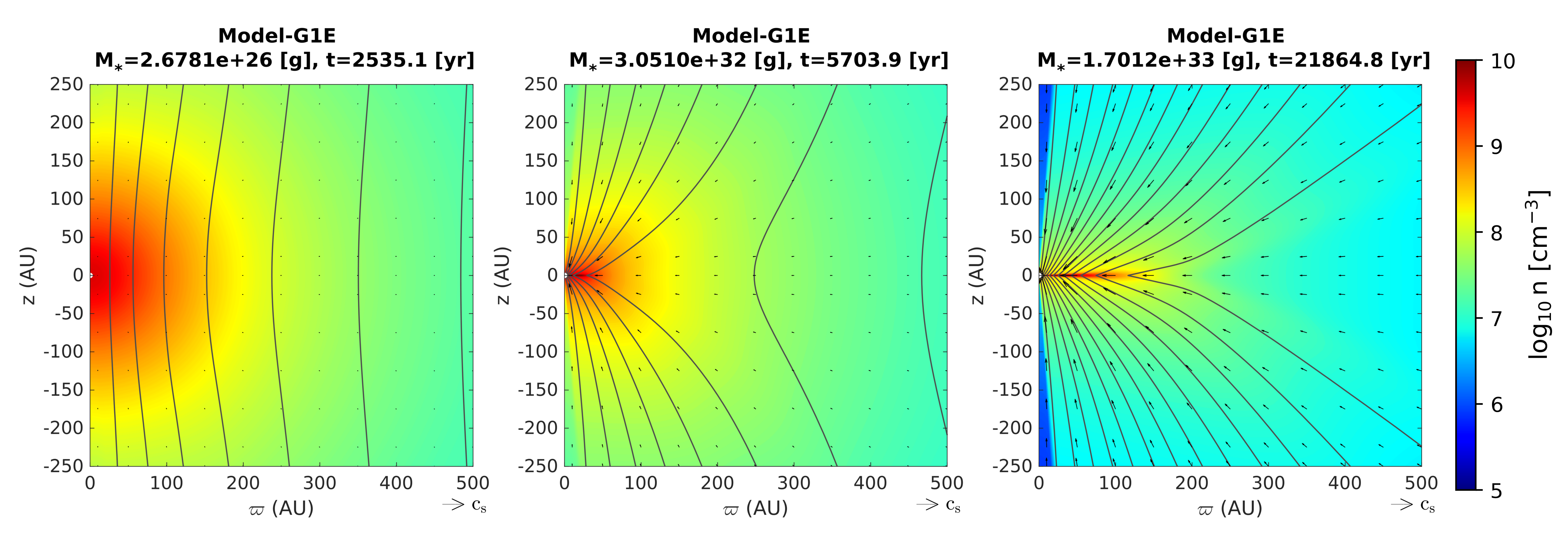}{\linewidth}{}
}
\caption{Time sequence of the magnetized isothermal collapse for Model-G1E (with a higher initial sound speed of $0.3\, \kms$) on scales of $500 \,  {\rm au} \times 500 \, {\rm au}$. 
A two-dimensional intensity map of number density with overlaid magnetic flux lines and poloidal velocity vectors.  
}
\label{fig:E0-E-s500}
\end{figure*}

\begin{figure*}[!ht]
\centering
\gridline{\fig{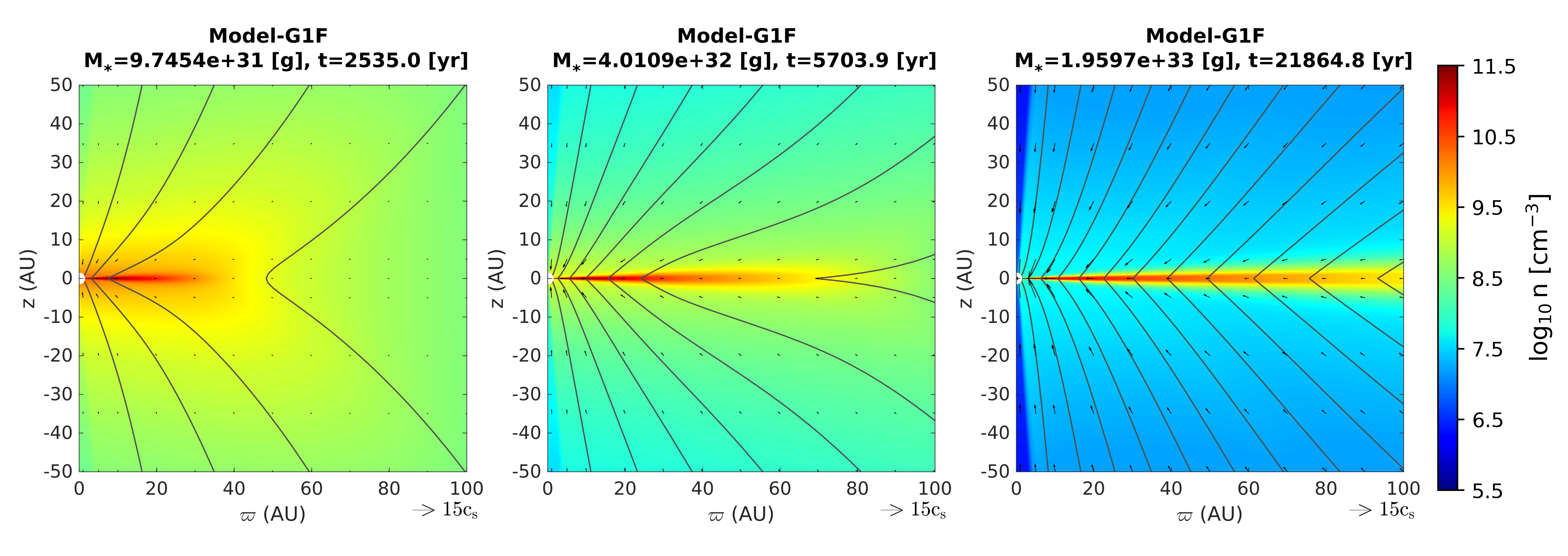}{\linewidth}{}
}
\vspace*{-0.8cm}
\caption{Time sequence of the magnetized isothermal collapse for Model-G1F (with higher initial magnetic fields as compared to Model-G1A) on scales of $100 \,  {\rm au} \times 100 \, {\rm au}$.  
A two-dimensional intensity map of number density with overlaid magnetic flux lines and poloidal velocity vectors. 
}
\label{fig:E0-F-s100}
\end{figure*}

Figure \ref{fig:E0-A1-s500} shows the intensity map of number density with overlaid magnetic flux lines and poloidal velocity vectors, illustrating the evolution of the isothermal magnetized collapse at distinct evolutionary time instances on scales $500 \,  {\rm au} \times 500 \, {\rm au}$ for Model-G1A, i.e., the fiducial isothermal model.  
Figure \ref{fig:E0-A1-s100} presents the same but on a zoomed-in scales of $100 \,  {\rm au} \times 100 \, {\rm au}$.  
It illustrates the collapse commencing with the initial conditions of a spherically symmetric configuration, featuring an initial density profile proportional to $r^{-2}$ \citep{Shu1977}, but with the vertical magnetic fields. 
As collapse progresses, infalling material starts to flatten out along the equatorial plane due to the influence of the magnetic fields. 
As a result, a magnetically supported disk-like structure, often known as pseudodisk starts to form in the equatorial plane \citep{GalliShu1993a,GalliShu1993b,Galli+2006}. 
The pseudodisk is likely produced by the pinching exerted by the tangential component of the Lorentz force that results from the dragging in of the cloud magnetic fields during the infall.
The pseudodisk extending up to few hundreds of au, plays out a significant role to channeling material towards the central point mass.

Figure \ref{fig:1D-densVr} shows the radial profiles of number density in the top panel as well as the radial component of the infall velocity $v_r$ in the bottom panel at distinct evolutionary time instances for Model-G1A. 
In the top panel, the black solid curve corresponds to the initial (referring to $t=0$) radial profile of the number density having a power-law behaviour of $\rho \propto r^{-2}$ with a flattening radius 
(refer to Sec.\ \ref{sec:ICs} for further details ).   
As the collapse proceeds, the density levels start to rise up, yielding a (nonideal) magnetic counterpart to the SIS collapse solutions of \cite{Shu1977}. 
As the runaway collapse ensues with the rapid infall, a point mass 
starts forming at the center of the collapse flow. 
The initiation of the rapid infalling
process progresses outward in the form of a radial expansion wave having a density profile consistent with a power law of $\propto r^{-3/2}$, whose tip of the head propagates at the magnetosonic speed. 
The bottom panel of Figure \ref{fig:1D-densVr} depicts that before the runaway collapse, 
infall speed stays at the subsonic. 
Once the inside-out collapse commences, within the expanding region of radial expansion wave, the material falls the fastest, at the speed of free-fall ($\sim r^{-1/2}$), while the outer parts of the envelope are also contracting, although at a slower rate than it does inside.

The Figure \ref{fig:masscompare-G1}(a) and (b)
illustrates time evolution of the mass of the central point mass (sink particle) 
and the corresponding mass accretion rate  
into that central sink particle ($\dot{M}_{\ast}$) in the units of $c_s^3/G$ for collapse models with  $\Gamma=$ 1, 1.2, 4/3, 1.4, 1.5. 
For Model-G1A, under isothermal magnetized collapse, $M_*$ grows linearly with time as seen in Figure \ref{fig:masscompare-G1}(a) 
and it describes a spatially constant mass accretion rate (in the units of $c_s^3/G$) with time as seen from Figure \ref{fig:masscompare-G1}(b). 
These features for the mass evolution of a central point mass are qualitatively consistent with the ones obtained in the SIS collapse \citep{Shu1977}, the counterpart under a hydrodynamic collapse.
For Model-G1A, the mass accretion rate has a scale of the order of $c_s^3/G$, being approximately 26 times higher because of the choice of our numerical parameters, for example, with the selection of a higher characteristic constant ($A=10.01$) of $\propto r^{-2}$ density profile of SIS collapse in the asymptotic limit (see Appendix \ref{sec:massApp} for further details) and the choice of a large supercritical prestellar core. 
In this section, we limit our discussion to the isothermal case only. Please refer to Sec.\ \ref{sec:polytropes} for the further discussion on the $M_*$ vs t curves (of Fig. \ref{fig:masscompare-G1}) corresponding to the nonisothermal collapse models.

Figure \ref{fig:G1A_vpol_s100} shows the dynamical evolution of poloidal velocity in the units of initial sound speed $c_s$ on scales of $100 \times 100 \, {\rm au}^2$ for Model-G1A 
at various epochs. 
It illustrates that material begins inflowing toward the center at marginally supersonic speeds soon after the onset of the collapse. 
The infall speed gradually increases by two order of magnitude within the inner regions.  
The enhanced infall speed is also linked to the mass reservoir available for accretion. 
This is because a larger mass reservoir exerts a stronger gravitational pull, accelerating the infalling material to highly supersonic speeds. 
In magnetized collapse, the pseudodisk becomes the primary means to channeling material towards 
the central sink particle in addition to the direct infall onto the upper and lower surfaces of the pseudodisk.  

Figure \ref{fig:G1A_vmsbypol_s6000} delineates the intensity map of the ratio of magnetosonic to poloidal velocity $v_{\rm ms}/v_{\rm pol}$ with overlaid magnetic flux lines and poloidal velocity vectors on scales of {\bf $6000 \times 6000 \, {\rm au}^2$}, demonstrating the magnetized inside-out collapse. 
The depicted white contour corresponding to $v_{\rm ms}= v_{\rm pol}$, is progressively expanding outward, featuring the radial expansion wavefront moving outward at the magnetosonic speed as the collapse continues from inside to outside, implying that the inner regions are collapsing at a faster rate than the outer ones in the earlier times.
The region within this white contour characterized by $v_{\rm ms} < v_{\rm pol}$, indicates the rapid infall even at the highly supersonic speed; (refer to Figure \ref{fig:G1A_vpol_s100}). 
Notably, just within the edge of the white contour, the magnetic field lines begin to bend. 
Whereas the region outside the white contour characterized by $v_{\rm ms} > v_{\rm pol}$, the magnetic field lines are 
still straight
indicating that collapse has not yet initiated in that region. 
Next, Figure \ref{fig:G1A_dens_s6000} shows the corresponding density maps along with the increasingly flared-out density structures featuring the magnetized inside-out collapse. 
The progression of bending of field lines right near the edge of white contour implies the magnetic decoupling driven by the nonideal MHD effects, facilitating collapse in the inner region while the outer shell, where collapse has yet to begin, remains unaffected.

Figure \ref{fig:G1A_Bpol_s100} presents the intensity map of the poloidal magnetic fields on scales of $100 \times 100 \, {\rm au}^2$ with overlaid poloidal velocity vectors as well as the magnetic flux lines in white color. 
It illustrates a gradual increase in the spatial map of $B_{\rm pol}$ (thus in magnetic pressure $\propto {\rm B^2 _{pol}}$) by one order of magnitude from around $100 \,{\rm au}$ further inward towards the central regions, shortly after the point mass
forms at the center. 
Eventually, the gradient becomes more pronounced and expands over a wider area  
with time as the collapse progresses.
This increase is attributed to the accumulation of magnetic fields along with freely infalling matter towards the center.

Figure \ref{fig:G1A_plasmabeta_s100} presents
intensity maps of plasma-$\beta$ ($= { P_{\rm gas}/P_{\rm mag}}$) on scales of $100 \times 100 \, {\rm au}^2$, overlaid with the poloidal velocity vectors and magnetic flux lines at selected epochs. 
It illustrates the dynamics of thermal pressure relative to magnetic pressure within the collapsing regions. 
As the collapse continues, the magnetic pressure appreciably becomes greater within the inner region as the intensity of the poloidal magnetic field increases (refer to Figure \ref{fig:G1A_Bpol_s100}). 
However, the thermal pressure seems to be greater than the magnetic pressure within the pseudodisk because of its higher accumulation of density.

\subsection{Consequences of the nonideal MHD effects} \label{sec:diffNImhdmodels}

Figure \ref{fig:E0-BD-s100} shows the intensity map of density with overlaid magnetic flux lines and poloidal velocity vectors, illustrating the evolution of isothermal magnetized collapse on scales of $100 \,  {\rm au} \times 100 \, {\rm au}$  
for Model-G1B and Model-G1D, highlighting the effects of variation in the ambipolar diffusion for a given Ohmic resistivity. 
The geometry of magnetic flux lines seem to be less pinched for Model-G1D as the magnetic tension is weakened compared to Model-G1B due to higher ambipolar diffusion. 
For Model-G1D, the pseudodisk grows smaller than that of the case with Model-G1B because higher ambipolar diffusivity is efficient at diffusing out the magnetic fields at a faster rate. 
Please refer to Figure \ref{fig:E0-A1-s100} for the respective comparison with Model-G1A. 
Note that, Model-G1B has a higher Ohmic resistivity than Model-G1A, but having the same measure of ambipolar diffusion in terms of neutral-ion coupling parametrization as in Model-G1A (refer to Table \ref{tab:MODELS}). 
Model-G1D has a higher neutral-ion coupling parametrization and a higher Ohmic resistivity than Model-G1A. 
From Figure \ref{fig:E0-BD-s100}, it is noticed that at an evolved stage (t $\sim$ 31.7\, Kyr), the magnetized collapse with Model-G1D reveals multiple branches in the infalling density structures above and below the pseudodisk than Model-G1B. 
This branching in the collapse flow can result from the enhanced decoupling of the magnetic fields (or the magnetic flux) from the infalling matter in the presence of higher ambipolar diffusion. 
Decoupling is stronger around the region of the separatrix than it is further inside. 
At scales of 90-100 au, the magnetic field lines are less pinched and more diffused more Model-G1D, deploying that magnetic field lines are being pulled at a slower rate than the matter is being pulled inward during the magnetized collapse under the presence of nonideal MHD effects.  
Whereas for model-G1B, we see more of a quasi-continuous density distribution for the same, that may develop stronger decoupling at a much later time.

Figure \ref{fig:ModelG1BD-dens}  
presents the intensity map of density with overlaid magnetic flux lines and poloidal velocity vectors on scales of $1000 \, {\rm au} \times 1000 \, {\rm au}$ at an advanced evolutionary stage when ${M_\ast \sim M_{\rm core}/3}$ for Model-G1B and Model-G1D. 
It prominently highlights the multiple branching in the density structures for Model-G1D 
at the scales of 250-500 au, 
consistent with the structures previously seen at smaller scales of $\sim$ 40 and 100 au in Figure \ref{fig:E0-BD-s100} (see the figure referring to $t\sim 21.9\, {\rm Kyr}$). 
These density branches are present even at the scales of 1000 au as depicted in Figure \ref{fig:G1A_dens_s6000} for Model-G1A. 
This feature of magnetized branching is
likely originating from efficient decoupling as a result of higher ambipolar diffusion.

Figure \ref{fig:ModelG1BD-Bpol} portrays the intensity maps of $B_{\rm pol}$ at the corresponding time instant for Model-G1B and Model-G1D, highlighting the distinction in magnetic field geometry due to variations in the strength of ambipolar diffusion. 
The higher the ambipolar diffusion (or nonideal MHD effects), the greater the decoupling between the matter and the magnetic fields as the field diffuses outward, thus allowing matter to flowing through the field lines without bending them significantly. 
In Figure \ref{fig:ModelG1BD-plasmabeta}, we see the variation in the corresponding intensity maps of plasma-$\beta$ for Model-G1B and Model-G1D as shown in the left and right column, respectively. 
The gradient in plasma-$\beta$ is stronger in Model-G1B than Model-G1D as the magnetic pressure is relatively larger within the inner region of about 200 au for the former, due to lesser diffusion of the magnetic fields.
In the plasma-$\beta$ map of Model-G1D, there is a kink-like structure at about 150 au in our model, which is consistent with the branching in the density structures for the same model, as seen in Figure \ref{fig:ModelG1BD-dens}. 


\subsection{Consequences of different initial $c_s$ and B-fields }
Figure \ref{fig:E0-E-s500} the intensity map of number density with overlaid magnetic flux lines and poloidal velocity vectors, illustrating the isothermal magnetized collapse on scales of $500 \times 500 \, {\rm au^2}$ for Model-G1E, the case with a higher initial sound speed of $c_s=0.3 \, {\rm km} \, {\rm s}^{-1}$ than that of Model-G1A. 
The enhanced thermal support significantly delays the collapse by a few ${\rm Kyr}$ as compared to Model-G1A(see Figure \ref{fig:E0-A1-s500} for further comparisons).

Figure \ref{fig:E0-F-s100} presents the intensity map of number density with overlaid magnetic flux lines and poloidal velocity vectors, illustrating the isothermal collapse starting with an initial higher magnetic field strength for Model-G1F on scales of $100 \times 100 \, {\rm au^2}$. 
This case shows an enhanced flattening in the equatorial plane due to the stronger magnetic fields as compared to Model-G1A. 
Greater pinching in the magnetic fields lines occurs due to the increased tension force resulting from the stronger magnetic field. 
The stronger magnetic fields delays the isothermal collapse process, resulting in a relatively slower initial growth of central point mass compared to Model-G1A, when compared with the corresponding timestamps for Model-G1F.  
See Figure \ref{fig:E0-A1-s100} for further comparisons with the fiducial model.

\subsection{B-field morphologies in magnetized collapse} \label{sec:separatrix}
\begin{figure}[!ht]
\epsscale{1}
\plotone{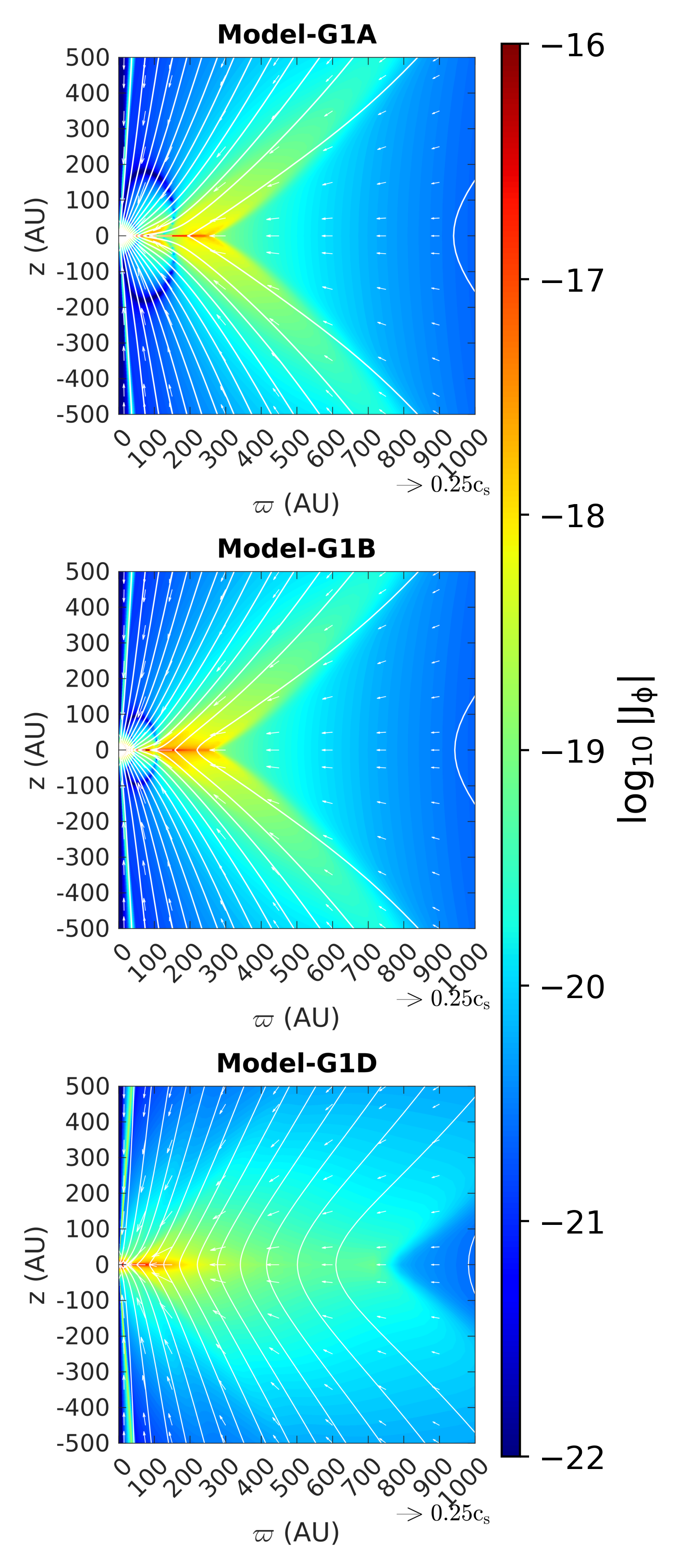}
\caption{A two-dimensional intensity map of the logarithmic of the absolute magnitude of current density $J_{\phi}\equiv(\nabla\times\vec{B})_\phi$, 
calculated in Lorentz-Heaviside units, on scales of $1000\, {\rm au} \,\times 1000 \, {\rm au}$, with overlaid magnetic flux lines and poloidal velocity vectors; evolution of the magnetic fields as shown for the models with increasing nonideal MHD strengths (from top to bottom): Model-G1A, Model-G1B, and Model-G1D referred to a time instance of $t\sim\,$21.9 Kyr.}
\label{fig:Model-G1ABDseparatrix}
\end{figure}

Over the course of magnetized collapse, the density distribution is not spherically symmetric but depends on the
mass loading of magnetic fields lines \citep{GalliShu1993a, GalliShu1993b, Galli+2006}.  
The pseudodisk shrinks with higher nonideal MHD strengths  due to dissipation/diffusion of the magnetic fields  (refer to Figure \ref{fig:E0-BD-s100}). 
We explore the dynamical evolution of magnetic fields with infall for the nonideal MHD collapse models as presented in Figure \ref{fig:Model-G1ABDseparatrix} that shows a two-dimensional intensity map of logarithmic absolute magnitude of current density $J_{\phi}$ in Lorentz-Heaviside units, on scales of $1000\, {\rm au} \,\times 1000 \,  {\rm au}$ for Model-G1A (top panel), Model-G1B (middle panel), and Model-G1D (bottom panel) with overlaid magnetic flux lines and poloidal velocity vectors. 
These models are chosen as test cases to examine the effects of variations in nonideal MHD strengths in the distribution of current density.
The intensity maps of $J_{\phi}$ 
highlight that the lower the nonideal MHD effects, the more pronounced the pinching of magnetic fields in the midplane.
We inspect the presence of finite thickness current sheets by calculating the electrical current density, $J_{\phi}$, which is a quantitative measure of the spatial variability in $\vec{B}$.
Our study presents the geometry of the magnetic fields in a collapsing magnetized cloud, where there exists a critical field line (often termed as separatrix) separating field lines that have been pulled into an accreting central point mass
from those intersecting the equatorial plane, tied to matter in the infalling envelope \cite[as introduced by][see their Fig. 1]{Galli+2006}. 

Here are a few words about the ideal MHD scenario, not covered in the set of simulations of our current study. 
The concentration of the poloidal magnetic fields lines trapped by the 
central point mass under flux-freezing (ideal MHD), independent on the details of the initial state, creates a divergent strong split-monopole configuration in which the magnetic fields strength increases as the inverse square of the distance from the center as anticipated by \cite{GalliShu1993b}.
By contrast to the ideal MHD results as mentioned in the literature, 
our results from the nonideal MHD collapse as depicted in Figure \ref{fig:Model-G1ABDseparatrix} shows that the pseudodisk cannot support infinitely large current leading to magnetic forces that balance the radial force of the gravity, thus preventing the formation of an ideal split-monopole configuration consisting of an infinite current sheet at the equatorial plane, often termed as the gravomagneto catastrophe \citep{shu1995,Galli+2006,AdamsShu2007}.

The pivotal instance of gravomagneto catastrophe refers to the transition from the prestellar to protostellar, often characterized by the developements of gradient in the density going from quasi-flat to $\propto r^{-3/2}$ as well as the inflow velocity rising from nearly zero (but nonzero) to the speed of free-fall (ballistic). 
Nonideal MHD effects play a crucial role in diffusing and dissipating the magnetic field, which in turn prevents the split-monopole configuration from occurring during the collapse by instead generating a finite-thickness current sheet at the equatorial plane with a finite value of $J_{\phi}$. 
Under the presence of higher nonideal MHD strengths (Ohmic resistivity and/or ambipolar diffusivity), it allows the numerics to better resolve the thickness of the current sheet at the equatorial plane.  
The consequences of higher nonideal MHD effects are seen in the 
the dynamical morphology of the magnetic fields  during the infall as seen from the respective intensity maps of $J_{\phi}$. 
Numerical resistivity can also mimic the physical effects of magnetic resistivity/diffusivity in smoothing out sharp gradients to promote numerical convergence. 
Therefore, the dissipation of dynamically important
levels of magnetic fields is a fundamental requisite for the formation of disks around young stars, which leads to the scenario of magnetized collapse under the presence of nonideal MHD effects.


\section{Results from the magnetized Collapse with stiff EOS} \label{sec:nonisothermal_collapse}
We extrapolate the study of magnetized collapse encompassing a wide range of EOSs with harder $\Gamma \, (> 1)$ to understand the consequences of stiffness in the EOSs on the infall ranging from the cloud scale down to 1 au in the near proximity of the central point mass
In the first part (Sec.\ \ref{sec:polytropes}), our results show the collapse cases with
$\Gamma=1.2$, 1.33 (= 4/3), 1.4, 1.5 
and acknowledge the criticality at $\Gamma=4/3$ complemented by the magnetized virial theorem (see the derivation in Sec.\ \ref{sec:virial}). 
In the latter part (Sec.\ \ref{sec:gamma5by3}), we study the detailed consequences with monoatomic adiabatic value $\Gamma=5/3=1.67$ as inspired by the literature and past studies and discuss the comparison in Sec \ref{sec:discussions}.

\subsection{Cases with $1 <\Gamma \leq 1.5$}\label{sec:polytropes}

\begin{figure*}[!ht]
\gridline{\fig{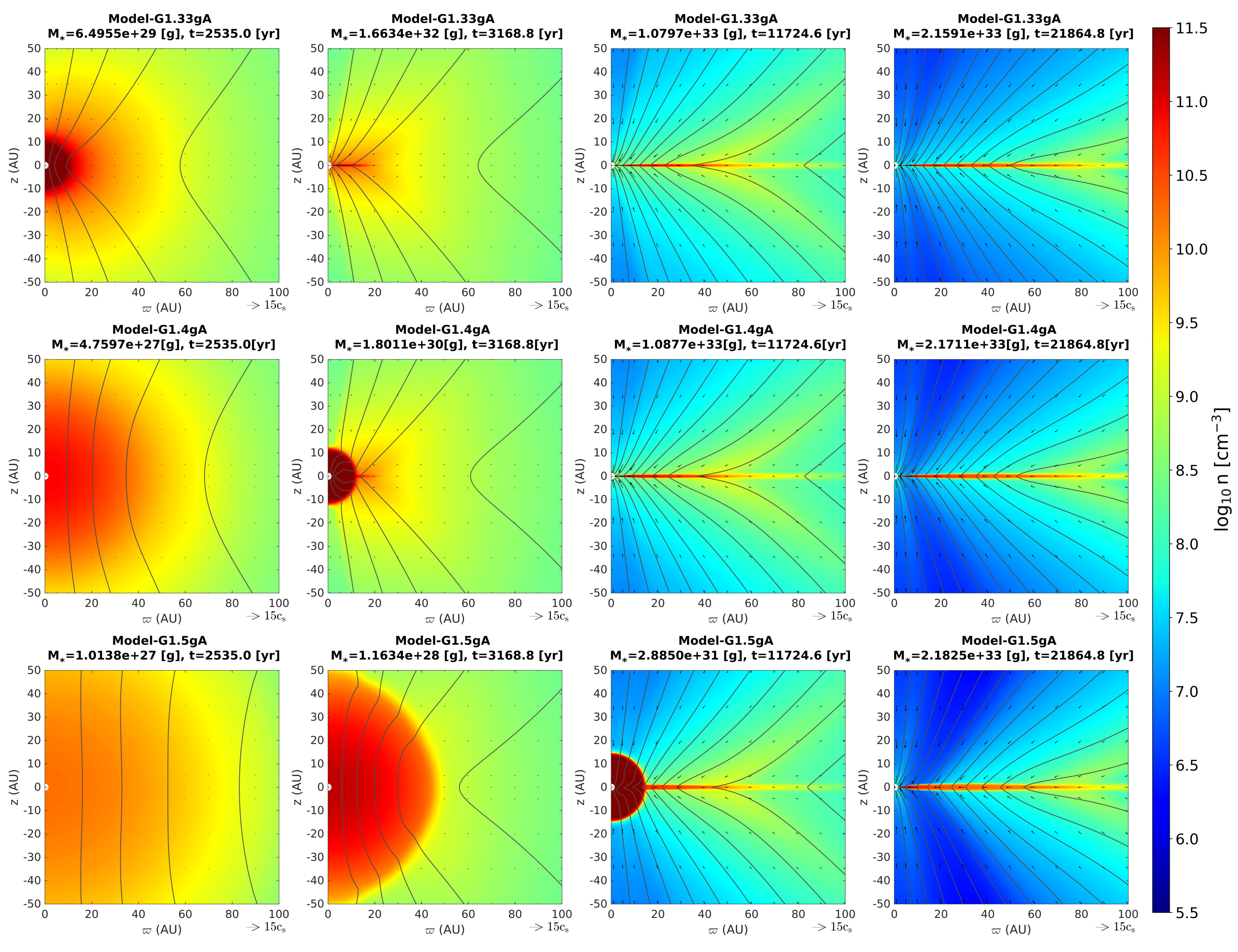}{\linewidth}{}
}
\caption{Time sequence of the magnetized non-isothermal collapse models with global polytropic indices: Top panel: ModelG1.33gA (with $\Gamma =1.33$), Middle Panel: ModelG1.4gA (with $\Gamma =1.4$), Bottom Panel: ModelG1.5gA (with $\Gamma =1.5$). 
A two-dimensional intensity map of number density depicted on
scale $100 \, {\rm au} \times 100 \, {\rm au}$ with overlaid magnetic flux lines and poloidal velocity vectors. 
All of the above cases are presented for the fiducial values of nonideal MHD effects, given the collisional coupling parametrization of the ambipolar diffusion is set at $3.15 \times 10^{-3}$ and Ohmic resistivity is set at $10^{17} \, {\rm cm}^2 \, {\rm s}^{-1}$.}
\label{fig:globalmodels}
\end{figure*}

In this section, we present the study for the magnetized (MHD) collapse cases with $\Gamma>1$ to up to $\Gamma=1.5$. 
First, we studied the cases with a global polytropic $\Gamma=$1.2 (Model-G1.2gA), 1.33 (Model-G1.33gA), 1.4 (Model-G1.4gA), and 1.5 (Model-G1.5gA)  for the same strengths of nonideal MHD effects, initial local sound speed, and initial magnetic fields, as in Model-G1A (see Sec.\ \ref{sec:IsothermalCollapse}). 
See Table \ref{tab:MODELS} for the model parameters.

The upper row of Figure \ref{fig:globalmodels} depicts the time sequence of the magnetized collapse models for Model-G1.33gA 
It shows that the infall happens continuously throughout the (nonideal) MHD collapse and as a result the central point mass steadily builds up the mass.  
The evolutionary stages of the magnetized collapse with $\Gamma$ no harder than 4/3 qualitatively behave in the similar way to that of an isothermal collapse (see Figure \ref{fig:E0-A1-s100} and Sec.\ \ref{sec:IsothermalCollapse}), indicating an EOS with a $\Gamma$ up to $4/3$ is soft enough to aiding in the direct mass growth of the central point mass. 
Whereas, for the respective nonisothoermal global collapse models with $\Gamma=1.4,1.5$, it is noticed that the collapse with a $\Gamma$ harder than $4/3$ allows the formation of a condensed sphere within the inner central region of the infalling collapse flow since the beginning of the magnetized collapse as seen in Figure \ref{fig:globalmodels}.

Figure \ref{fig:masscompare-G1} shows the time evolution of the 
mass of the central point mass (sink particle) 
and the corresponding mass accretion rate 
into the central sink particle 
(in the units of $c_s^3/G$) for different nonisothermal collapse models along with the reference line corresponding to the isothermal fiducial model. 
The mass evolution curve for $\Gamma=1.2$ qualitatively follows the same trend as seen with a choice of $\Gamma=1$ (Figure \ref{fig:masscompare-G1}a). 
The respective curve with $\Gamma=1.33$ shows a subtle deviation from the SIS collapse at the beginning, followed by a linear increase in $M_*$ over time, similar to the SIS model.
Notably, for $\Gamma=1.4$ and 1.5, a pronounced vertical jump is noticed in the temporal profile of $M_*$ at the beginning of the collapse, followed by a linear mass growth as that of SIS. 
This vertical jump in the $M_*$ corresponds to the shrinking of the condensed spheres, which were initially present at the center of the collapse flow and did not allow the direct infall into the central sink particle. 
As this vertical jump appears, the evolution of $M_*$ vs $t$ deviates from that of SIS collapse model,  subsequently aligning with SIS as these condensed spheres shrink. 
See Fig. \ref{fig:mass-Gamma1.67} and Sec. \ref{sec:gamma5by3} for further discussion on the evolution of $M_*$ vs $t$ for $\Gamma=5/3$. 
Such distinctions in the evolution of $M_*$,
as observed in our results, delineate that a $\Gamma$ harder than 4/3 is no longer able to provide sufficient cooling to allow the direct 
mass growth of the central point mass, as complemented by the magnetized virial theorem (Eq. \ref{eq:Gammavirial}).

As a result of the magnetized collapse, the pseudodisk starts to grow in the equatorial plane from the outer edge of these condensed spheres. 
Such condensed spheres supported by the enhanced thermal pressure, temporarily inhibit the direct infall into the central sink particle
and do not allow the direct mass growth of that sink particle. 
The lifetime of these condensed spheres depends on the stiffness of the EOS, with harder EOS allowing longer survival (see further discussions in Sec. \ref{sec:discussions}). 
As their self-gravitational pull overcomes thermal and magnetic pressure, these spheres gradually shrink, enabling the central point mass to resume growing.

From Figure \ref{fig:globalmodels}, it is also noticed that the size of such condensed spheres depends on the choice of $\Gamma$. 
For a stiffer $\Gamma$, the resulting change in pressure for a given density change is greater, allowing the gas to support a larger volume against the gravitational collapse. 
The gas is less gravitationally compressible for a harder polytropic EOS than it is for a softer EOS, so it resists the compression more effectively. 
This can be realized from the balance between the self-gravity and thermal pressure within these (transient) condensed spheres that leads to the variation in the degree of compression with increasing gas pressure as $\Gamma$ becomes harder.
Therefore for a given mass, with a harder EOS it tends to occupy a larger volume compared to a system with a softer EOS, which causes the initial volume or radius/size of the condensed spheres larger with harder polytropic EOSs (see further in Sec. \ref{sec:discussions}).

In addition to the above, we examine the magnetized collapse cases for harder EOSs using a broken power-law in the form of $\Gamma=1 \ {\rm for}\, \rho < \rhostiff$ and $\Gamma=1.2, 1.33, 1.4, 1.5 \ {\rm for}\,  \rho \geq \rhostiff$, given $\rhostiff=10^{-15}\, {\rm g \, {cm}^{-3}}$. 
Here, $\rho_{\rm stiff}$ serves as a measure for the size of such a condensed sphere, implying the volume within which $\rho \geq \rhostiff$ is appeared as condensed spheres. 
See further discussions on $\rho_{\rm stiff}$ in Sec.\ \ref{sec:rhostiff}.
Please see Table \ref{tab:MODELS} for the model parameters of Model-G1.2S15A, Model-G1.33S15A, Model-G1.4S15A, and Model-G1.5S15A.
By numerically setting $\rhostiff$ to a smaller value allows the evolution to follow similar trends as in the global cases, but with subtle changes that may be triggered by the low density threshold that does not cause
any significant differences in the dynamical sequence of the collapse.

\subsection{Cases with $\Gamma=5/3$}
\label{sec:gamma5by3}

In this section, we present a comprehensive analysis of the magnetized collapse for a hard EOS with $\Gamma=5/3=1.67$ (labeled as ``G1.67'' in the corresponding figures) using a broken power-law in the form of $\Gamma=1 \ {\rm for}\, \rho < \rhostiff$ and $\Gamma=5/3 \ {\rm for}\,  \rho \geq \rhostiff$, given $\rhostiff$ controls the level of stiffening in density, which can be considered as an equivalent magnetic counterpart to the Larson's hydrodynamic collapse \citep{Larson1969}.

\subsubsection{Effects of $\Gamma=5/3$ on the collapse dynamics}
\label{sec:gamma5by3S14}
\begin{figure*}[!ht]
\gridline{\fig{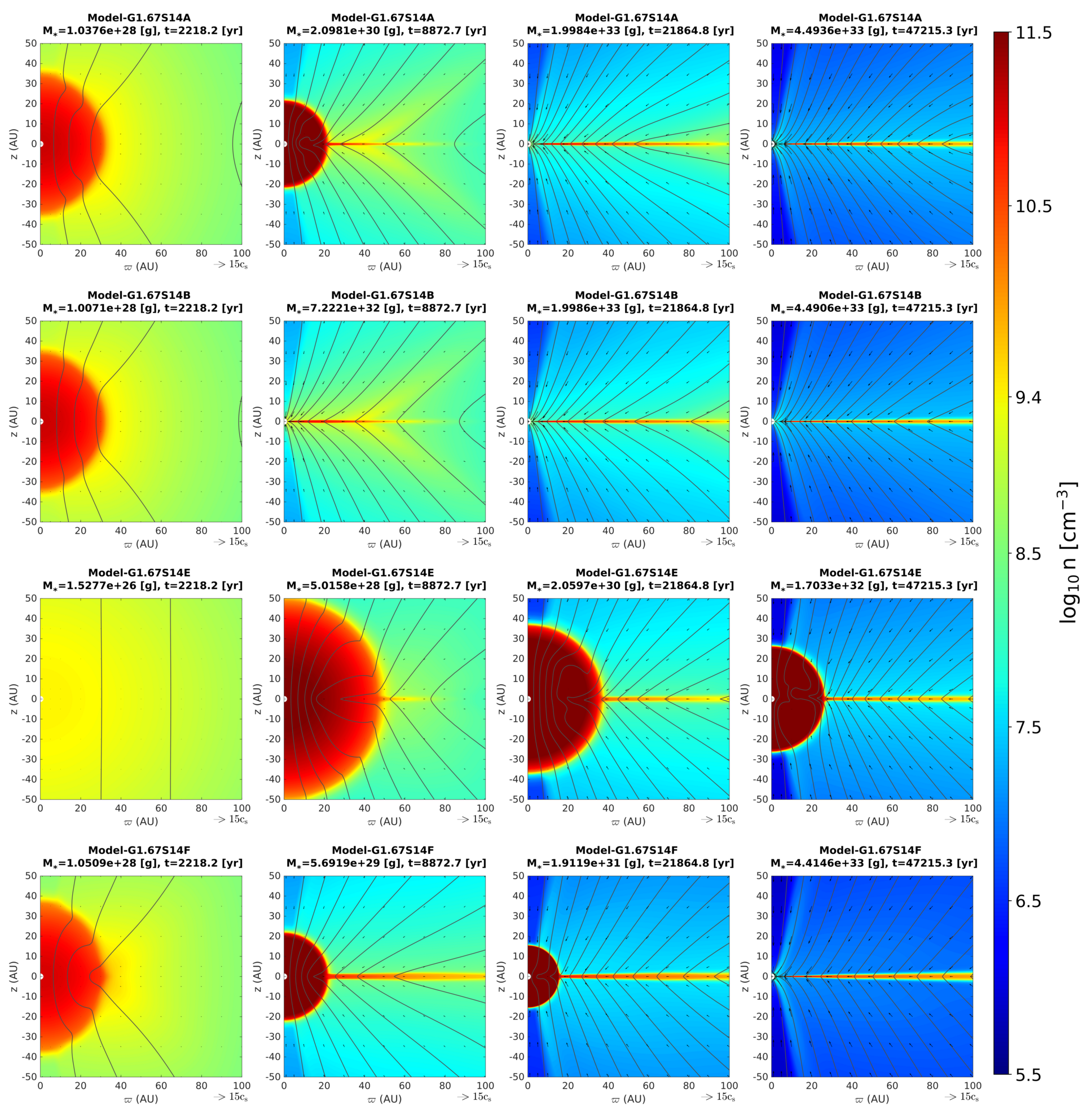}{\linewidth}{}
}
\vspace{-0.7cm}
\caption{Time sequence of the magnetized non-isothermal collapse models with $\Gamma=5/3$ and a $\rhostiff$ of $10^{-14}\, {\rm g}\cm^{-3}$ ($\nstiff = 10^{10.5} \, \cm^{-3}$): Model-G1.67S14A in the first row, i.e., the model with fiducial nonideal MHD effects;  Model-G1.67S14B in the second row, i.e., the model with a higher Ohmic resistivity than the former; Model-G1.67S14E in the third row, i.e., the model with a higher initial $c_s$ than Model-G1.67S14A; Model-G1.67S14F in the fourth row, i.e., the model with a higher B-fields than Model-G1.67S14A. 
A two-dimensional intensity map of number density with overlaid magnetic flux lines and poloidal velocity vectors on scales of $100 \,  {\rm au} \times 100 \, {\rm au}$ at distinct time instances.}
\label{fig:g1.67S14ABEF}
\end{figure*}

Figure \ref{fig:g1.67S14ABEF} depicts the time evolution of the magnetized collapse with an EOS of $\Gamma=5/3$, for a $\rho_{\rm stiff}$ of  $10^{-14} \, {\rm g}\, {\rm cm}^{-3}$. 
Because of greater stiffening of the EOS with $\Gamma=5/3$, there appears a quasi-stable condensed sphere within the innermost region of infalling collapse flow that stops the material to further falling into the center. 
The first row of Figure \ref{fig:g1.67S14ABEF} presents the evolutionary states for Model-G1.67S14A, the case with the same strengths of the nonideal MHD effects as in Model-G1A. 
The second row of Figure \ref{fig:g1.67S14ABEF} shows the evolution for Model-G1.67S14B, the case with higher Ohmic resistivity than Model-G1.67S14A. 
The higher Ohmic resistivity makes a substantial change in its dynamical evolution by efficiently dissipating the magnetic field. 
This further allows self-gravity to overpower the infall, causing the quasi-thermal balance inside the condensed sphere to gradually break down such that the condensed sphere shrinks (within 1 au) faster with time. 
This aids the central point mass to form and grow in mass efficiently due to the enhanced dissipation/diffusion of the magnetic fields caused by the nonideal MHD effects at play.   
Realization of the collapse dynamics within 1 au is beyond the scope of the current study.

The third row of Figure \ref{fig:g1.67S14ABEF}
depicts the time evolution of the magnetized collapse for Model-G1.67S14E, the model with a higher initial sound speed of $0.3 \, {\rm km}\, {\rm s}^{-1}$ compared to Model-G1S14A (refer to the first row of Figure \ref{fig:g1.67S14ABEF} for comparison). 
Due to greater thermal support against the collapse provided by the initial higher sound speed,  collapse starts at a later time $t\,\sim$ 4436.3~yr and the appearance of the condensed sphere is larger in size and longer-lived (although eventually shrinks) than it is in Model-G1.67S14A 
that causes a significant delay in the collapse. 
Lastly, the fourth row of Figure \ref{fig:g1.67S14ABEF} depicts the time evolution of the magnetized collapse for Model-G1S14F, the case with higher initial magnetic fields than Model-G1.67S14A. 
Because of stronger initial magnetic fields strength, there is a slightly enhanced flattening in the condensed sphere at a very early epoch, soon after the collapse commences. 
In addition to that, the magnetic field lines are more pinched due to greater tension and the pseudodisk starts to grow faster and larger in the equatorial plane than the respective case with Model-G1.67S14A. 
Hence, the condensed sphere takes a bit longer to shrink, which is reflected on the slight delay in the mass growth of the protostar.

\begin{figure*}[ht!]
\plotone{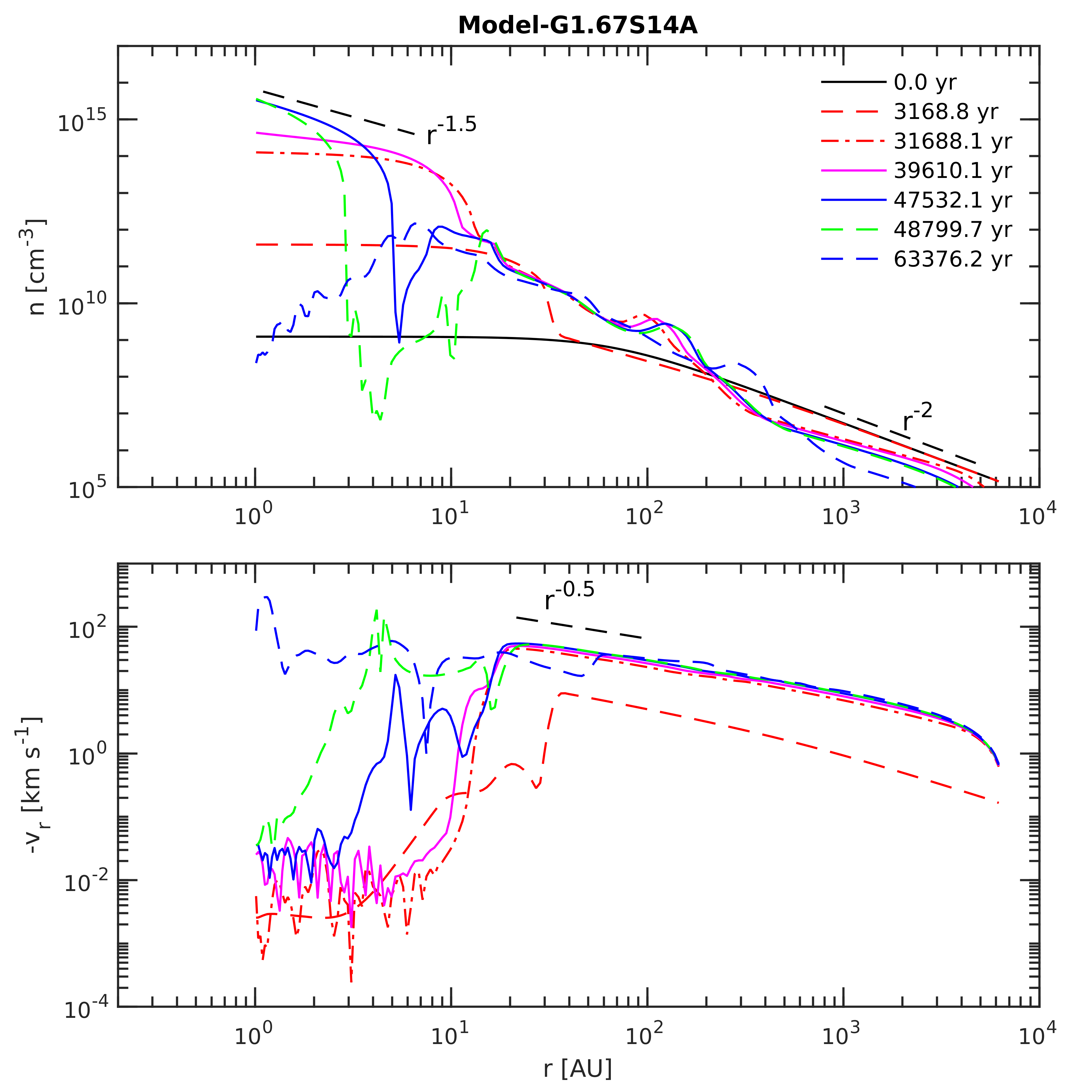}
\caption{Radial profiles of number density and radial velocity at the equatorial plane corresponding to different evolutionary time for Model-G1.67S14A. 
Refer to Figure \ref{fig:1D-densVr} to compare the infall characteristics with the fiducial isothermal model.}
\label{fig:1D-radialS14}
\end{figure*}
Figure \ref{fig:1D-radialS14} shows the radial profiles of number densities in the top panel as well as the radial infall velocity ${\rm v}_{\rm r}$ in the bottom panel at distinct evolutionary times for the nonideal MHD collapse with Model-G1.67S14A. 
Initially the density goes as $\rho \propto r^{-2}$ in the outer parts with a flattening across the inner regions once the collapse commences as similar to the SIS collapse \citep{Shu1977} and also for the magnetized isothermal collapse (see Sec.\ \ref{sec:IsothermalCollapse} and Figure \ref{fig:1D-densVr}).
But now, due to the formation of a condensed sphere caused by the stiffening of EOS to a $\Gamma=5/3$, the infall ceases for a substantial period as seen from the radial profiles of infall velocity at earlier times.  
Once the condensed sphere shrinks by about a factor of $\sim 2-3$ relative to its  initial size, the infall becomes dominant over the support provided by the thermal and/or magnetic pressure. 
As a result, material continues to be transported to the center that eventually leads to the 
central point mass formation.
Hence, there developes a density profile of $\rho \propto r^{-3/2}$ referring to the radial expansion wave within the innermost region of their density profiles after the point mass is formed at the center.

\begin{figure*}[!ht]
\gridline{\fig{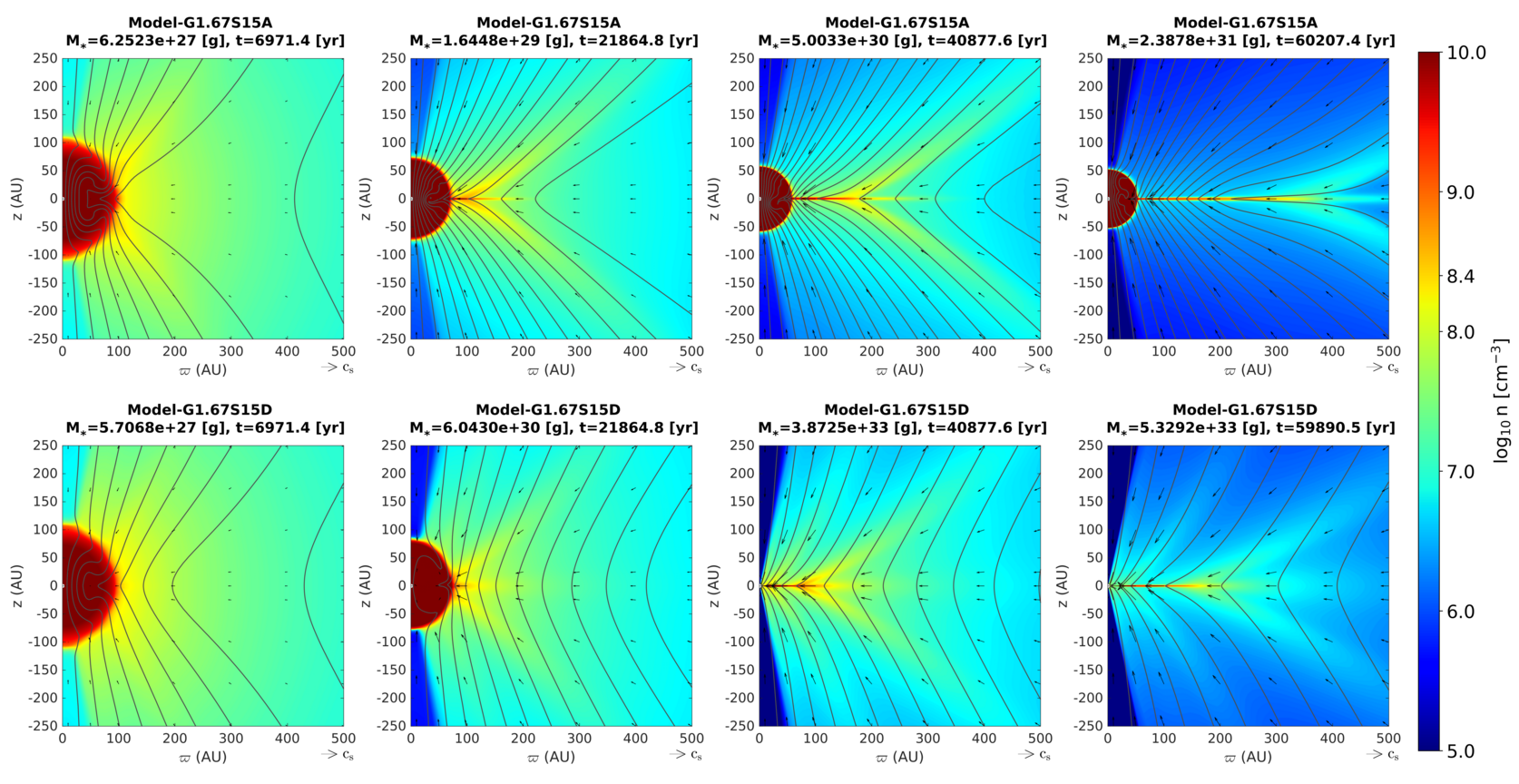}{\linewidth}{}
}
\vspace{-0.7cm}
\caption{Time sequence of the magnetized non-isothermal collapse models for Model-G1.67S15A (top panel) and Model-G1.67S15D (bottom panel) associated with a $\rhostiff$ of $10^{-15} \, {\rm g}\cm^{-3}$ ($\nstiff = 10^{8.4}\, \cm^{-3}$) showing the effects of different nonideal MHD strengths. 
A two-dimensional intensity map of number density with overlaid magnetic flux lines and poloidal velocity vectors on scales of $500 \,  {\rm au} \times 500 \, {\rm au}$. 
}
\label{fig:g1.67S15AD}
\end{figure*}

\begin{figure*}[!ht]
\gridline{\fig{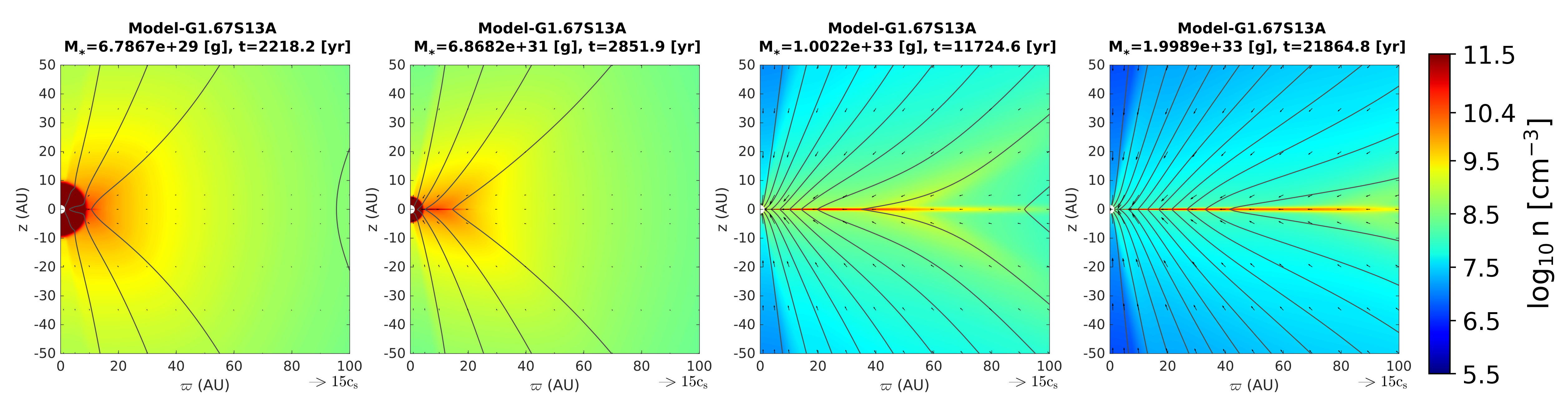}{\linewidth}{}
}
\vspace{-0.7cm}
\caption{Time sequence of the magnetized non-isothermal collapse models: Model-G1.67S13A associated with a $\rhostiff$ of $10^{-13} \, {\rm g}\cm^{-3}$ ($\nstiff = 10^{10.4}\, \cm^{-3}$) on scales of $100 \,  {\rm au} \times 100 \, {\rm au}$ at different evolutionary times. 
A two-dimensional intensity map of number density with overlaid magnetic flux lines and poloidal velocity vectors. 
}
\label{fig:g1.67S13A}
\end{figure*}


\subsubsection{Consequences of different $\rho_{\rm stiff}$}\label{sec:rhostiff}

We study the effects of $\rhostiff$ on the MHD collapse with the stiff EOS of $\Gamma=5/3$. 
Figure \ref{fig:g1.67S15AD} shows the collapse cases with a $\rhostiff$ of $10^{-15} \, {\rm g}\cm^{-3}$.
Top row of Figure \ref{fig:g1.67S15AD} show the respective time evolution of MHD collapse for Model-G1.67S15A, the model with the same strengths of nonideal MHD effects as in Model-G1A. 
From our exploration, we notice $\rhostiff$ serves as a measure of size of the quasi-stable condensed sphere appearing within the central collapse flow. 
With the decrement in $\rhostiff$ by one order of magnitude results in the formation of a larger condensed sphere than that of seen in Model-G1.67S14A. 
Because of the occurrence of the larger condensed sphere, the majority of the infalling material is being engulfed by it, maintaining a quasi-equilibrium state for a longer time. 
This prevents the material from falling further inward, thus inhibiting the formation of a point mass at the center.
The condensed sphere in this case seems to shrink in size relatively slowly and to be longer-lived than in Model-G1.67S14A, causing an insignificant growth to $M_{\ast}$. 
The bottom row of Figure \ref{fig:g1.67S15AD} shows the respective case with Model-G1.67S15D, the model with the higher nonideal MHD effects than in Model-G1.67S15A. 
The condensed sphere shrinks down at a much faster rate 
due to stronger diffusion/dissipation of the magnetic fields, thus enabling the self-gravitational contraction to overcome the magnetic and thermal support. 
The pseudodisk starts to grow from the outer boundary of the condensed sphere in the equatorial plane for both the cases. 
However, the extension of pseudodisk is smaller in the latter case due to stronger nonideal MHD effects. 
Note that, after the condensed sphere shrinks completely, the dynamics of collapse resembles the case of $\Gamma$ softer than 4/3.
In addition to this, branching in the infalling density structures are also observed (also see Figure \ref{fig:E0-BD-s100}) in this case at a much evolved stage due to enhanced magnetic decoupling.

On the other hand, Figure \ref{fig:g1.67S13A} shows the respective time evolution of MHD collapse for Model-G1.67S13A, the model with the same strengths of nonideal MHD effects as in Model-G1A, but with a higher $\rhostiff$ of $10^{-13}\massden$. 
Because of the increment in the $\rhostiff$, 
the condensed sphere decreases in size by a factor of $\sim2$ even at the early times and starts to shrink down faster than that of Model-G1.67S14A.
After the condensed sphere is completely shrunken within 1 au, the collapse resembles the case of with a $\Gamma$ softer than 4/3 (see Sec.\ \ref{sec:IsothermalCollapse} and \ref{sec:polytropes}).

\begin{figure*}[ht!]
\gridline{\fig{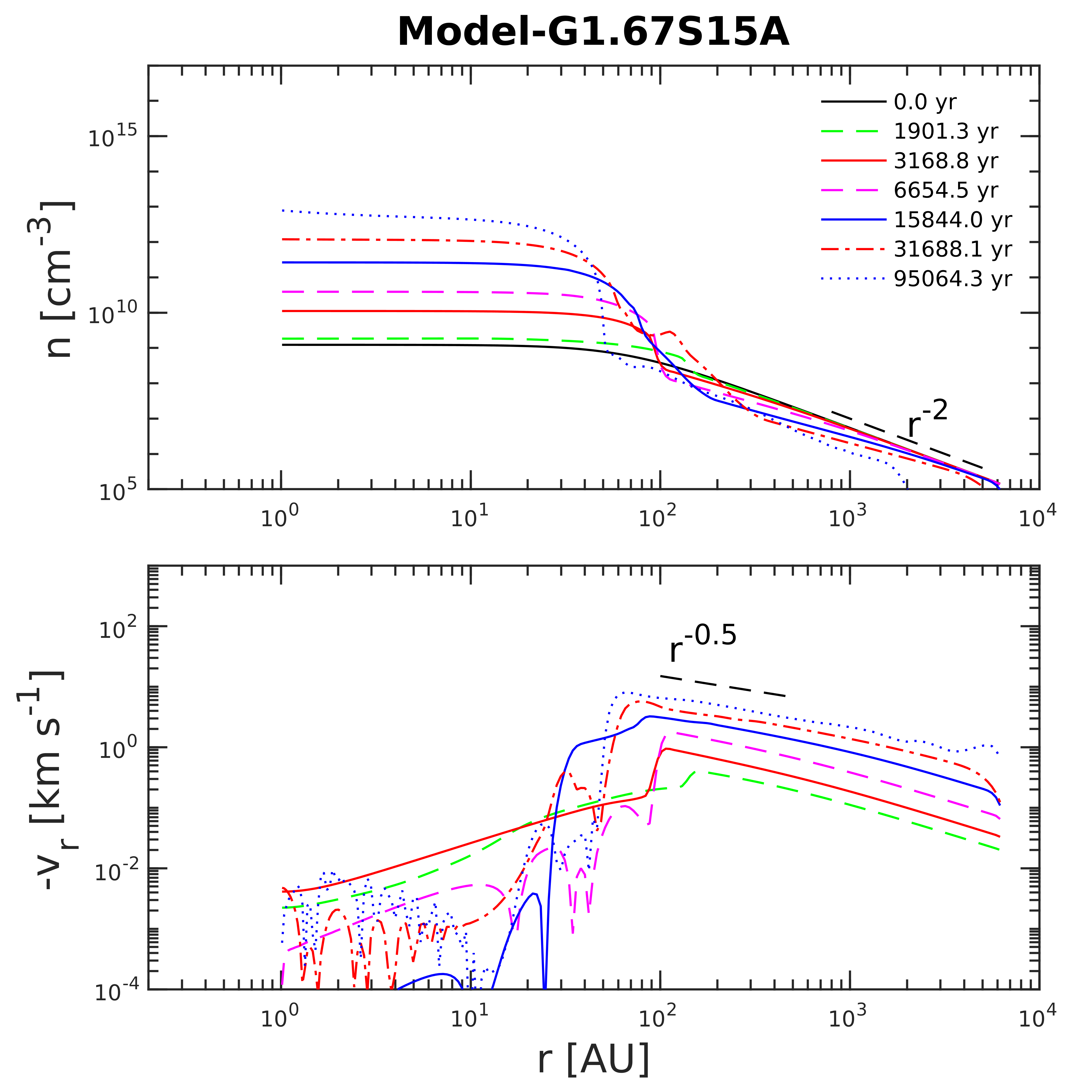}{0.55\textwidth}{(a)}
\hspace{-0.35cm}
\fig{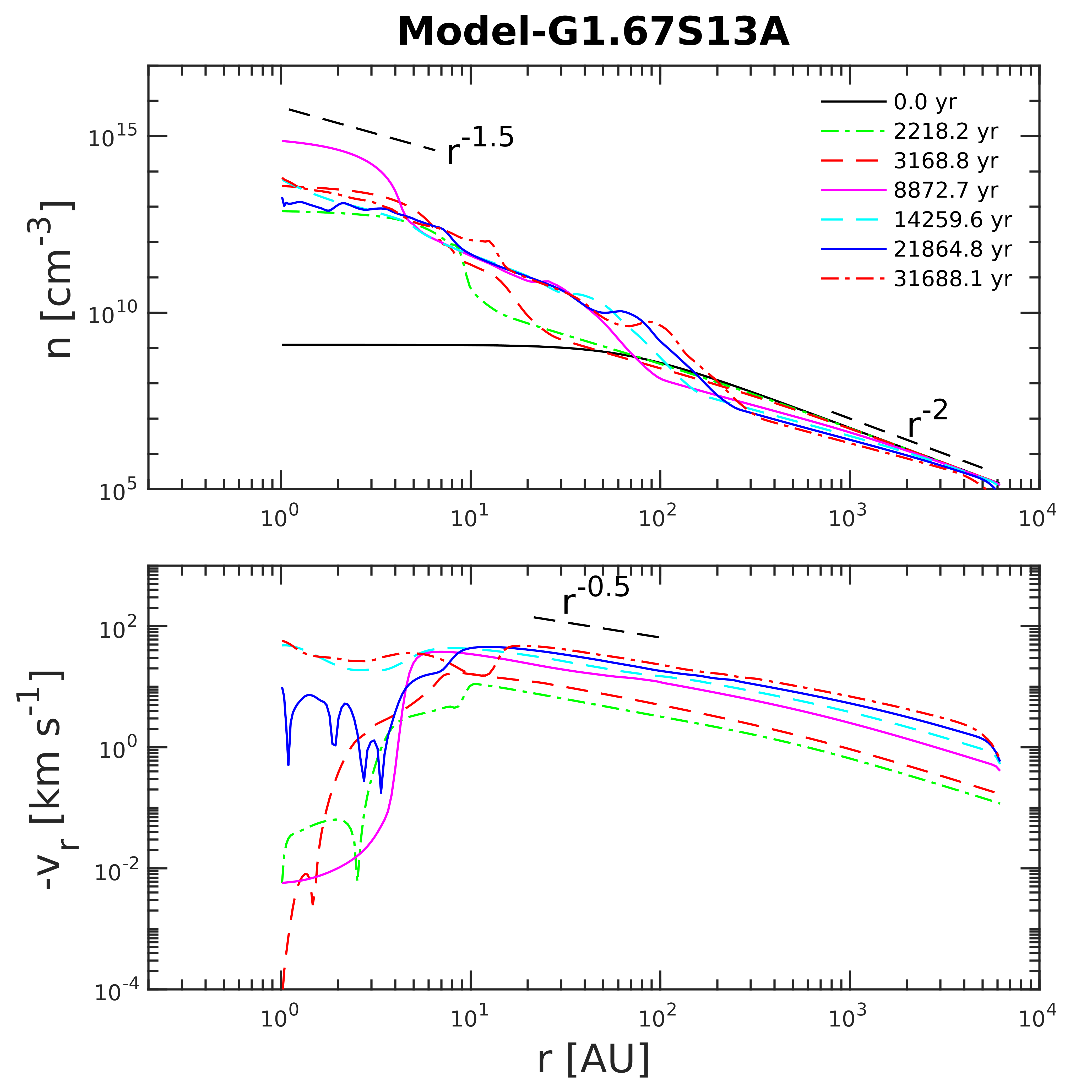}{0.55\textwidth}{(b)}
}
\vspace{-0.2cm}
\caption{Radial profiles of number density and radial infall velocity at the equatorial plane corresponding to distinct evolutionary time instances for Model-G1.67S15A (left panel) and Model-G1.67S13A (right panel).}
\label{fig:1D-radS1513}
\end{figure*}

\begin{figure*}[!ht]
\gridline{\fig{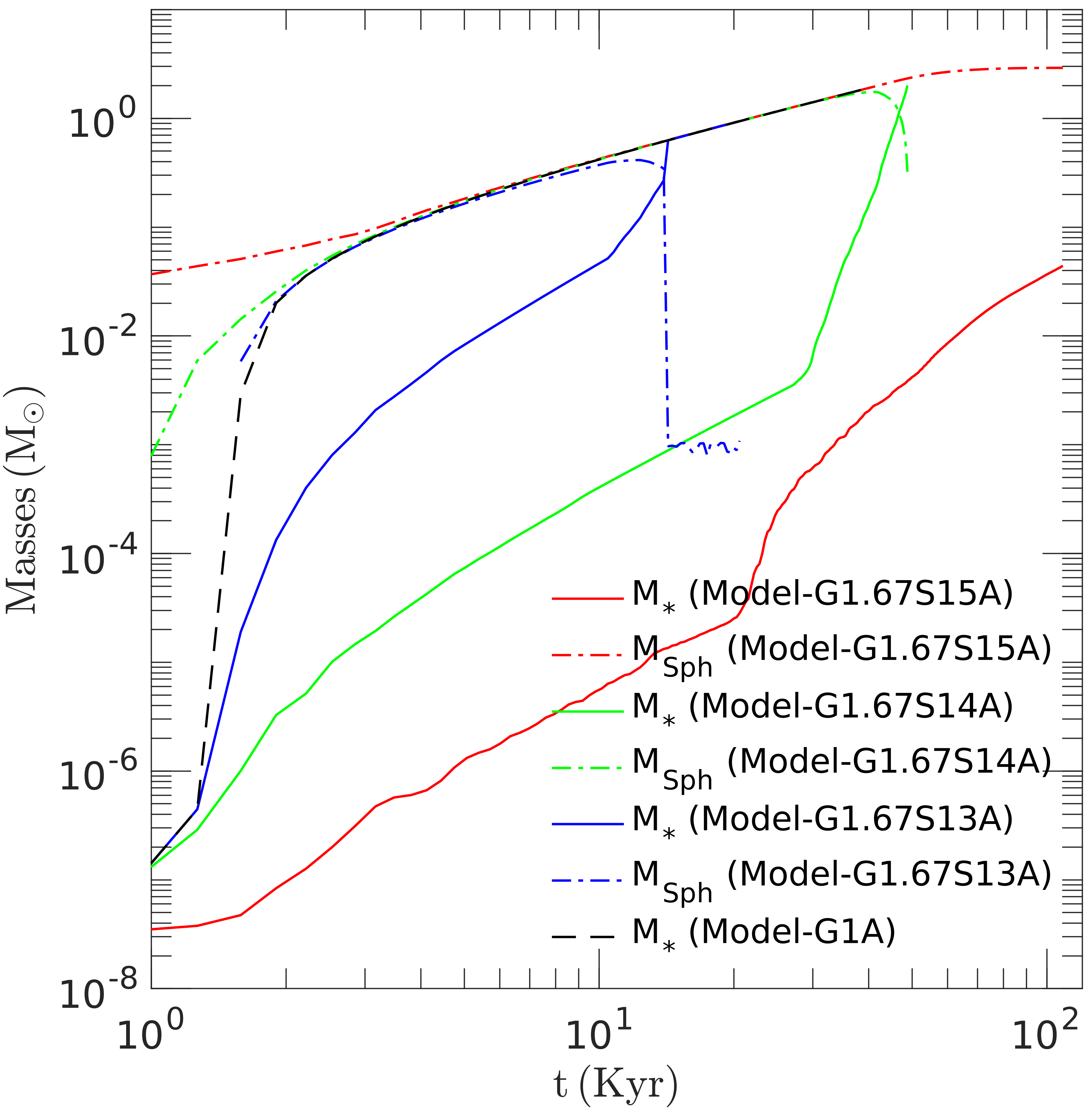}
{0.45\textwidth}{(a)}
\fig{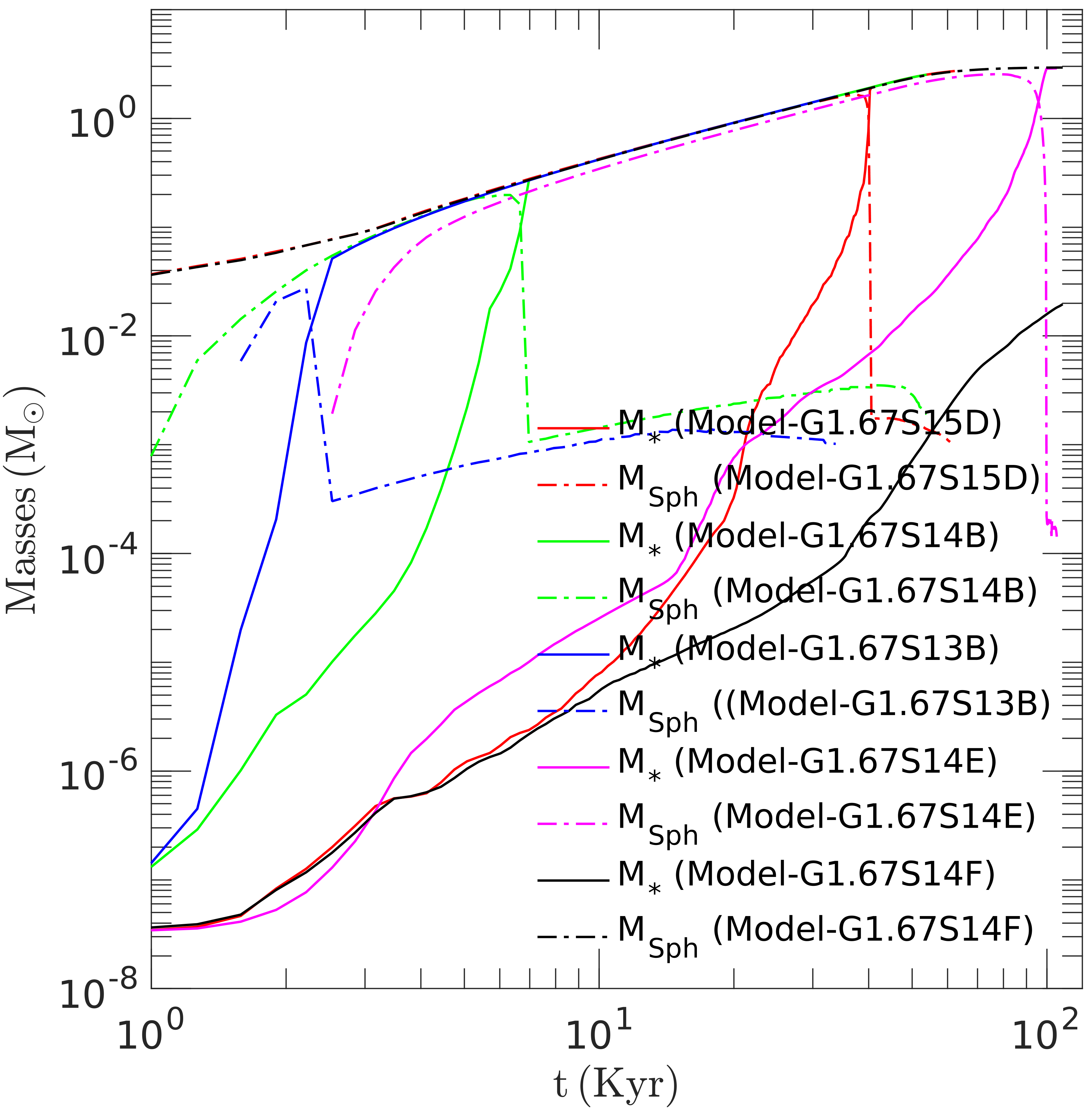}{0.45\textwidth}{(b)}
}
\caption{Mass of the central sink particle point mass($M_{\ast}$) and mass of the condensed sphere (${\rm M}_{\rm sph}$) i.e., related to mass contained with $\rho > \rho_{\rm stiff}$,  with time for the non-isothermal magnetized collapse models. 
(a): models with the fiducial nonideal MHD strength for different $\rhostiff=$ \, $10^{-15}\, {\rm g. {cm}^{-3}}$ (Model-G1.67S15A), $10^{-14} \, {\rm g. {cm}^{-3}}$ (Model-G1.67S14A), and $10^{-13}\, {\rm g. {cm}^{-3}}$ (Model-G1.67S13A); (b): the corresponding models with a higher nonideal MHD effects (Model-G1.67S15D, Model-G1.67S14B, Model-G1.67S13B) for the respective cases of $\rho_{\rm stiff}$, and additionally for a model with a higher initial $c_s$ (Model-G1.67S14E), and a higher initial magnetic field (Model-G1.67S14F). } 
\label{fig:mass-Gamma1.67}
\end{figure*}

The top and bottom panels of Figure \ref{fig:1D-radS1513}(a) and (b) show the radial profiles of number density and the radial velocity ${\rm v}_{\rm r}$ at distinct evolutionary times for MHD collapse with Model-G1.67S15A (left column) and Model-g1.67S13A (right column), respectively. 
In Figure \ref{fig:1D-radS1513}(a), corresponding to Model-G1.67S15A, there is no noticeable developements in the gradient of the density profile other than $\rho \propto r^{-2}$ in the outer parts with a flattening across the inner regions once the collapse commences. 
This states that the infalling material is prevented from moving further inward; instead, it is consumed within the quasi-equilibrium condensed sphere that acts as a barrier to the infall, thereby stopping the 
central point mass to form and grow.  
The radial profile of $v_r$ shows that at the scales of $\sim 5000 \, {\rm au}$ the material gradually begins to infall nearly at a subsonic speed. 
Subsequently, the infall becomes supersonic and eventually halts by rigid sonic wall formed at the outer boundary of such a condensed sphere that prevents further transporting the matter to the center.

On the contrary, Figure \ref{fig:1D-radS1513}(b) corresponding to Model-G1.67S13A, shows that the choice of a higher $\rho_{\rm stiff}$ accelerates the collapse significantly due to a smaller condensed sphere in the first place that shrinks soon after the collapse commences. 
It seems that the condensed sphere behaves more like a leaky sphere, not as rigid as observed in cases with lower $\rho_{\rm stiff}$ (for example, with Model-G1.67S15A or Model-G1.67S14A) than for Model-G1.67S13A. 
The density level rises within the inner region of $\lesssim 10 \, {\rm au}$ at a relatively faster rate than that with lower $\rho_{\rm stiff}$ and the radial expansion wave starts propagating outward developing a power-law behavior of $\rho \propto r^{-3/2}$, eventually aiding the formation of the point mass 
at the center.
At an evolved stage, after the condensed sphere shrinks down, the infall velocity profile qualitatively behaves in the similar way as in the isothermal magnetized collapse with Model-G1A (see Sec.\ \ref{sec:IsothermalCollapse} and Figure \ref{fig:1D-densVr}).


The Figure \ref{fig:mass-Gamma1.67}, in general, illustrates the mass evolution over time for the protostar, $M_{\ast}$  (presented by the solid line) and the condensed sphere, $M_{\rm sph} (\rho > \rho_{\rm stiff})$ (presented by the dashed-dotted line of the same color) for the magnetized collapse models with $\Gamma=5/3$. 
In Figure \ref{fig:mass-Gamma1.67}(a), we observe scenarios with the variation in $\rho_{\rm stiff}$. 
For Model-G1.67S15A, 
negligible growth is seen in the temporal evolution of the central sink particle 
mass ($M_{\ast}$). 
However, as the stiffening threshold increases for example, with  Model-G1.67S14A and Model-G1.67S13A, 
the evolution of 
point mass accelerates along with a noticeable enhancement occurring 
upon the collapsing of such condensed spheres into the sink particle.
Eventually, the time evolution of $M_{\ast}$ merges with that of the case with the fiducial isothermal magnetized collapse at a much earlier time as seen in Figure \ref{fig:mass-Gamma1.67}(a).

Furthermore, with increasing strengths of nonideal MHD effects, 
in Figure \ref{fig:mass-Gamma1.67}(b), we notice a significant increase in 
the mass of the sink particle at an earlier time compared to its corresponding lower nonideal MHD counterparts.
For Model-G1.67S15D, 
the higher resistivity/diffusivity leads to a notable acceleration in $M_{\ast}$ at an earlier stage. 
Similarly, for Model-G1.67S14B and Model-G1.67S13B, i.e., the models with higher Ohmic resistivity than in Model-G1.67S14A and Model-G1.67S13A, respectively, time evolution of $M_{\ast}$ increases at a much faster rate.
On the other hand, for Model-G1.67S14E, growth of $M_{\ast}$ delays than in Model-G1.67S14A as a consequence of greater thermal support due to initial choice of higher sound speed.
For Model-G1.67S14F, the enhanced magnetic fields acts as a counterbalance to the gravitational collapse, thereby slowing down the initial accretion process 
into the central point mass.


\subsubsection{Evolution of infall velocity and B-fields}

\begin{figure*}[ht!]
\gridline{\fig{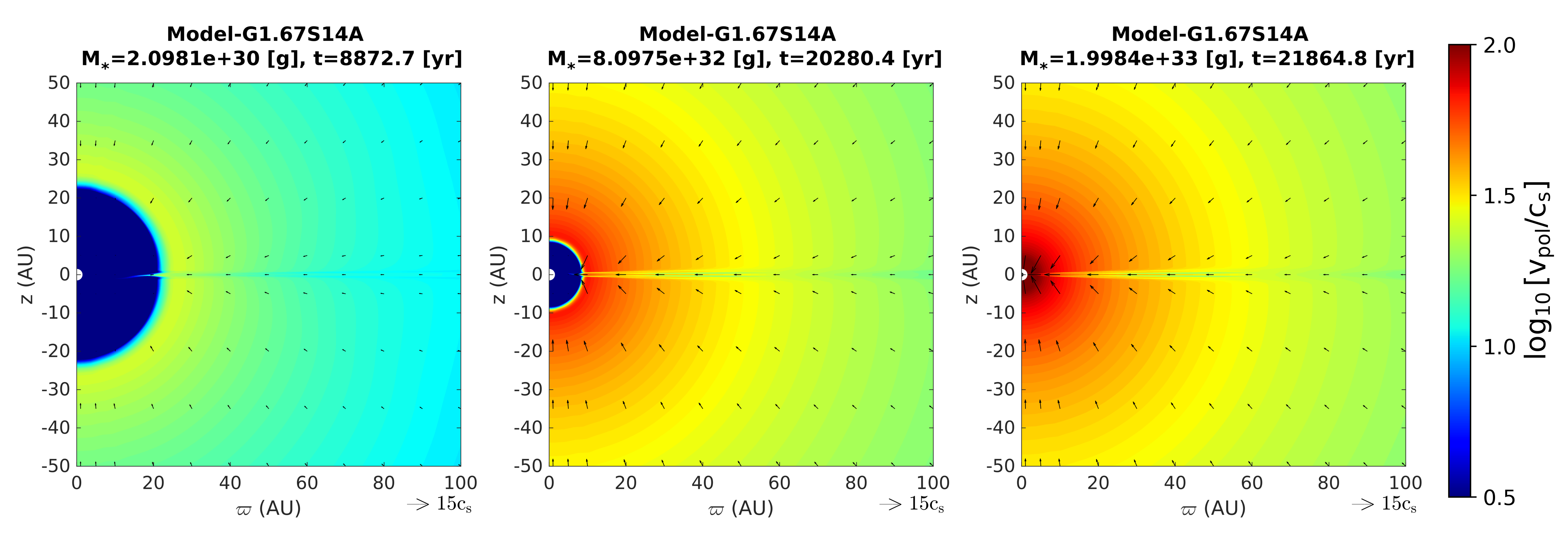}{\linewidth}{}
}
\vspace{-0.2cm}
\caption{Dynamical evolution of 
normalized poloidal velocity w.r.t to the initial sound speed $v_{\rm pol}/c_s$ (given $c_s=0.2 \kms$) with overlaid poloidal velocity vectors in black color as depicted for the magnetized isothermal collapse with Model-G1.67S14A on scales of $100 \,  {\rm au} \times 100 \, {\rm au}$ at distinct evolutionary time instances.}
\label{fig:G1.67S14vpol}
\end{figure*}

\begin{figure*}[ht!]
\gridline{\fig{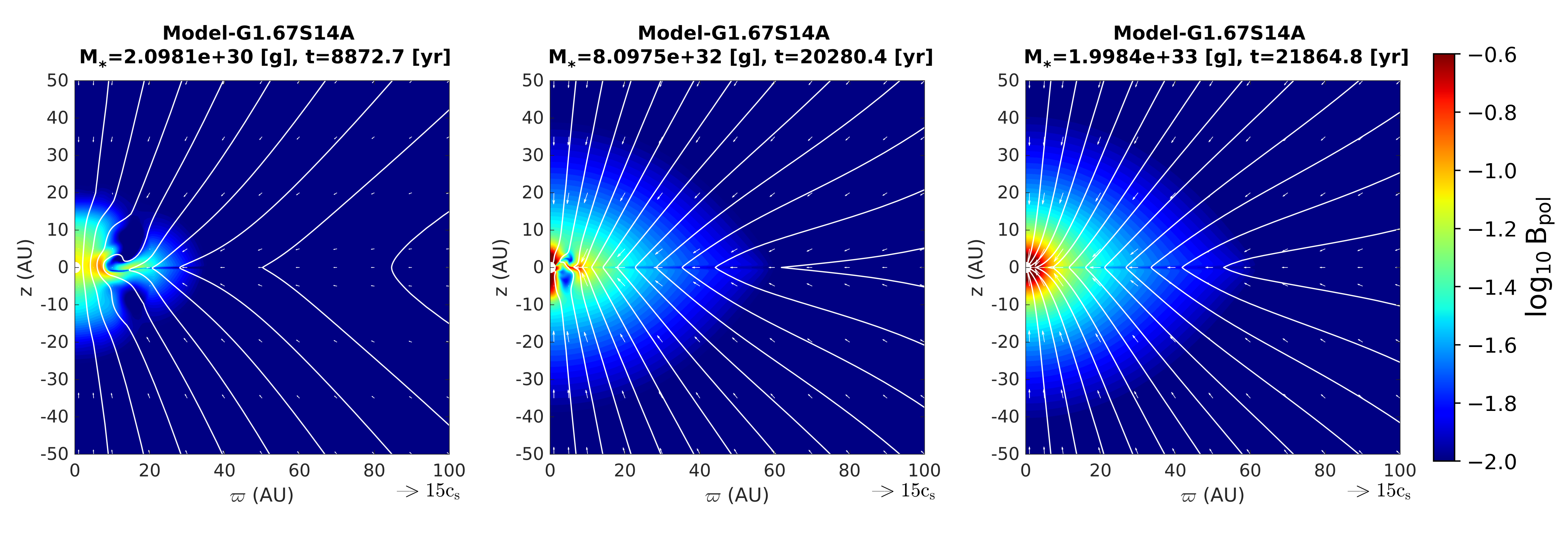}{\linewidth}{}
}
\vspace{-0.2cm}
\caption{Dynamical evolution of poloidal magnetic fields (expressed in Lorentz-Heaviside units) with overlaid poloidal velocity vectors as well as the magnetic flux lines in white color as depicted for the magnetized isothermal collapse with Model-G1.67S14A on scales of $100 \,  {\rm au} \times 100 \, {\rm au}$ at distinct evolutionary time instances. }
\label{fig:G1.67S14Bpol}
\end{figure*}

\begin{figure*}[ht!]
\gridline{\fig{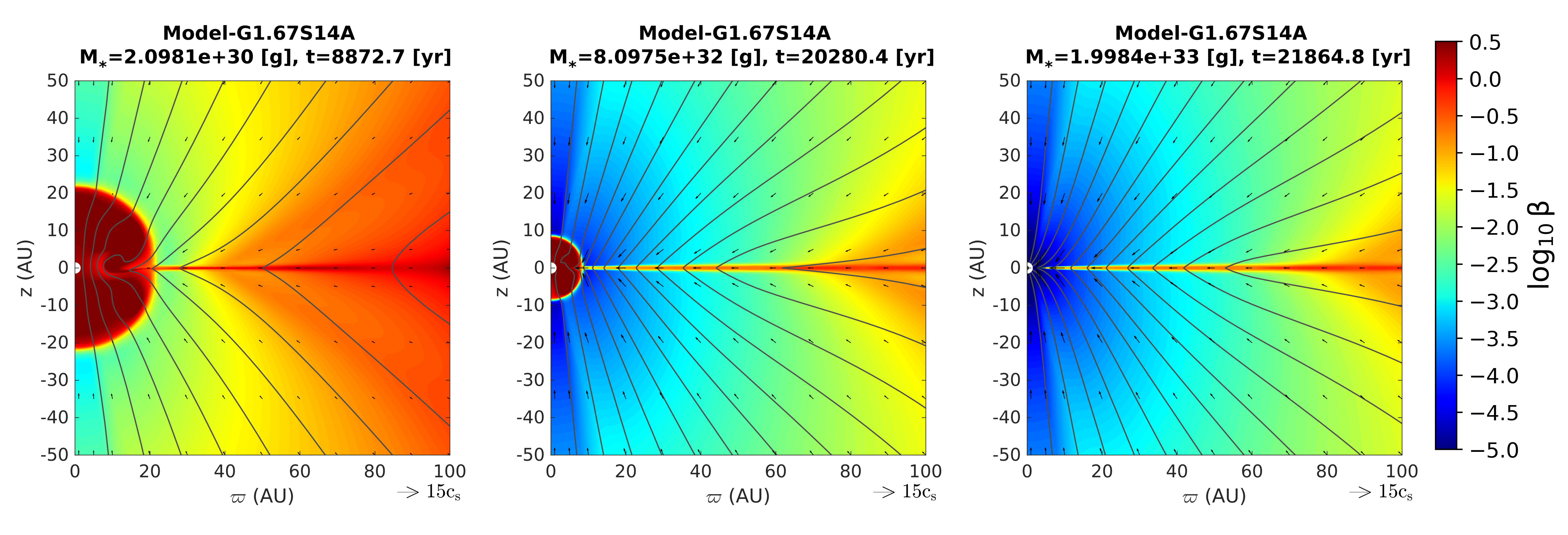}{\linewidth}{}
}
\vspace{-0.2cm}
\caption{Dynamical evolution of plasma-$\beta$ with overlaid poloidal velocity vectors as well as the magnetic flux lines in black color as depicted for the magnetized isothermal collapse with Model-G1.67S14A on scales of $100 \,  {\rm au} \times 100 \, {\rm au}$ at distinct evolutionary time instances. }
\label{fig:G1.67S14vAlfven}
\end{figure*}

In this section, we discuss the collapse dynamics in terms of the poloidal infall velocity and intensity of the poloidal magnetic fields. 
Figure \ref{fig:G1.67S14vpol} shows the intensity map of poloidal velocity normalized to the units of initial $c_s$
at different evolutionary stages for Model-G1.67S14A (see Sec.\ \ref{sec:gamma5by3S14}). 
Please refer to the first row of Figure \ref{fig:g1.67S14ABEF} for the corresponding density maps. 
In Model-G1.67S14A, the enduring presence of the long-lived condensed sphere is noteworthy. 
Infall persists beyond it at the outer layers, abruptly halting at its edge to create an impenetrable sonic wall, a trait not present in the isothermal magnetized model. 
Initially, this transition occurs from supersonic to subsonic speeds. 
Over time, it evolves into a shift from highly supersonic to subsonic speeds, accentuating the prominence of the shock surface at the outer edge of the condensed sphere. 
For Model-G1.67S14A, once the condensed sphere diminishes over time prompts a sharp increase in the $M_{\ast}$ (see Figure \ref{fig:mass-Gamma1.67}(a)).

Figure \ref{fig:G1.67S14Bpol} shows the intensity map of poloidal magnetic fields ($B_{\rm pol}$) for Model-G1.67S14A, 
at different evolutionary stages of MHD collapse that reveals a development in the spatial map of  $B_{\rm pol}$ that starts to forming at the edge of the condensed sphere in the equatorial plane, where the pseudodisk meets its outer boundary. 
With the shrinking of the condensed sphere, the $B_{\rm pol}$ intensifies, allowing it to pull the magnetic fields anchored to the matter further inward. 
As the pseudodisk forms from the outer edge of the condensed sphere, it enables transporting more matter and magnetic fields to the it. 
As a result, the magnetic field lines start to pinching from near the outer boundary of the such condensed spheres, which were all the way straightened out towards the inner part. 
The variation in the accumulation of the magnetic flux eventually leads to a spatial distortion in $B_{\rm pol}$ that gives rise to a magnetic plummes connecting the pseudodisk to the outer edge of the condensed sphere for channeling matter and magnetic fields towards the inner region.  
During this process, 
the condensed sphere that was in quasi-equilibrium earlier, loses its stability against the collapse due to more matter infalling into it, causing further contraction. 
Upon the shrinking of the condensed sphere (within 1 au),  there is a gradual increase in $B_{\rm pol}$ from outer to inner region, similar to that of the isothermal magnetized collapse case (refer to Figure \ref{fig:G1A_Bpol_s100}).

Figure \ref{fig:G1.67S14vAlfven} shows the intensity map of plasma-$\beta$ 
that illustrates the dynamic interplay between gas pressure and magnetic pressure for the collapse with Model-G1.67S14A. 
Within the condensed sphere and pseudodisk, 
plasma-$\beta$ exhibits the relative increment of thermal pressure over magnetic pressure.
The magnetic pressure substantially exceeds gas pressure as the condensed sphere gradually shrinks and poloidal magnetic fields intensity increases in the central regions. 
Once the condensed sphere completely shrinks within 1 au, the distribution of plasma-$\beta$ becomes similar to that of the case of isothermal collapse (see Figure \ref{fig:G1A_plasmabeta_s100}), following the same features, where the magnetic pressure is greater than the thermal pressure within the inner region.



In a series of ancillary test cases, we studied the magnetized collapse cases with an extremely stiff EOS with $\Gamma=2$ as listed in Table \ref{tab:MODELS}. 
The different exploration study cases\textemdash, such as variations in $\rhostiff$, different strengths of nonideal MHD effects, different initial sound speed, and magnetic field strength\textemdash are qualitatively the same as those used for the respective cases with $\Gamma=5/3$. 
The latter part of Table \ref{tab:MODELS} presents the visual type of spheres found with $\Gamma=2$. 
Because of the significantly greater stiffening in EOS with $\Gamma=2$, there appears to be a more rigid sonic wall, leading to a much larger condensed sphere in each of the cases compared to the respective cases with $\Gamma=5/3$.

\section{Discussions}\label{sec:discussions} 
Our numerical exploration of the magnetized gravitational collapse of a prestellar cloud core encompassing a wider range of EOSs provides insights on constraining the choice of EOSs that facilitates necessary cooling, which in turn, allows 
the direct mass growth of the point mass at the center.
As an outcome of the magnetized collapse, the pseudodisk forms in the equatorial plane, through which the material is channeled to the central point mass.

The choice of an isothermal EOS i.e., with $\Gamma=1$ leads to the most natural outcome for the magnetized inside-out collapse in forming the central point mass
(refer to Sec \ref{sec:IsothermalCollapse} and Fig.\ \ref{fig:E0-A1-s500}, \ref{fig:1D-densVr}, \ref{fig:masscompare-G1} for further details)   
where the sink particle 
steadily builds up its mass, at a rate of the order of $c_{\rm s}^3/G$. 
The dimensional mass accretion rate for the case with $\Gamma=1$ obtained from our studies seems to be approximately 26 times higher the theoretical value of $c_{\rm s}^3/G$ likely due to several factors, such as the choice of numerical factor associated with the characteristic constant used in the initial density profile of SIS collapse model (see further in Appendix \ref{sec:massApp}), the initial mass of the prestellar core, and 
a large magnetically supercritical core (see further in Appendix \ref{sec:lambdaAPP}). 

It is noticed that as the runaway isothermal magnetized inside-out collapse commences,
upon the point mass formation,
a radial expansion wave having a power law of $r^{-3/2}$ 
begins (see Figure \ref{fig:1D-densVr}). 
The head of the expansion wave propagates nearly at the magnetosonic speed as shown in Fig. \ref{fig:G1A_vmsbypol_s6000}. 
This behaviour is consistent with the depiction as first demonstrated in \cite{AdamsShu2007} with a focus on the pivotal stage, transitioning from the prestellar to protostellar phase. 
In addition to that, the evolution of the large scale spatial distribution of the ratio of the magnetosonic speed to that of poloidal and density as shown in Fig. \ref{fig:G1A_vmsbypol_s6000} and \ref{fig:G1A_dens_s6000}, attributes the expanding outward progression of the radial expansion wavefront with the ongoing inside-out collapse, as characterized by the white contour presenting $v_{\rm ms}= v_{\rm pol}$ (see further in Sec.\ \ref{sec:IsothermalCollapse}). 
The edge of the white contour basically separates the region of rapid infall associated with the pulled-in magnetic field lines from the region associated with the straight field lines where the collapse has not begun yet at least at the rate observed inside, indicating the magnetic decoupling due to the nonideal MHD effects. 
In the process of the magnetized collapse, in particular, under the presence of higher nonideal MHD effects, there several density branches in the collapse flow are noticed at an evolved stage of collapse at around scales ranging from 1000 au down to 50-100 au, that may again result from the enhanced decoupling of magnetic fields (refer to Fig.\ \ref{fig:E0-BD-s100} and Fig.\ \ref{fig:ModelG1BD-dens}).

The semi-analytical treatment of \cite{AdamsShu2007} demonstrates that in the case of a magnetized collapse under the condition of flux-freezing, the collapse is likely to encounter a gravomagneto catastrophe (often described as the formation of a split-monopole at the center); however, this issue is avoided in the presence of nonideal MHD effects. 
Due to the inherent nature of any analytical work, it remains unaffected by the constraints and limitations imposed by any computational grids. 
Our numerical results incorporating ambipolar diffusion and Ohmic dissipation show that the presence of nonideal MHD effects prevents the formation of such split-monopole during the magnetized collapse by 
resolving the current sheet of finite thickness (see further in Sec.\ \ref{sec:separatrix}). 
We noticed that the higher the magnitude of nonideal MHD effects, the better the resolution of the current sheet, the more effective the diffusion and dissipation of magnetic fields, a decoupling leading to the avoidance of the gravomagneto catastrophe from happening, 
while allowing the collapse to continue.

In addition to the exploration of the magnetized isothermal (i.e., an EOS with $\Gamma=1$) collapse, we further extrapolate our investigations for the nonisothermal ($\Gamma>1$) magnetized collapse. 
Our numerical findings complemented by the magnetized virial theorem (refer to Eq.\  \ref{eq:Gammavirial} and Sec.\ \ref{sec:virial}) reveal that, the nonisothermal magnetized collapse models with a $\Gamma$ no harder than 4/3, 
allow the direct mass growth of the central sink particle without producing any transient condensed spheres (see Fig.\ \ref{fig:globalmodels} and Sec.\  \ref{sec:polytropes}). 
Under such collapse cases, the fundamental infall features facilitate
the central sink particle to steadily build up its mass with a spatially constant mass accretion rate that has a scale of the order of $c_{\rm s}^3/G$ similar to the isothermal counterpart (see Fig.\ \ref{fig:masscompare-G1}).

Furthermore, the numerical magnetized collapse models with a $\Gamma$ harder than 4/3 yield a condensed sphere at the center of collapse flow (see Fig.\ \ref{fig:globalmodels}, \ref{fig:g1.67S14ABEF}, \ref{fig:g1.67S15AD}, \ref{fig:g1.67S13A} and Sec.\ \ref{sec:polytropes}, \ref{sec:gamma5by3}), which does not allow the direct infall into the sink particle 
and it causes a reduction in its the mass accretion rate, thus delaying the collapse. 
This is also noticed from the initial vertical jump in the profile of $M_*$ vs $t$ for the respective collapse models with a  $\Gamma$ harder than 4/3 (e.g., refer to Fig.\ \ref{fig:masscompare-G1} for the cases with $\Gamma=1.4\,,1.5$ and Fig.\ \ref{fig:mass-Gamma1.67} for the case with $\Gamma=5/3$). 
The stiffer the $\Gamma$, the more pronounced this vertical jump becomes, resulting in a greater delay in the direct mass growth of the central sink particle.
Although suggested by the magnetized virial theorem (Eq. \ref{eq:Gammavirial}), our numerical results show that such condensed spheres even obtained with a global $\Gamma$ harder than 4/3 are not eternal; and they eventually collapse into the central sink particle.
The theoretically predicted lifetime of such condensed spheres with the initial sizes of $\sim$~10-100~au varies on the order of 1000-30000 yr. 
This depends largely on several aspects, including physical processes such as nonideal MHD effects and numerical factors, such as the choice of $\rhostiff$ and the use of a sink particle (physical inner boundary) 
in the computational domain that allows the infalling matter to sink in, neither of these accounted for by the theoretical considerations of the magnetized virial theorem as stated in Sec.\ \ref{sec:virial}. 
For a given magnetic field strength, the harder the EOS, the longer these spheres survive before collapsing into the sink particle.
Typically, the value of $\rhostiff$ coincides with the outer boundary of such condensed spheres and acts as a measure of the size for these condensed spheres.  
For a harder polytropic EOS, the size of such condensed spheres becomes larger as a harder EOS tends to occupy a greater volume by providing more resistance against the gravitational compression (e.g., Fig.\ \ref{fig:globalmodels}).

In the context of collapse, the production of such condensed spheres resembles the characteristics of the numerical results, appearing as ``first hydrostatic cores'' (or first cores), obtained in works associated with the choice of certain barotropic EOS with a $\Gamma$ as hard as of 5/3, as found in a fraction of the literature on the HD/MHD collapse simulations \citep[][etc.]{Larson1969, Tomisaka1995, Tomisaka1996, Tomisaka1998, Tomisaka2002, Machida+2004, Machida+2005a, Machida+2005b,  Machida+2008a, Dapp+2012, Wurster2016, Wurster+2021}, including the following works on the radiation hydrodynamics and/or radiation magnetohydrodynamics \citep[e.g.,][etc.]{Masunaga+1998, MasunagaInutsuka2000, WhitehouseBate2006, Tomida+2010a, Vaytet+2012, Vaytet+2013, Vaytet+2018, Wurster2016, Wurster+2018, Wurster+2021}.
Most of these steep-$\Gamma$ works
have been explored with a choice of $\Gamma=5/3$ for $\rho>\rhostiff$, invoked through a rationale of high opacity along with very large column density, conditions expected by them to lead to sufficient radiation heating to halt the collapse, followed at even smaller radii or higher densities by a softening of the EOS to a barotropic index of $4/3$ or even $7/5$ for continuing the further mass growth of the central point mass sink particle.
From our current exploration, we found that the  evolution of the magnetized collapse with a $\Gamma$ no harder than 4/3 is qualitatively similar to that of an isothermal magnetized collapse.
Thus the choice of a soft isothermal EOS with a $\Gamma$ no stiffer than 4/3 is well justified, because until a sufficient mass (or density) accumulates to form a star at the center, there may not no apparent physical 
reason for stiffening the EOS, which is also in conjunction with the observed establishment of abundant protostars typically with sizes comparable to that of several solar radii.

During the magnetized collapse, the density structures develop (as seen in Fig. \ref{fig:E0-BD-s100}, \ref{fig:ModelG1BD-dens}) in various dense regions of the infalling flow, including the regions above and below  
the pseudodisk and the central point mass. 
These regions can be further resolved by Adaptive Mesh Refinement (AMR) techniques for magnetized flows, such as those presented in e.g.\ Athena++ \citep{athena_AMR} and Sadhana \citep{sadhana_AMR}, which can focus the numerical resolution into the regions of specific interest, bearing particularly distinct magnetic physics. 
In the current numerical setup, the density distribution around the wedge along the north and south poles may appear to be lower than the rest of the regions as the collapse progresses. 
The numerical configuration has been designed this way to avoid the generation of ``hot zones" where the Courant conditions demand such a small timestep that the simulation halts prematurely. 
Our current numerical setup ensures the matter to concentrate near the equatorial plane as a natural outcome of the magnetized collapse. 
In addition to that, choice of constant Ohmic resistivity may seem to overestimate the field strength above and below the equatorial plane. However, as this region is nearly current-free, it does not significantly affect our results.

The process of transforming a piece of interstellar cloud to a star involves changes in density of more than twenty orders of magnitude during the phase of runaway collapse with the isothermal EOS, a heavy-duty computational chore at that time of Larson-Pentson solution \citep{Larson1969,Pentson1969} was challenging to carry out accurately in such a computational grid setup that did not incorporate a central sink particle even in its simplest spherically symmetric context, particularly in cases where the numerical resolution is lacking \citep[refer to \S\ 19 of][]{ShuBookVol2}. 
The uncertainty in grid resolution can impact the appearance/existence of such theoretical condensed spheres that hinders the direct infall into the central sink particle and delays the collapse. 
The numerical results obtained by \cite{Larson1969} diverging from the analytical predictions of \cite{Shu1977}, may indicate the requirement for a more refined approach to deal with the radiation accretion shock at the protostar's surface \citep[see further in \S\ 19 of][]{ShuBookVol2}, such as the study of the
inner envelope structure as demonstrated in \cite{Stahler+1980a,Stahler+1980b,Stahler+1981} using semianalytical methods or even the observed radiation flux from the astrochemical lines owing to non-isothermality. 
However, the exploration of this radiation accretion shock is beyond our scope of the current study.

Outside the purview of this work, and following \cite{Stahler+1980a,Stahler+1980b,Stahler+1981},
at the scales of protostar's surface, which is likely several solar radii, the heating from the accumulated material within may begin to significantly influence the EOS by transitioning to the regime of non-isothermality. 
However, this regime of non-isothermality may remain confined to within the protostar's surface only, similar to the case of a chromosphere and certainly does  not impede the mass infall during collapse. 
As the protostar evolves, its mass and luminosity increase significantly, it may then become pertinent to consider factors like radiation energy, opacity, 
deuterium burning, and other relevant processes. 
The high-density gas primarily composed of molecular hydrogen, is supposed to form atomic hydrogen within near proximity of the protostar's surface itself once it attains sufficient density and temperature for undergoing dissociation and may finally settle into the zero age main sequence (ZAMS). 


\section{Conclusions} \label{sec:conclusions}
In this study, we present the results of numerical experiments conducted using two-dimensional axisymmetric nonideal MHD (incorporating ambipolar diffusion and Ohmic dissipation) simulations of magnetized collapse of a non-rotating prestellar cloud core encompassing a wide range of EOSs, 
with the aim of constraining the 
choice of EOS for 
allowing the direct mass growth of the central point mass (sink particle).  
We primarily investigate the impact of soft and hard EOS on the infall along with the dynamical evolution of magnetic fields during the collapse from the molecular cloud scale (of $\sim \,0.1 \,{\rm pc}$) down to $\sim\, 1 \,{\rm au}$ in the near proximity of the central point mass.  
The key findings of our study are listed below: 
\begin{enumerate}
    \item Our numerical results complemented with magnetized virial theorem (Eq. \ref{eq:Gammavirial} and further details in Sec.\ \ref{sec:virial}), reveal that the magnetized gravitational collapse models with a $\Gamma$ softer than 4/3 facilitates the dynamical inside-out collapse of the prestellar cloud core 
     that allows the direct growth of the central point mass.
    
    \item The central point mass for the collapse cases with a $\Gamma$ softer than 4/3, steadily builds up its mass from the infalling envelope, in which the mass accretion rate 
    into the sink particle has a scale of the order of $c_{\rm s}^3/G$ (Figure \ref{fig:masscompare-G1}). 
    Consequently, as a result of the direct 
    mass growth found with the soft nonisothermal models (with a $\Gamma<4/3$), the overall dynamical sequence of collapse (Figure \ref{fig:globalmodels} and Sec.\ \ref{sec:polytropes}) qualitatively follow the similar infall features as that of the isothermal magnetized collapse (Figure \ref{fig:E0-A1-s100} and Sec  \ref{sec:IsothermalCollapse}).
    
     \item From the criticality at $\Gamma=4/3$ as obtained from the magnetized virial theorem (Eq. \ref{eq:Gammavirial} and Sec.\ \ref{sec:virial}), it can be shown that only a $\Gamma$ harder than $4/3$ 
     does not provide sufficient cooling to allow the direct growth of the central point mass (see Figure \ref{fig:globalmodels} and Sec.\ \ref{sec:polytropes}; Figure \ref{fig:g1.67S14ABEF} and Sec.\ \ref{sec:gamma5by3}) and it causes a reduction in the mass accretion rate into the central sink particle, delaying the collapse. 
     This is noticed from the initial sharp vertical jump in the profile of $M_*$ vs $t$ for the collapse models with a $\Gamma$ harder than 4/3 (e.g., Fig.\ \ref{fig:masscompare-G1} and \ref{fig:mass-Gamma1.67}). 

    \item In the magnetized collapse, a pseudodisk, a highly flattened quasi-equilibrium dense structure forms in the equatorial plane resulting from the dragging in of the cloud magnetic field, that plays out a fundamental role to channeling material towards the 
    central point mass from the freely-infalling envelope. In addition to that, during the collapse with stronger nonideal MHD effects, density branching appears above and below the pseudodisk (see Figure \ref{fig:E0-BD-s100} and \ref{fig:ModelG1BD-dens}) due to the magnetic decoupling during the infall, delineating the streamlines of collapse flow anchored to the magnetic fields.
    
\end{enumerate}

The choice of an appropriate EOS is crucial for facilitating the collapse as it determines the amount of energy radiated away and ensures sufficient cooling in order to allow the 
direct mass growth of the point mass like protostar itself.
Our comprehensive study on the magnetized gravitational collapse can potentially offer fundamental insights on the significance of EOSs in allowing star formation.

\section*{Acknowledgments}
We thank the anonymous referee for the insightful comments that
improved the manuscript.
We thank Zhi-Yun Li and Susana Lizano for their valuable discussions. The authors acknowledge support for the CHARMS project from the Institute of Astronomy and Astrophysics, Academia Sinica (ASIAA), and from the National Science and Technology Council (NSTC) in Taiwan through grants 112-2112-M-001-030- and 113-2112-M-001-008-. The authors acknowledge the in-house access to high-performance facilities in ASIAA. We also thank the National Center for High-performance Computing (NCHC) of National Applied Research Laboratories (NARLabs) in Taiwan for providing computational and storage resources.


\appendix

\section{Virial theorem under an equilibrium configuration for a non-magnetized cloud} \label{sec:VirialChandra}

In this section, we portray the classical virial theorem under an equilibrium configuration for a non-magnetized cloud as discussed in detail by \cite{Chandrasekhar1939}. 
See further on p. 51-53 in \S\ 2 of \cite{Chandrasekhar1939}. 
For a non-magnetized cloud of particles under the consideration of an equilibrium configuration, let's consider $\mathcal{T}$ and $\mathcal{W}$ are total kinetic energy the total gravitational potential energy.
Note that, in this section we follow the notation as \cite{Chandrasekhar1939}'s. 
Here, $\mathcal{T}$ includes both thermal and bulk components. 
The bulk component happens to be zero under an equilibrium configuration. 
For such a cloud of particles in the limit of negligible external forces, under the consideration of the steady state, it can be written as follows 
\begin{equation}
    2\mathcal{T} + {\mathcal{W}} =0 \, .
    \label{eq:virial1}
\end{equation} 
Now, when the virial equation \ref{eq:virial1}
is applied to an ideal gas configuration in a gravitational equilibrium where $PdV$ work done arises from separating the pair of particles to infinity against the gravitational attraction. 
The contribution to the kinetic energy, $d\mathcal{T}$ of molecular motion is related to the internal energy, $dU_{\rm int}$. 
Thus for the whole configuration, 
\begin{equation}
\mathcal{T} = \frac{3}{2}(\Gamma -1)U_{\rm int} \, ,
\label{eq:eqA}
\end{equation} 
Therefore, rewriting the virial theorem (Equation \ref{eq:virial1}) using Equation \ref{eq:eqA} yields
\begin{equation}
    3(\Gamma -1)U_{\rm int} + \mathcal{W} = 0 \, .
    \label{eq:eqB}
\end{equation} 
Now, if $E_{\rm tot}$ is the total energy of the cloud of particles, implying $U_{\rm int} +\mathcal{W} = E_{\rm tot}$, then 
using Equation \ref{eq:eqB} to the definition of $E_{\rm tot}$ is reduced to the following 
\begin{equation}
    E_{\rm tot} = - (3\Gamma -4) \, U_{\rm int} = \frac{3\Gamma -4}{3(\Gamma-1)} \, \mathcal{W} \, . 
\label{eq:energyVirial}    
\end{equation} 
From Equation \ref{eq:energyVirial}, it is evident that for blob of gas with $\Gamma=4/3$ yields $E_{\rm tot}=0$, implying that a tiny radial expansion of the gas blob, accordingly would change its mass corresponding to a quasi-transition going from one equilibrium configuration to an adjacent configuration of equilibrium without the change of its total energy. 
It can be further realized that at $\Gamma=4/3$, a transition going from the regime of stability referring to $\Gamma >4/3$ (with $E_{\rm tot}<0$) to going to the regime of instability referring to $\Gamma <4/3$ (with $E_{\rm tot}>0$) must set in \citep[See further on p. 52-53 in \S\ 2 of][]{Chandrasekhar1939}.  
Therefore, there exists a criticality at $\Gamma=4/3$. 
Thereby for $\Gamma > 4/3$, Equation \ref{eq:energyVirial} corresponds to a negative $E_{\rm tot}$, implying that  the total energy in such a steady state is even less than that of in a state of diffusion at infinity. 
If such gas configuration with $\Gamma > 4/3$ contracts so that gravitational potential energy changes by an amount $\Delta\mathcal{W}$ (note that, $\Delta\mathcal{W} < 0$), followed by the corresponding changes of $\Delta E_{\rm tot}$ and $\Delta U_{\rm int}$ in the total energy and internal energy, respectively.
Hence, $- \Delta E_{\rm tot}$, the amount of energy lost by radiation can be quantified as from Equation \ref{eq:energyVirial} which follows
\begin{equation}
    - \Delta E_{\rm tot} = - \frac{3\Gamma -4}{3(\Gamma-1)} \Delta \mathcal{W} \, ,
\end{equation}
which is positive for a contraction  
It is straightforward to realize that the increase in the internal energy due to contraction of such a gas blob comes from the $PdV$ work done by the gravitational contraction. 
The fraction worth of $(3 \Gamma -4)/3(\Gamma-1)$ is lost in radiation and the remaining $\left[1 - (3 \Gamma -4)/3(\Gamma-1) \right]=1/3(\Gamma-1) $  can be used up in raising the temperature of the gas blob.
Hence, it follows that the stable (against the gravitational contraction) gas spheres can be obtained for only $\Gamma>4/3$. 
Whereas, no stable gas spheres form for $\Gamma<4/3$. 
Interestingly enough, Equation \ref{eq:energyVirial} yields that for $\Gamma=1$, $\mathcal{W}=0$ for any prescribed E, i.e., no stable configuration is possible, which can be realized from the SIS collapse as first introduced by \cite{Shu1977} in the context of star formation.


\section{Magnetic parametrization in the numerical setup}
\label{sec:lambdaAPP}
Given the initial density profile having the form of $\rho \propto r^{-2}$, we choose the following magnetic flux ($\Phi_{\rm mag}$ in Lorentz-Heaviside units) profile for setting up the initial vertical magnetic fields in ZeusTW code as follows
\begin{equation}
    \Phi_{\rm mag} = f_{\rm s} r_{\rm c}^2 \left[\left(r_{\rm max}- \sqrt{r_{\rm max}^2- \varpi^2} \right)  - r_{\rm c}\, {\rm arctan}\left(\frac{r_{\rm max}}{r_{c}}\right) + \sqrt{r_{\rm c}^2 + \varpi^2} \ {\rm arctan} \,\sqrt{\frac{r_{\rm max}^2 - \varpi^2}{r_{\rm max}^2 + \varpi^2}}
    \right]  \, ,
\label{eq:PhiEqun}
\end{equation}
where $f_s = 2 \sqrt{G/\pi} \rho_0 (\rho_{f}^{e_{\rho}/2})\mu_{\rm param}^{-1}$, 
$\rho_{f} = 1+ (r_{\rm max}/r_c)^2$, and $e_{\rho} = 2$. See Sec.\ \ref{sec:ICs} for the numerical values of $\rho_0$, $r_c$, and $r_{\rm max}$ used in our numerical setup.

The analytical form of the initial vertical magnetic field ($B_z$) profile as derived from Equation \ref{eq:PhiEqun} follows
\begin{equation}
B_z = \frac{f_s r_c^2}{\sqrt{r_{c}^2 + \varpi^2}} {\rm arctan}\,\sqrt{\frac{r_{\rm max}^2 - \varpi^2}{r_{\rm max}^2 + \varpi^2}} \, ,
\label{eq:BzEqun}
\end{equation}
where $B_z$ is defined in the Lorentz-Heaviside units. 
In this work, all the collapse simulations are performed for a moderately supercritical prestellar core for the sake of computational efficiency, where the magnetic parameter $\mu_{\rm param}$ used in the code is scaled by a $\pi$, implying $\lambda =\pi \mu_{\rm param}$.  
In our numerical setup, 
$\lambda$ refers to the definition of the normalized mass-to-flux ratio as defined in Equation (\ref{eq:lambdaEq}) (see in Sec.\ \ref{sec:ICs}).

\section{Treatment of Ambipolar diffusion in the numerical setup}
\label{sec:ambipolarApp}

In this section, we describe the numerical implementation of ambipolar diffusion (in Lorentz-Heaviside units) for setting up the current nonideal MHD simulations using the ZeusTW code. 

Since molecular clouds are lightly ionized, the neutrals predominantly feels the Lorentz force
through collisions with ions that slip past the neutrals in the
process known as ambipolar diffusion. 
By definition, ambipolar diffusivity can be defined as 
\begin{equation*}
    \eta_{\rm AD} = v_{\rm A}^2 \tau_{\rm ni} 
    = \frac{B^2}{\gamma_{\rm in} \rho_{\rm i}\rho} \, ,
\end{equation*}
where neutral-ion collision time $\tau_{\rm ni}=1/(\gamma_{\rm in} \rho_{\rm i})$, neutral-ion collisional coupling constant $\gamma_{\rm in}=3.5\times 10^{13} \, {\cm}^3 \, {\rm g}^{-1} \, {\rm s}^{-1}$, Alfv\'en speed $v_{\rm A}={B/\rho^{1/2}}$, $\rho=\rho_{i}+\rho_{n} \approx \rho_n$, given $\rho_i$, $\rho_n$ being the volume density of ions, neutrals. 
Furthermore, $\rho_{\rm i}= C \rho^{1/2}$, 
where $C = 3\times 10^{-16} \,{\cm}^{-3/2}\,{\rm g}^{-1/2}$ implying a fractional ionization $n_{i}/n_{n} \sim 10^{-7}$ for $n_{n}\sim 10^4 \, {\rm cm}^{-3}$, within the limits of $10^{-8}$ to $10^{-6}$ set by the observations of molecular cloud cores \citep[see further in \S 27 of][]{ShuBookVol2}. 
Setting the drag force ($-f_{\rm d}$) exerted by the neutrals on the ions equal to the Lorentz force ($f_{\rm L}$) felt by ions allows to solve for the drift velocity ($\vec{v}_{\rm d}$) that follows
\begin{equation}
    \vec{v}_{\rm d} \equiv \frac{1}{\gamma_{\rm in} \rho \rho_{i}} (\vec{\nabla} \times \vec{B}) \times \vec{B}  = q_{\rm AD} f_{\rm L} \, ,
\label{eq:driftvelocity}
\end{equation}
where $q_{\rm AD} = 1/(\gamma_{\rm in} C \rho^{3/2})$. Thus ambipolar diffusivity can also be rewritten as $\eta_{\rm AD} =q_{\rm AD} B^2$.

In our current numerical setup, we define $q_{\rm AD}$ as $q_{\rm AD, basic}$ that has the following form
\begin{equation*}
    q_{\rm AD, basic} = \frac{1}{\tilde{\gamma}_{\rm AD} \rho \tilde{\rho}_{\rm i}} \, ,
\end{equation*}
where
\begin{equation*}
\tilde{\rho}_{\rm i} = \rho_{\rm AD, i0} \left(\frac{\rho}{\rho_{\rm AD, n0}} \right)^{\alpha_{\rm AD}} \, ,
\end{equation*}
where $\rho_{\rm AD, i0} =1$, 
$\rho_{\rm AD, n0}=1$, 
$\alpha_{\rm AD}=0.5$, 
$\tilde{\gamma}_{\rm AD} = 3.15\times 10^{-2}$. 
Here, $\tilde{\gamma}_{\rm AD}$ is equivalent to the factor $\gamma_{\rm in} C$ as explained in Equation \ref{eq:driftvelocity} \citep[see further in][]{ShuBookVol2}.

Because of the high computational cost arising from the grid resolution and the CFL (Courant–Friedrichs–Lewy) condition, a ceiling for the computation of the ambipolar diffusion ($q_{\rm AD, ceiling}$) is used in ZeusTW code such that the effective ambipolar coefficient ($q_{\rm AD}$) in the code is calculated as follows  
\begin{equation*}
    q_{\rm AD} = {\rm min} (q_{\rm AD, basic}, q_{\rm AD, ceiling}) \, ,
\end{equation*}  
where
\begin{equation*}
    q_{\rm AD, ceiling} = \frac{\mathcal{C} (\Delta x)^2}{4 \Delta t_{\rm AD, floor} B^2} \,  ,
\end{equation*}
where the Courant number $\mathcal{C}=0.5$, $\Delta x$ is the minimum spatial distance given the axisymmetric spherical polar coordinates, and $\Delta t_{\rm AD, floor}$ provides an effective maximum to the value of ambipolar diffusion coefficient.


\section{Notes on the dimensionless coefficient of mass infall rate}\label{sec:massApp}

From the self-similarity solutions for the self-gravitating isothermal flow \citep[see Eq. (11) and (12) of][]{Shu1977}, 
the initial density distribution is taken to have the following form under asymptotic behaviours, where $\rho(r, 0) = c_s^2 A/( 4 \pi G)r^{-2} $. 
In the context of spatially constant mass infall rate of the form, $\dot{M}_\ast = m_0 {c_{\rm s}}^3/ G$, the constant $A$ in the asymptotic solutions is linked to the coefficient of the dimensionless mass infall rate, $m_0$ as explained for the SIS model \citep[see further in Sec. II(b) of][]{Shu1977}. 
Using the solutions in self-similar form, the corresponding asymptotic solution can be extrapolated at the initial instant at an extremely large self-similar coordinate, $x=r/(c_s t)$ that states, self-similar density $\alpha \sim A/x^2$, velocity $v \sim - (A-2)/x$, dimensionless mass infall rate $m\sim Ax$ as $x \rightarrow \infty$, where $A$ is a constant whose value must be greater than 2 in order to achieve the collapse (inflow) due to the local imbalance between gravity and pressure gradients. 
As a consequence of greater $M_{\rm core}$
in our numerical models, it results in higher values of $A=10.01$, for which the normalized mass accretion rate increases by a factor of $\sim 26$ compared to the conventional hydrodynamic case with $A=2.00+$. 
However, the resulting dimensionless coefficient $m_0$ for the magnetized cases with $A=10.01$, is still slightly larger than its respective purely hydrodynamic counterpart for the same $A$. 
The difference is not significantly large because of the choice of higher $M_{\rm core}$ as it expedites the computational runtime for the purposes of this numerical study.

\bibliography{myref}{}
\bibliographystyle{aasjournal}

\end{document}